\documentclass[aps,superscriptaddress,twocolumn,pre,longbibliography,floatfix]{revtex4-1}

\usepackage{amsmath,amsthm,amsfonts,amssymb,bm}
\usepackage{graphicx,color}
\usepackage{times,braket}
\usepackage[colorlinks=true,citecolor=blue,filecolor=blue,linkcolor=blue,urlcolor=blue]{hyperref}
\usepackage[caption=false]{subfig}
\usepackage{soul}
\usepackage{booktabs}

\allowdisplaybreaks

\begin{document}
\title{Optimal strategies for transient and equilibrium quantum thermometry using Gaussian and non-Gaussian probes}
\author{Asghar Ullah}
\email{aullah21@ku.edu.tr}
\affiliation{Department of Physics, Ko\c{c} University, 34450 Sar\i yer, Istanbul, T\"urkiye}
\author{M. Tahir Naseem}
\affiliation{Faculty of Engineering Science, Ghulam Ishaq Khan Institute of Engineering Sciences and Technology, Topi 23640, Khyber Pakhtunkhwa, Pakistan}
\author{\"Ozg\"ur E. M\"ustecapl\i o\u{g}lu}	
	\affiliation{Department of Physics, Ko\c{c} University, 34450 Sar\i yer, Istanbul, T\"urkiye}
	\affiliation{T\"UBİTAK Research Institute for Fundamental Sciences (TBAE), 41470 Gebze, T\"urkiye}
    
\date{\today} 
\begin{abstract}

We study temperature estimation using quantum probes, including single-mode initial states and two-mode states generated via stimulated parametric down-conversion in a nonlinear crystal at finite temperature. We explore both transient and equilibrium regimes and compare the performance of Gaussian and non-Gaussian probe states for temperature estimation. In the non-equilibrium regime, we show that single-mode non-Gaussian probe states—such as Fock, odd cat, and Gottesman-Kitaev-Preskill states—can significantly enhance the speed of estimation, particularly at short interaction times. In the two-mode setting, entangled states such as the two-mode squeezed vacuum, NOON state, and entangled cat state can enable access to temperature information at earlier times.
 In the equilibrium regime, we analyze temperature estimation using two-mode squeezed thermal states, which outperform single-mode strategies. We evaluate practical measurement strategies and find that energy-based observables yield optimal precision, population difference observables provide near-optimal precision, while quadrature-based measurements are suboptimal. The precision gain arises from squeezing, which suppresses fluctuations in the population difference.

\end{abstract}	
\maketitle
\section{Introduction}

Accurate temperature estimation plays a vital role in both foundational studies of quantum thermodynamics~\cite{DePasquale2016, Goold_2016} and the development of emerging quantum technologies~\cite{Levitin2022, Sarsby2020, individualProbes}. Quantum thermometry, which lies at the interface of quantum metrology and thermodynamics~\cite{DePasquale2018, Mehboudi_2019, Montenegro2025}, uses quantum parameter estimation to infer temperature with minimal disturbance to the system. A wide variety of quantum probes have been proposed, including single qubits~\cite{PhysRevA.91.012331, individualProbes, UllahSpinchains, PhysRevA.101.032112, PhysRevA.110.032605, PhysRevA.84.032105, PhysRevResearch.5.043184, PhysRevA.86.012125, Razavian2019, PhysRevA.99.062114, Singlequbit}, spin chains~\cite{Abiuso_2024, Aybar2022criticalquantum, RingQuant, Topological2025, ullah2025configuration}, harmonic oscillators~\cite{Campbell_2018, PhysRevA.96.062103, Campbell_2017, Ullah_2025MTCSs}, and many-body systems~\cite{PhysRevApplied.22.024069, Mok2021, PhysRevB.98.045101, Montenegro2025, PhysRevLett.132.240803}. Across these settings, performance is commonly benchmarked using the quantum Cramér–Rao bound, which is determined by the quantum Fisher information (QFI)~\cite{Mehboudi_2019}.
Numerous platforms, including collisional models and ultracold gases, have been explored for implementing quantum thermometry~\cite{PhysRevLett.123.180602, Purdy, PhysRevLett.122.030403}. Yet a key challenge persists: to engineer minimal, efficient quantum probes that can extract thermal information both rapidly and with high precision~\cite{individualProbes, Mehboudi_2019, PhysRevB.98.045101}.

One widely used strategy in quantum thermometry is to allow the probe to thermalize with the target system and then infer the temperature from its equilibrium properties, often through energy measurements~\cite{PhysRevA.82.011611, PhysRevA.97.063619, Carlos, Nguyen}. In this regime, the optimal probe effectively reduces to a two-level system with a non-degenerate ground state and a highly degenerate excited state~\cite{individualProbes}. Crucially, the choice of the initial state has no impact on the final outcome, leaving the advantages of quantum state preparation unexploited. In contrast, non-equilibrium thermometry extracts temperature information from the probe's transient dynamics, before it fully equilibrates~\cite{PhysRevA.84.032105, PhysRevA.98.050101, Ravell, PhysRevA.109.023309, PhysRevA.99.062114, PhysRevE.110.024132}. This approach allows both the initial state and the time evolution to serve as metrological resources. As a result, it can enable faster and potentially more precise temperature estimation, particularly when combined with tailored probe preparation and optimized measurement strategies~\cite{PhysRevA.109.L060201, Mirkhalaf_2024}.

In continuous-variable systems, Gaussian states have long served as the natural and widely used choice of quantum probes~\cite{RevModPhys.84.621}, thanks to their mathematical tractability and well-structured phase-space properties. These features have enabled a wide range of analytical studies in quantum thermometry~\cite{Mirkhalaf_2024, Cenni2022thermometryof, PhysRevA.110.052421, 5wn8-d9ks, Frigerio_2022}, which focused primarily on Gaussian probes and specific measurement strategies. A systematic side-by-side comparison of Gaussian and non-Gaussian probes across both transient and equilibrium regimes is still lacking. In this work, we provide the first such benchmark, analyzing the performance of different probe states under experimentally feasible measurement strategies and identifying regimes where non-Gaussian features may offer advantages. While most previous studies remained within the Gaussian framework, recent advances suggest that non-Gaussian states can reveal distinct quantum features that are inaccessible to Gaussian probes, offering potential advantages in precision parameter estimation~\cite{Deng2024, Oh_2020, PhysRevA.100.032318, PhysRevLett.134.180801, santos2025}. For instance, single-mode Fock states have been shown to outperform squeezed vacuum states—previously regarded as optimal—in estimating the loss parameter of bosonic channels~\cite{ PhysRevLett.98.160401, PhysRevA.79.040305}. More recently, single-mode squeezed vacuum states have been shown to surpass the ultimate precision bounds attainable by classical Gaussian states in non-equilibrium thermometry~\cite{Mirkhalaf_2024}. For a fixed probe energy, these states achieve the maximum QFI among all single-mode Gaussian probes, establishing them as optimal within this class.
These findings naturally raise a broader question: {\it{for a fixed energy, can non-Gaussian states—such as Fock states—outperform energy-matched squeezed vacuum states as initial probes in extracting temperature information from dissipative quantum systems?}}

Motivated by this question, we study quantum thermometry in both transient and equilibrium regimes using single- and two-mode probes. Our goal is to minimize the time required for accurate temperature estimation while maximizing precision through optimized choices of initial states and measurement observables. While non-Gaussian advantages in metrology are known, our contribution systematically quantifies their short-time advantage in thermometry under fixed energy constraints~\cite{Deng2024, Oh_2020, PhysRevA.100.032318, PhysRevLett.134.180801, santos2025}. We adopt a fixed energy constraint as a resource-aware metric, relevant for size, weight, and power (SWAP) considerations in practical sensors. The core of our protocol is based on the output of a stimulated parametric down-conversion (PDC) process~\cite{PhysRevA.91.053801}, where a strong coherent pump drives a nonlinear crystal at finite temperature, and the signal and idler modes are seeded with non-vacuum states. In contrast to spontaneous PDC, which begins with vacuum inputs and generates a two-mode squeezed vacuum state (TMSVS), stimulated PDC allows flexible probe-state engineering, producing a broad class of both Gaussian and non-Gaussian outputs~\cite{Kolkiran2008,  Roux2021} compared to spontaneous PDC~\cite{PhysRevLett.98.160401, PhysRevA.79.040305}. This enhanced flexibility enables precise state engineering for thermometric applications~\cite{Roux2021}.

In the non-equilibrium regime, we begin with single-mode probes and show that non-Gaussian states—including Fock, cat, and GKP states—can outperform Gaussian probes such as squeezed vacuum states. For two-mode probes, non-Gaussian entangled states such as NOON and entangled cat states enable faster and more precise temperature estimation than all classical Gaussian probes, though they do not surpass the performance of the TMSVS. In the equilibrium regime, we first analyze single-mode squeezed thermal states and find that they provide less estimation precision compared to thermal states. Extending to two-mode squeezed thermal probes reveals a clear advantage over both single-mode squeezed and thermal probes. To assess practical measurement strategies, we compute the classical Fisher information (CFI) for quadrature detection, energy measurements (optimal), and population-difference observables. We show that squeezing enhances thermometric precision by suppressing population-difference fluctuations—an effect well known in quantum metrology and here adapted to temperature estimation. Experimental platforms for generating squeezed states with tunable parameters are reviewed in Refs.~\cite{Andersen_2016, Takeno:07, PhysRevLett.117.110801, PhysRevLett.104.251102}.

The remainder of this paper is organized as follows. In Sec.~\ref{CQET}, we briefly review the essential concepts of classical and quantum estimation theory. The physical model underlying our thermometric protocol is introduced in Sec.~\ref{model}. Section~\ref{transient} presents our results for non-equilibrium thermometry, followed by the equilibrium analysis in Sec.~\ref{equilibrium}. A tabular summary of the key findings is provided in Sec.~\ref{summary}. Finally, we conclude in Sec.~\ref{conc} with a discussion of the main results and their implications. The calculations of QFI for an initial Fock state and a squeezed vacuum state are given in Appendices~\ref{QFI:Fock} and~\ref{R_short}, respectively. We elaborate on non-Gaussian characteristics of single-mode probe states in Appendix~\ref{App:Kutosis}. The CFI for single-mode and two-mode squeezed thermal states is discussed in Appendix~\ref{quadrature:sm} and ~\ref{population}, respectively.

\section{Classical and Quantum Estimation Theory} \label{CQET}

Accurate parameter estimation is central to quantum metrology. In the context of quantum thermometry, the task is to infer an unknown temperature \( T \) from measurements performed on a quantum probe that has interacted with a thermal environment. This section briefly outlines the classical and quantum estimation frameworks relevant to our analysis.

In classical estimation theory, for any unbiased estimator $\hat{T}$ of the temperature $T$, based on $m$ independent and identically prepared measurements, the Cramér--Rao bound (CRB) sets a fundamental limit on the estimation precision~\cite{Helstrom1969,Paris2009}:
\begin{equation}
\operatorname{Var}(\hat{T}) \;\ge\; \frac{1}{m\,F_C(T)} ,
\label{eq:CRB}
\end{equation}
where $F_C(T)$ denotes the CFI associated with the probability distribution of measurement outcomes.
For a probability distribution \( \{ p_k(T) \} \) over measurement outcomes \( \{ k \} \), the CFI is given by
\begin{equation}
F_C(T) = \sum_k \frac{1}{p_k(T)} \left( \frac{\partial p_k(T)}{\partial T} \right)^2.
\end{equation}
The CFI thus quantifies the sensitivity of measurement outcomes to changes in the unknown parameter. 
In quantum estimation theory, the CFI is maximized over all possible quantum measurements, resulting in the QFI, \( F_Q(T) \), which sets the ultimate precision bound:
\begin{equation}
\mathrm{Var}(\hat{T}) \geq \frac{1}{m F_Q(T)}.
\end{equation}
For a quantum state \( \rho(T) \) that depends smoothly on the parameter \( T \), the QFI is defined as
\begin{equation}
F_Q(T) = \mathrm{Tr}\left[ \rho'(T) L_T \right],
\end{equation}
where \( \rho'(T) = \partial \rho(T)/\partial T \), and \( L_T \) is the symmetric logarithmic derivative (SLD), implicitly defined by
\begin{equation}\label{eq:QFILog}
\frac{\partial \rho(T)}{\partial T} = \frac{1}{2} \left( \rho(T) L_T + L_T \rho(T) \right).
\end{equation}

In thermometry, the quantum probe acquires temperature-dependent features through interaction with a thermal bath. The QFI then depends on the probe's temperature-evolved state and serves as a figure of merit for optimizing both the initial state and measurement observable. In equilibrium thermometry, the QFI depends only on the final thermal state, making the choice of initial state irrelevant. However, in non-equilibrium scenarios, where the probe is measured before full thermalization, the initial state significantly influences the estimation precision. This may allow for enhanced metrological performance through judicious probe preparation and timing \cite{PhysRevA.109.L060201, Mirkhalaf_2024}.
In this work, we use the QFI to benchmark the temperature sensitivity of various probe configurations and dynamics in two distinct regimes. In the transient regime, we compare Gaussian and non-Gaussian initial states, while in the equilibrium regime, we evaluate the performance of thermal, single-mode, and two-mode squeezed thermal states to identify quantum advantages over classical strategies.
\begin{figure}
    \centering
    \includegraphics[scale=0.5]{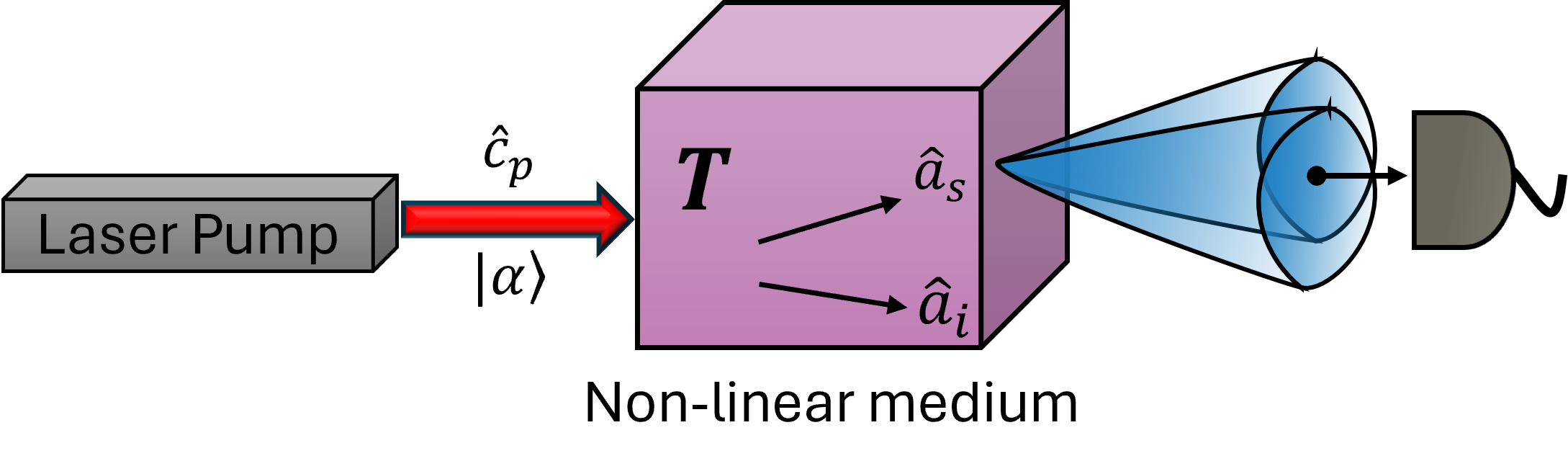}
    \caption{Schematic of parametric down conversion process: a coherent pump $|\alpha_p\rangle$ enters a non-linear $\chi^{(2)}$ medium, which is at temperature $T$, converting pump photons into signal ($a_s$) and idler ($a_i$) modes. A detector is used to perform measurements on the two modes for estimation of the temperature $T$.}
    \label{fig:1}
\end{figure}
\section{Framework for Thermometry}\label{model}

We consider two distinct frameworks for quantum thermometry: (i) a single-mode probe model used as a baseline reference, and (ii) a two-mode probe generated via stimulated PDC. In both cases, we investigate the sensitivity of temperature estimation under various input probe states and measurement observables.

\subsection{Benchmark: Single-mode probe model}

As a baseline, we start with a minimal single-mode bosonic system coupled to a thermal reservoir at a temperature \(T\). A weakly driven cavity mode, described by the annihilation operator \( \hat{a} \), interacts with the thermal environment. The drive is eliminated via a displacement transformation \( \hat{a} \rightarrow \hat{a} + c_p \), where \( c_p \) is the classical drive amplitude. The effective Hamiltonian reduces to (we consider $\hbar=1$)
\begin{equation}
    \hat{H}_1 = \omega \hat{a}^\dagger \hat{a},
    \label{eq:SingleMode}
\end{equation}
describing a linear harmonic oscillator of frequency \( \omega \). While structurally simple, this model serves as a benchmark for evaluating the performance of more complex probes. In our analysis, we consider a variety of initial states, including squeezed, Fock states, cat, and GKP states, and assess their thermometric performance.

\subsection{Two-mode PDC probe model}

To explore enhanced thermometric performance, we consider a two-mode probe generated via stimulated PDC (see Fig.~\ref{fig:1}). A nonlinear crystal held at temperature \( T \) is pumped by a strong coherent field \( \hat{c} \), which undergoes down-conversion into signal and idler photons in modes \( \hat{a}_s \) and \( \hat{a}_i \), respectively. Injecting non-vacuum states into the signal and idler modes enables the preparation of a wide range of Gaussian and non-Gaussian two-mode states.
The full Hamiltonian takes the form
\begin{equation}
    \hat{H}_2 = \hat{H}_0 + \hat{H}_I,
\end{equation}
where the free part is
\begin{equation}
    \hat{H}_0 =  \omega_s \hat{a}_s^\dagger \hat{a}_s +  \omega_i \hat{a}_i^\dagger \hat{a}_i,
\end{equation}
and the interaction Hamiltonian is given by~\cite{Agarwal_2012, Scully_Zubairy_1997}
\begin{equation}
    \hat{H}_I = g \left( \hat{c}^\dagger \hat{a}_s \hat{a}_i + \hat{c} \hat{a}_s^\dagger \hat{a}_i^\dagger \right),
\end{equation}
where \( g \) is the coupling constant. Applying the classical pump approximation, we replace \( \hat{c} \rightarrow \alpha_p = |\alpha_p| e^{i\varphi} \). 
We then introduce an effective interaction constant \(\xi = g \alpha_p\), which we assume to be real (\(\phi = 0\)) for simplicity.
As a notational simplification, we identify the signal and idler modes as \( \hat{a} \equiv \hat{a}_s \) and \( \hat{b} \equiv \hat{a}_i \), the total Hamiltonian becomes~\cite{Agarwal_2012,Scully_Zubairy_1997} 
\begin{equation}
    \hat{H}_2 =   \omega_s \hat{a}^\dagger \hat{a} + \omega_i \hat{b}^\dagger \hat{b} \
    +   \xi\left( \hat{a}^\dagger \hat{b}^\dagger + \hat{a} \hat{b}\right).
    \label{eq:squeezeH}
\end{equation}
The above Hamiltonian describes a two-mode squeezed system, which we diagonalize via the following Bogoliubov transformation:
\begin{align}
    \hat{A} &= \cosh r \, \hat{a} +  \sinh r \, \hat{b}^\dagger, \\
    \hat{B} &= \cosh r \, \hat{b} + \sinh r \, \hat{a}^\dagger,
\end{align}
where the squeezing parameter $r$ takes the following form
\begin{equation}\label{eq:squeezing}
    r=\frac{1}{2}\tanh^{-1}{\Big(\frac{2\xi}{\omega_s+\omega_i}\Big)}.
\end{equation}
The transformed Hamiltonian reads
\begin{equation}
    \tilde{H}_2 = \tilde{\omega}_+ \hat{A}^\dagger \hat{A} + \tilde{\omega}_- \hat{B}^\dagger \hat{B} + E_0,
    \label{diagHamil}
\end{equation}
where 
\begin{equation}
\tilde{\omega}_{\pm} = \frac{\Omega \pm (\omega_s - \omega_i)}{2},
\quad \text{with} \,\,
\Omega = \sqrt{(\omega_s + \omega_i)^2 - 4\xi^2}
\end{equation}  
and \( E_0 \) is a constant vacuum energy shift. In the symmetric case $\omega_s = \omega_i = \omega$, we have $\tilde{\omega}=\sqrt{\omega^2-\xi^2}$.

\section{Non-equilibrium thermometry}\label{transient}
In this section, we first examine the role of various single-mode Gaussian and non-Gaussian states in non-equilibrium quantum thermometry, using the single-mode case as a baseline, and emphasizing how their initial preparations impact the speed of temperature sensing.
We then extend our analysis to two-mode probes, investigating how different entangled and separable initial states influence the rate at which thermal information is acquired.
\subsection{Single-mode probe states}
\subsubsection{Gaussian states}

We consider a single-mode bosonic system described by quadrature operators \( \hat{x} \) and \( \hat{p} \), and define the phase-space vector \( \hat{R} = (\hat{x}, \hat{p})^T \). The state of the system is fully characterized by its first moment \( \mathbf{d}_t = \langle \hat{R} \rangle_t \) and covariance matrix
\begin{equation}
\sigma_t = \frac{1}{2} \langle \{ \hat{R}, \hat{R}^{T} \} \rangle_t - \mathbf{d}_t \mathbf{d}_t^{T} .
\end{equation}
Under a standard thermal damping channel, the dynamics preserve the Gaussian character of the state. The evolution of the first and second moments takes the form~\cite{PhysRevA.63.032312, Göran}:
\begin{equation}
    \mathbf{d}_t = X_t \mathbf{d}_0, \qquad \sigma_t = X_t \sigma_0 X_t^T + Y_t,
\end{equation}
where \( \mathbf{d}_0 \) and \( \sigma_0 \) are the initial first moment and covariance matrix, respectively. The matrices \( X_t \) and \( Y_t \) encode the dissipative evolution and are given by
\begin{equation}
    X_t = e^{-\gamma t/2} O_t, \qquad Y_t = \left(1 - e^{-\gamma t}\right) \sigma_T,
\end{equation}
with \( \gamma \) the damping rate and \( O_t \in \text{SO}(2) \) a rotation matrix, which can be taken as the identity in the interaction picture.
The thermal noise covariance matrix is
\begin{equation}
    \sigma_T = \nu I_2, \qquad \nu = \coth\left( \frac{\omega}{2T} \right),
\end{equation}
where \( \omega \) is the oscillator frequency, \( T \) is the bath temperature, and \( I_2 \) is the \( 2 \times 2 \) identity matrix.
Assuming \( O_t = I_2 \), the covariance matrix evolves simply as
\begin{equation}\label{CM:time}
    \sigma_t = e^{-\gamma t} \sigma_0 + \left(1 - e^{-\gamma t} \right) \nu I_2.
\end{equation}
Thus, any initially Gaussian state---thermal, coherent, or squeezed---remains Gaussian throughout the evolution. This justifies the use of covariance matrix techniques to analyze thermometric performance.

For a general single-mode Gaussian state with first moment \( \mathbf{d}_T \) and covariance matrix \( \sigma_T \), the QFI with respect to temperature \( T \) is given by~\cite{PhysRevA.88.040102}:
\begin{align}
F_Q(T) &= \frac{1}{2} \frac{\mathrm{Tr} \left[ \left( \sigma_T^{-1} \, \partial_T \sigma_T \right)^2 \right]}{1 + \mu_T^2} 
+ \frac{2 \, (\partial_T \mu_T)^2}{1 - \mu_T^4} \notag \\
&\quad + (\partial_T \mathbf{d}_T)^T \sigma_T^{-1} \, \partial_T \mathbf{d}_T,
\end{align}
where \( \mu_T = 1/\sqrt{\det \sigma_T} \) is the purity of the state.
In our thermometry protocol, the displacement vector \( \mathbf{d}_T \) does not depend on temperature, i.e., \( \partial_T \mathbf{d}_T = 0 \), which implies that the third term vanishes. Therefore, the QFI reduces to
\begin{equation}\label{eq:QFI}
F_Q(T) = \frac{1}{2} \frac{\mathrm{Tr} \left[ \left( \sigma_T^{-1} \, \partial_T \sigma_T \right)^2 \right]}{1 + \mu_T^2} 
+ \frac{2 \, (\partial_T \mu_T)^2}{1 - \mu_T^4},
\end{equation}
which depends solely on the covariance matrix and its temperature dependence.
As a specific example, consider an initial squeezed vacuum state with covariance matrix \( \sigma_0 = \mathrm{diag}(r, 1/r) \), where \( r \geq 1 \) quantifies the squeezing in terms of quadrature variances (i.e., \( r = e^{2s} \), with \( s \) the standard squeezing parameter). Using Eq.~\eqref{CM:time} for the evolved covariance matrix and Eq.~\eqref{eq:QFI} for the QFI, one obtains~\cite{Mirkhalaf_2024}
\begin{equation}
    F_Q (\sigma_t; t) = \frac{(1 - e^{-\gamma t})^2 \, (\partial_T \nu)^2 \big( 2 + [\sigma_t]_{11} + [\sigma_t]_{22} \big)}{2 \big( [\sigma_t]_{11} [\sigma_t]_{22} - 1 \big)}.
\end{equation}
This formalism allows us to quantitatively assess the temperature sensitivity of single-mode Gaussian probes under dissipative dynamics. We will later use this benchmark to compare the performance of non-Gaussian states such as Fock and cat states under similar conditions.

\subsubsection{Non-Gaussian states}

Non-Gaussian initial states can give rise to rich dynamical features under thermal evolution, including the generation of non-Gaussianity and enhanced sensitivity to temperature. To analyze these effects, we solve the full Lindblad master equation, which enables a direct treatment of non-Gaussian dynamics. The QFI is computed directly from the time-evolved density matrix using the symmetric logarithmic derivative formalism introduced in Eq.~(\ref{eq:QFILog}). 

For a single bosonic mode, whose Hamiltonian is given in Eq.~\eqref{eq:SingleMode}, interacting with a thermal bath at temperature \( T \), the system evolves according to the standard Lindblad master equation~\cite{Breuer, Seifoory:17} (throughout, we set the Boltzmann constant \( k_B = 1 \)):
\begin{equation}\label{sm:master_equation}
\frac{d\rho}{dt} = -i\left[ \hat{H}_1, \rho \right] 
+ \gamma (\bar{n} + 1) \mathcal{D}[\hat{a}] \rho 
+ \gamma \bar{n} \mathcal{D}[\hat{a}^\dagger] \rho.
\end{equation}
Here \( \bar{n} = ( e^{\omega / T} - 1 )^{-1} \) is the mean thermal occupancy, $\gamma$ denotes the system-bath coupling, and the standard dissipator is defined as
\begin{equation}
\mathcal{D}[\hat{L}] \rho = \hat{L} \rho \hat{L}^\dagger - \frac{1}{2} \left\{ \hat{L}^\dagger \hat{L}, \rho \right\}.
\end{equation}
While we do not incorporate decoherence mechanisms like cavity losses in this analysis, the availability of analytical solutions is a significant advantage of our model.
\paragraph{Fock states:} We first consider Fock states, which are generated by repeated application of the creation operator on the vacuum:

\begin{equation}
|n_0\rangle = \frac{(\hat{a}^\dagger)^{n_0}}{\sqrt{n_0!}} |0\rangle.
\end{equation}
For an initial state \( \rho(0) = |n_0\rangle\langle n_0| \), the time-dependent QFI can be computed exactly by evolving the full density matrix under Eq.~\eqref{sm:master_equation} and using Eq. (\ref{eq:QFILog}). The resulting closed-form expression reads (see Appendix~\ref{QFI:Fock}):
\begin{equation}
F_Q(T; t) = \sum_{r=0}^{\infty} \sum_{n,n' = n_{\min}}^r
    \frac{e^{\gamma t} S_{r,n} S_{r,n'}}{F(T; t)^2 \, p_r(T; t)} \, 
    \Theta_n^{(r)} \, \Theta_{n'}^{(r)}.
\end{equation}
\paragraph{GKP states:} More generally, non-Gaussian states span a broad family that includes coherent superpositions of Gaussian states. 
A widely studied example is the GKP states, which have garnered attention in recent years~\cite{PhysRevA.95.052352, PRXQuantum.2.030204}. However, preparing high-fidelity GKP states remains experimentally challenging.
The canonical form of the (approximate) GKP state is a superposition of displaced squeezed vacuum states~\cite{PhysRevA.64.012310}:
\begin{equation}
    |\text{GKP}\rangle \approx \sum_k e^{-2\pi k \delta} \hat{D}(2k\sqrt{\pi}) \hat{S}(r) |0\rangle,
\end{equation}
where \( \hat{D}(\alpha) = e^{\alpha \hat{a}^\dagger - \alpha^* \hat{a}} \) is the displacement operator, \( \hat{S}(r) = \exp\left[ \frac{r}{2} (\hat{a}^2 - \hat{a}^{\dagger 2}) \right] \) is the squeezing operator, and \( \delta \) characterizes the Gaussian envelope.
\paragraph{Cat states:}  Another prominent class of non-Gaussian states is defined as coherent superpositions of two oppositely displaced coherent states~\cite{PRXQuantum.2.030204}:
\begin{equation}
|\text{c}\rangle_{\pm} = \frac{1}{\sqrt{\mathcal{N}}} (|\alpha\rangle \pm |-\alpha\rangle),
\end{equation}
where \( \mathcal{N} = 2(1 \pm e^{-2\alpha^2}) \) is a normalization constant, and \( |\alpha\rangle = \hat{D}(\alpha) |0\rangle \). Unlike GKP states, cat states are relatively easier to realize experimentally. Cat states have already been demonstrated across a variety of physical platforms, including atomic systems, optical cavities, circuit QED architectures, and mechanical resonators \cite{science.272.5265.1131, PhysRevLett.77.4887, Gao2010, science.adf7553}.

For the GKP and cat non-Gaussian states, closed-form analytical expressions for the QFI are generally intractable. Therefore, we numerically compute the QFI using the general SLD-based formula given in Eq. (\ref{eq:QFILog}), following the implementation detailed in Ref.~\cite{PhysRevResearch.4.043057}. This method allows us to benchmark the thermometric power of GKP and cat states under dissipative evolution. To perform numerical calculations, each bosonic mode is represented in the Fock basis with a finite Hilbert space of dimension $N = 60$ for a single mode, and we similarly verified that the QFI converges for the two-mode case as well. In the following subsection, we compare the QFI for Gaussian and non-Gaussian states in the transient regime, highlighting the conditions under which non-Gaussianity yields a metrological advantage.

\begin{figure}[t!]
    \centering
    \includegraphics[scale=0.41]{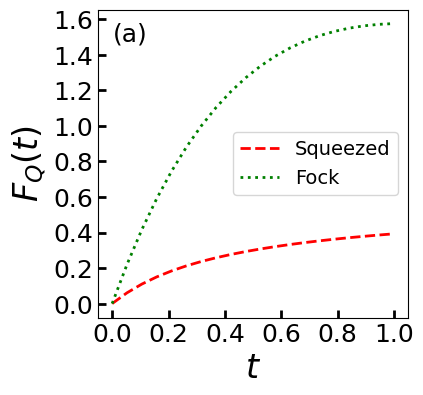}
    \includegraphics[scale=0.41]{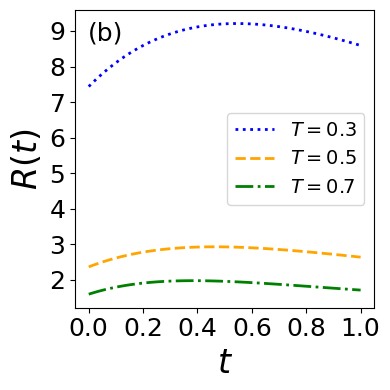}
    \caption{(a) Non-equilibrium QFI $F_Q(t)$ as a function of the interaction time $t$ for initial squeezed vacuum state (red dashed curve) and Fock state (green dotted curve). (b) shows the time-dependent ratio \( R(t) = F_Q^{\mathrm{Fock}}(t) / F_Q^{\mathrm{SVS}}(t) \), comparing the QFI of the Fock state to that of the squeezed vacuum state at different values of temperature $T$. The parameters are set to $\omega=1$, $T=0.4$, $\gamma=0.2$, and $n_0=4$. The two states have the same initial energy for which $r=\sinh^{-1}(\sqrt{n_0})$.}
    \label{fig:Fock+SVS}
\end{figure}

\subsubsection{Performance comparison of Gaussian and non-Gaussian single-mode probe states}

To establish a fair baseline for performance comparisons, we begin by identifying the optimal single-mode Gaussian state for temperature estimation under a fixed energy constraint. Any such state can be expressed as
\begin{equation}
    \rho_G = \hat{D}(\alpha) \hat{S}(r) \rho_{\text{th}}(\bar{n}) \hat{S}^\dagger(r) \hat{D}^\dagger(\alpha),
\end{equation}
where \( \hat{D}(\alpha) \) is the displacement operator, \( \hat{S}(r) \) is the squeezing operator, and \( \rho_{\text{th}}(\bar{n}) \) is a thermal state with mean photon number \( \bar{n} \). The goal is to decide how to distribute the available energy between displacement, squeezing, and thermal noise.
As shown in Eq.~(\ref{eq:QFI}), the displacement term does not contribute to the QFI, as its temperature derivative vanishes under thermal damping. Moreover, as also intuitively expected, it is of no use to spend energy in preparing a thermal state \cite{PhysRevLett.98.160401}; that is, for any given input energy, the optimal Gaussian probe state is pure. Therefore, the optimal strategy is to allocate all energy to squeezing, making the squeezed vacuum state the best Gaussian probe under energy constraints.

However, non-Gaussian states may outperform these Gaussian states under a fixed energy constraint. To explore this, we consider the Fock state \( |n_0\rangle \) and compare its QFI with that of the Gaussian squeezed vacuum state, as illustrated in Fig.~\ref{fig:Fock+SVS}(a). The squeezing parameter is chosen as \( r = \sinh^{-1}(\sqrt{n_0}) \), ensuring that both states have the same energy. As shown for \(n_0=4\) and $r=1.443$ in Fig.~\ref{fig:Fock+SVS}(a), the Fock state outperforms the SVS at earlier times. Fig.~\ref{fig:Fock+SVS}(b) shows the time-dependent ratio $R(t)$ of the QFI of the Fock state to that of SVS, defined as
\begin{equation}
    R(t) = \frac{F_Q^{\mathrm{Fock}}(t)}{F_Q^{\mathrm{SVS}}(t)}.
\end{equation}
This ratio characterizes the relative performance between the two states over a short time for a fixed initial energy. One can see that the ratio increases when the temperature is low enough ($T=0.3$), while for higher temperatures ($T=0.7$), this quantity decreases. For longer times, the ratio \( R(t) \) is plotted as a function of time for different temperature values in Fig.~\ref{fig:R_long} (see Appendix~\ref{QFI:Fock}). The Fock state outperforms the energy-matched SVS, particularly at early times and low temperatures. As time progresses, \( R(t) \) approaches unity, reflecting the fact that both initial states evolve toward the same thermal state, which yields identical QFI.

To analytically understand $R$, we examine the ratio $R= F_Q^{\mathrm{Fock}}(t) / F_Q^{\mathrm{SVS}}(t)$, in the short-time limit. In this regime, the QFI for the Fock state and the SVS expands to (see Appendices~\ref{QFI:Fock} and \ref{R_short}):
\begin{equation}
\begin{aligned}
F_Q^{\mathrm{Fock}}(t)\approx \frac{\gamma t}{2}(\partial_T\nu)^2\frac{\nu(2n_0+1)+1}{\nu^2-1},\\
F_Q^{\mathrm{SVS}}(t)\approx\frac{\gamma t}{2}(\partial_T\nu)^2\frac{(2n_{sv}+1)^2}{\nu(2n_{sv}+1)-1},
\end{aligned}
\end{equation}
where $\nu=2\bar{n}+1$, $n_0$ denotes the Fock number and $n_{sv}$ is the initial mean photon number for SVS, defined as $n_{sv}=\sinh^2{r}$.  Imposing the condition of equal energy between the two probes, i.e., $n_0=n_{sv}=n$, and performing some algebra yields the simplified form of $R$ (see Appendix~\ref{R_short}):
\begin{equation}
    R \approx \frac{\nu^2-\frac{1}{(2n+1)^2}}{\nu^2-1}.
\end{equation}
This expression shows that for any nonzero probe energy $n>1$, the QFI of the Fock state exceeds that of an equal-energy SVS for temperature estimation in the short-time regime.

\begin{figure}
    \centering
    \includegraphics[scale=0.65]{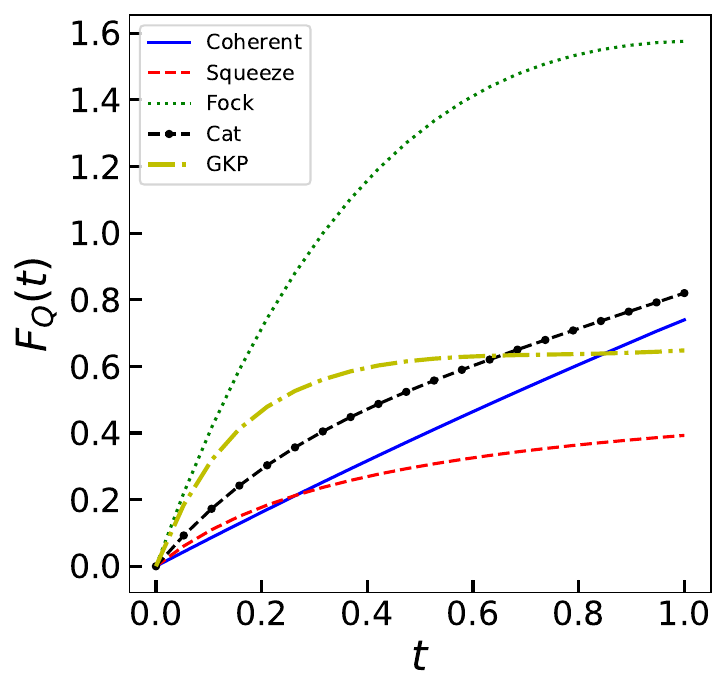}
    \caption{Non-equilibrium QFI $F_Q(t)$ for different initial preparations of the single probe state as a function of the interaction time $t$ for estimation of the bath temperature $T$. The probe states are prepared with parameters (coherent amplitude $\alpha$, squeezing $r$, photon number $n_0$, etc.) chosen such that all states have the same average energy $E_0 =4.5$. The other parameters are set to $\omega=1$, $T=0.4$, and $\gamma=0.2$.}
    \label{fig:Smode}
\end{figure}

To systematically compare the metrological performance of different Gaussian and non-Gaussian states, 
in Fig.~\ref{fig:Smode}, we plot the QFI as a function of time \( t \) for various initial states, including the coherent state \( |\alpha\rangle \) and the squeezed vacuum state \( \hat{S}(r)|0\rangle \). These are compared to non-Gaussian states such as the Fock state, the odd cat state \( |c\rangle_- \), and the GKP state. Under a fixed energy constraint, we numerically select parameters (e.g., squeezing $r$, displacement $\alpha$, Fock number $n_0$) to attain the target energy (e.g., $E_0=4.5$). This ensures a fair comparison across probe states. For the squeezed vacuum, this yields $r = \sinh^{-1}{\sqrt{E_0}} \approx 1.443$. 
The results in Fig.~\ref{fig:Smode} indicate that non-Gaussian states can access a significant portion of thermal information in the short-time regime. However, the maximum QFI is typically achieved at later times—possibly near thermal equilibrium—non-Gaussian, nonclassical states enable the extraction of a substantial fraction of that information much earlier.
At the fixed temperature \( T = 0.4 \), the squeezed vacuum state performs worse than the coherent state in terms of QFI. While squeezed states can reduce fluctuations along a specific quadrature, such noise suppression does not translate into enhanced thermometric precision at this temperature. In transient thermometry, this can align with the bath-induced noise channel, allowing better resolution of temperature-dependent features. In equilibrium, however, the exponential growth
of photon-number variance under squeezing outweighs the precision gain, making squeezed thermal states less effective than plain thermal states. This suggests that, in the non-equilibrium regime and for moderate temperatures, certain Gaussian states, such as coherent states, can outperform squeezed vacuum states.

Among the states considered, Fock states \( |n_0\rangle \) with \( n_0 = 4 \) outperform other probe states in the short-time regime, with precision improving as the photon number $n_0$ increases (see Appendix~\ref{R_short} for more details). The GKP and cat states also prove highly effective for rapid temperature estimation compared to Gaussian states, further highlighting the metrological advantage of non-Gaussian resources. This advantage stems from the fact that Non-Gaussian states such as Fock and cat states have sharper photon-number statistics and stronger nonclassical features (e.g., sub-Poissonian noise, negative Wigner functions), while GKP states derive their sensitivity from their structured, grid-like phase-space distribution. At very short interaction times, the thermal bath has only partially imprinted temperature information onto the probe. States that start with strong, well-defined non-Gaussian features exhibit greater sensitivity to small thermal fluctuations than Gaussian states, which are more “smooth” in phase space. In addition to computing the QFI, we quantify the non-Gaussian character of various initial probe states evolving under the master equation~\eqref{sm:master_equation} using a kurtosis-based measure, as detailed in Appendix~\ref{App:Kutosis}.
\subsection{Two-mode probe states}

We now turn our focus to a two-mode system used as a probe. We assume that the signal and idler modes, seeded with non-vacuum inputs, are coupled to a common thermal bath at temperature \(T\), as illustrated in Fig.~\ref{fig:1}. The time evolution of the density matrix \(\rho\) for these two modes is then governed by the Lindblad master equation~\cite{Seifoory:17}
\begin{equation}\label{2m:me}
\begin{aligned}   
\frac{d\rho}{dt} = -i[\hat{H}_2, \hat{\rho}] + \sum_{j=a,b}\left(\gamma_j(\bar{n}_j+1)\mathcal{D}[\hat{j}] + \gamma_j\bar{n}_j\mathcal{D}[\hat{j}^\dagger]\right)\\
+ \left(\Gamma_{1}\mathcal{D}[\hat{b}, \hat{a}^\dagger]
+ \Gamma_2\mathcal{D}[\hat{b}^\dagger, \hat{a}] \right),
\end{aligned}
\end{equation}
where
\begin{equation}
\Gamma_1 = \gamma\sqrt{(\bar{n}_b + 1)(\bar{n}_a + 1)}, \quad
\Gamma_2 = \gamma\sqrt{\bar{n}_b \bar{n}_a}.
\end{equation}
In simulating the master equation~\eqref{2m:me}, we set $\omega_a=\omega_b$, implying that $\bar{n}_a=\bar{n}_b=\bar{n}$. The goal is to use the two-mode system as a quantum probe for temperature estimation and explore whether suitable initial states enable faster information retrieval than in the single-mode case.
\subsubsection{Initial Gaussian and non-Gaussian states}
We classify the initial two-mode probe states into Gaussian and non-Gaussian categories. The two modes are prepared in separable coherent Gaussian states:
\begin{equation}
    |\psi_{\mathrm{coh}}\rangle = |\alpha\rangle_s \otimes |\alpha\rangle_i,
\end{equation}
where each single mode is in a coherent state \(|\alpha\rangle\). Next, we consider the TMSVS, an entangled Gaussian state given by~\cite{RevModPhys.77.513, RevModPhys.81.299}
\begin{equation}
    |\psi_{\mathrm{TMSVS}}\rangle = \hat{S}_2(r) |0,0\rangle_{s,i},
\end{equation}
where \(|0,0\rangle_{s,i} = |0\rangle_s \otimes |0\rangle_i\) is the two-mode vacuum state and \(\hat{S}_2(r)\) is the two-mode squeezing operator defined as
\begin{equation}
    \hat{S}_2(r) = \exp \left[ r \left( \hat{a}^\dagger_s \hat{a}^\dagger_i - \hat{a}_s \hat{a}_i \right) \right],
\end{equation}
with squeezing parameter \(r \geq 0\).
\begin{figure}[t!]
    \centering
         \includegraphics[scale=0.65]{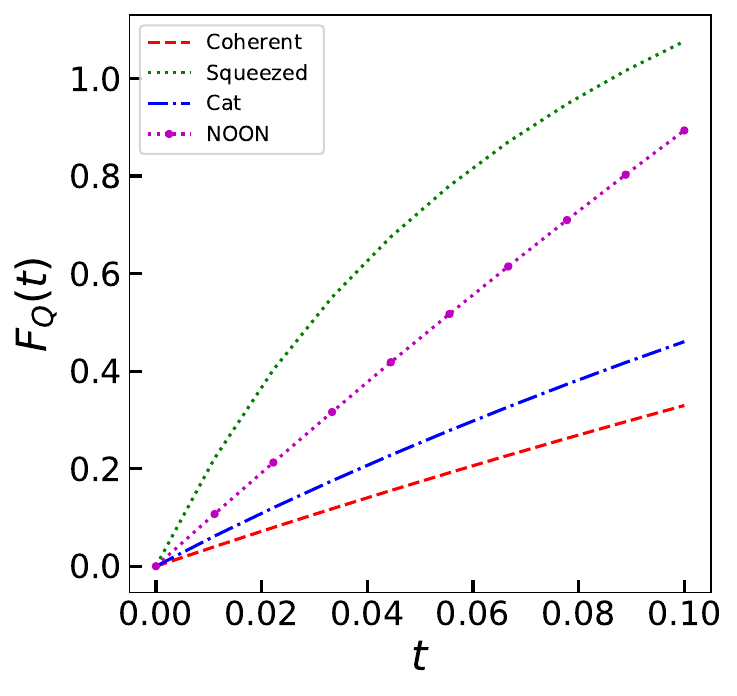}    
         \caption{Non-equilibrium QFI for different initial Gaussian and non-Gaussian states as a function of interaction time $t$ for estimation of bath temperature $T$. We use both modes as a probe to measure the temperature of the nonlinear crystal. The parameters are set to $\omega=1$, $\alpha_p=4$, $g=0.08$, $T=0.4$, and $\gamma=0.2$. The rest of the parameters are set to achieve the fixed target energy $E_t=6$.}
    \label{fig:2Modes}
\end{figure}

For non-Gaussian states, we consider two-mode cat states, formed as superpositions of coherent states, which are entangled and non-Gaussian. These states are valuable resources in quantum metrology, as they can enhance precision in parameter estimation~\cite{PhysRevA.66.023819}. Let \( |\alpha\rangle \) and \( |-\alpha\rangle \) denote single-mode coherent states with amplitude \( \alpha \) and define the tensor product states.
\begin{equation}
    |\phi_+\rangle = |\alpha\rangle_s \otimes |\alpha\rangle_i, \qquad
|\phi_-\rangle = |{-\alpha}\rangle_s \otimes |{-\alpha}\rangle_i,
\end{equation}
where subscripts \( s \) and \( i \) label the signal and idler modes, respectively.
Using these, the two-mode even and odd entangled cat states are defined as~\cite{Sanders_2012, Ourjoumtsev2007}
\begin{align}
|\psi_+\rangle &= \frac{1}{\mathcal{N}_+} \left( |\phi_+\rangle + |\phi_-\rangle \right), \\
|\psi_-\rangle &= \frac{1}{\mathcal{N}_-} \left( |\phi_+\rangle - |\phi_-\rangle \right),
\end{align}
where \( \mathcal{N}_\pm \) are normalization constants ensuring \( \langle\psi_\pm|\psi_\pm\rangle = 1 \).  We also include the NOON state, defined by maximal path entanglement for photon number \(N\), Such as
\begin{equation}
    |\psi\rangle_{NOON} = \frac{1}{\sqrt{2}} \left( |N\rangle_s \otimes |0\rangle_i + |0\rangle_s \otimes |N\rangle_i \right).
\end{equation}
NOON states have been extensively employed in proof-of-concept experiments as a fundamental resource for enhancing precision in quantum metrology~\cite{Dowling2008, Ono2013, Wolfgramm2013, D'Ambrosio2013}. NOON states, while not directly produced in our setup, can be engineered in similar PDC-based platforms using coherent stimulation or interferometric schemes~\cite{Dowling2008, Kolkiran2019}.
\subsubsection{Performance comparison}

Fig.~\ref{fig:2Modes} shows the results for two-mode initial states, highlighting how different preparations affect the speed of temperature estimation. Both modes are initialized identically and serve jointly as a probe for the bath temperature \( T \). These initial states are prepared with the same energy, fixed to a target value of \(E_t = 6\). Specifically, for the coherent state, the displacement amplitude is set to \(\alpha = \sqrt{E_t/2}\), while for the TMSV state, the squeezing parameter is chosen as \(r = \sinh^{-1}(\sqrt{E_t/2})\). These Gaussian probes are compared with non-Gaussian NOON and cat states prepared under the same energy constraint.
\begin{figure}[t!]
    \centering
        \includegraphics[scale=0.65]{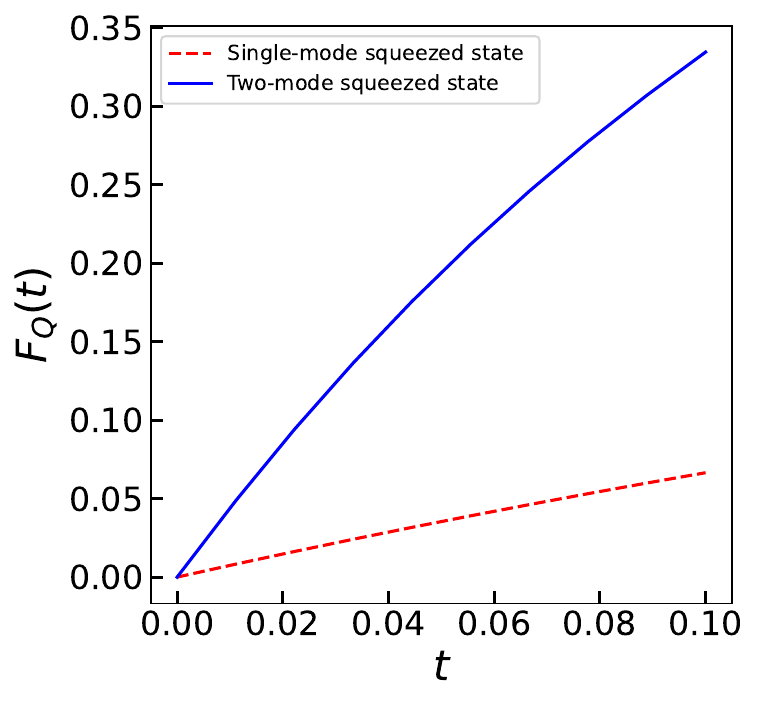}
    \caption{Non-equilibrium QFI as a function of time for a single-mode squeezed initial state (red dashed) and two-mode squeezed state (solid blue) for the parameters set to $r=1.0$ and $T=0.8$. The rest of the parameters are the same as in Figs.~\ref{fig:Smode} and ~\ref{fig:2Modes}.}
    \label{fig:singleVStwo}
\end{figure}

We note that the precision, as quantified by the QFI, does not significantly improve when using two-mode non-Gaussian states (such as NOON or cat states in Fig.~\ref{fig:2Modes}) compared to single-mode non-Gaussian states like Fock or GKP states in Fig.~\ref{fig:Smode}; in fact, it is slightly reduced. However, two-mode probes with initial states such as TMSVS and NOON states enable faster retrieval of temperature information because joint measurements on both modes provide access to information at earlier times than in the single-mode case. Figure~\ref{fig:2Modes} shows that the two-mode non-Gaussian probes, such as NOON and entangled cat states, do not surpass the performance of the TMSVS. This can be attributed to several factors: unlike the TMSVS, which distributes correlations efficiently across both modes, NOON and cat states concentrate energy into discrete photon-number sectors and are thus highly fragile to thermal losses. Their advantage relies on delicate interference features that are rapidly degraded by decoherence, whereas Gaussian entanglement already provides correlations close to the optimal bound under fixed-energy constraints. Moreover, the measurement strategies we consider—energy and population-difference observables—are naturally aligned with Gaussian probes and may not fully capture the non-classical features of NOON or cat states.  Compared to single-mode SVS (red dashed curve in Fig.~\ref{fig:singleVStwo}), the QFI for a TMSVS (see blue curve in Fig.~\ref{fig:singleVStwo}) is substantially higher for the same parameters \( r = 1.0 \) and \( T = 0.8 \). This figure clearly shows that the QFI for the two-mode squeezed state far exceeds that of the single-mode case, highlighting the advantage of using two-mode probes in quantum thermometry.

Although non-Gaussian probes such as Fock, cat, and GKP states exhibit superior metrological performance in principle, their experimental realization remains highly challenging. Preparing high-fidelity non-Gaussian states typically requires strong nonlinearities or conditional measurements, and their characteristic nonclassical features (e.g., Wigner negativity and superpositions) are highly fragile under decoherence and thermal noise, which could reduce their metrological advantage. By contrast, Gaussian squeezed states can be routinely generated with high purity in optical and microwave platforms, whereas cat/GKP states are still at the proof-of-principle stage. Nevertheless, recent progress in optical parametric down-conversion, trapped-ion, and circuit-QED setups has enabled the creation of early-generation cat and small-grid GKP states; however, scaling these to the sizes required for robust thermometry remains an open challenge. Our results, therefore, emphasize the importance of aligning theoretical precision advantages with such potential constraints. Gaussian probes are easier to prepare and more robust against decoherence, but they encode thermal information more smoothly, requiring longer interaction times to reach optimal precision. Non-Gaussian probes—both single-mode and two-mode—are more fragile yet can capture temperature signatures much earlier owing to their highly nonclassical photon statistics and phase-space structure.

Although non-equilibrium probes allow rapid temperature estimation, their ultimate precision is limited, motivating the equilibrium analysis presented in the next section.

\section{Equilibrium thermometry}\label{equilibrium}
In this section, we compare two thermometric schemes: single-mode and two-mode probes. We begin with the single-mode case using squeezed thermal states in equilibrium with a thermal bath, and then extend to two-mode squeezed thermal states, showing improved estimation precision. All parameters are normalized by the mode frequency $\omega$ and expressed in dimensionless units.
\subsection{Suboptimal and practical measurements}\label{sec:CFI}

In practice, we often evaluate the CFI with respect to experimentally accessible observables $\hat{X}$. 
For a Hermitian observable with the expectation value $\langle \hat{X} \rangle$ and variance $(\Delta \hat{X})^2$, the so-called CFI is defined as~\cite{PhysRevLett.125.080402, PhysRevA.96.062103} 
\begin{equation}\label{CFI:Formula}
    F_C(T) = 
    \frac{\big( \partial_T \langle \hat{X} \rangle \big)^2}{ (\Delta \hat{X})^2 },
\end{equation}
which reflects that using a single observable $\hat{X}$ yields a
\emph{sub-optimal} measurement. When the observable has zero mean but a temperature-dependent variance, temperature information can still be extracted from the variance $(\Delta\hat{X})^2$. In such cases, one can effectively treat the squared deviation, \(\hat{O} = (\hat{X}_\theta - \langle \hat{X}_\theta \rangle)^2\), as the estimator. The corresponding CFI then reads as~\cite{PhysRevResearch.4.023191}  
\begin{equation}\label{CFI:var}
    F_C(T) = \frac{\left(\partial_T \langle (\Delta \hat{X}_\theta)^2 \rangle \right)^2}{\langle (\Delta \hat{X}_\theta)^4 \rangle - \langle (\Delta \hat{X}_\theta)^2 \rangle^2}.
\end{equation}
The CFI computed from \(\hat{X}_\theta\)—via either its mean (Eq.~\eqref{CFI:Formula}) or variance (Eq.~\eqref{CFI:var})—provides a lower bound on the achievable precision in temperature estimation. Hence, the reported CFI is interpreted as a fundamental lower bound rather than the exact optimal precision.

\begin{figure}[t!]
    \centering
    \includegraphics[scale=0.82]{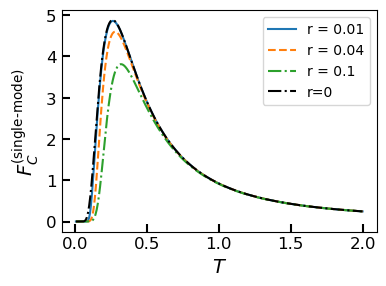}    
    \caption{CFI as a function of temperature $T$ for a single-mode squeezed thermal state using mean photon number $\langle\hat{n}\rangle$ as an observable. We set $\omega=1$.}
    \label{fig:sm_mph}
\end{figure}
\subsection{Squeezed thermal states for single-mode probes}
To assess the advantage of two-mode squeezing, we consider a benchmark case: a probe with a single-mode squeezed thermal state, where only one bosonic mode is subjected to squeezing. The single-mode Hamiltonian is given by $\hat{H}_1 = \omega \, \hat{a}^\dagger \hat{a}$. Applying a squeezing operation \( \hat{S}(r) = \exp\left[ \frac{r}{2} (\hat{a}^2 - \hat{a}^{\dagger 2}) \right] \) with a squeezing parameter $r$ to a thermal state at temperature \( T \), we obtain the single-mode squeezed thermal state
\begin{equation}\label{sqTS}
    \rho_{\mathrm{sq}} = \hat{S}(r)\, \rho_{\mathrm{th}}\, \hat{S}^\dagger(r),
\end{equation}
where $ \rho_{\mathrm{th}}$ is a thermal state in the Fock basis.
The mean photon number  and variance in the squeezed thermal state are given by
\begin{equation}\label{var:sm}
\begin{aligned}
\langle \hat{n} \rangle &= \sinh^2 r + \cosh(2r)\, \bar{n},\\
(\Delta\hat{n})^2 &= \sinh^2 r\,(\sinh^2 r + 1) + \cosh^2(2r)\, \bar{n}(\bar{n} + 1).
\end{aligned}
\end{equation}
Using the mean photon number \( \langle \hat{n} \rangle \) as an observable, the corresponding CFI using Eq.~\eqref{CFI:Formula} is given by
\begin{equation}\label{CFI:sm}
F_C^{(\text{single-mode})} = \frac{\cosh^2(2r)}{(\Delta \hat{n})^2}
\left( \frac{\partial \bar{n}}{\partial T} \right)^2,
\end{equation}
where \begin{equation}
\frac{d\bar{n}}{dT} 
= \frac{\omega\, e^{\omega/T}}
       {T^2 \left(e^{\omega/T}-1\right)^2}.
\end{equation}
We plot the CFI for the single-mode squeezed thermal state, as given by Eq.~\eqref{CFI:sm}, in Fig.~\ref{fig:sm_mph} for various values of the squeezing strength \( r \). It is evident that the CFI attains its maximum at weak squeezing. Notably, as the squeezing strength \( r \) approaches zero, the CFI converges to the QFI, indicating that the lower bound~\eqref{CFI:Formula} on the estimation precision can be saturated by using the mean photon number \(\langle \hat{n} \rangle\) as the measurement observable. Conversely, increasing \( r \) leads to a slight reduction in the peak value of the CFI, while its overall qualitative behavior remains largely unchanged.

In the limit \( r \to 0 \), the expression~\eqref{CFI:sm} reduces to the QFI of a single-mode thermal state at equilibrium, which is given by~\cite{PhysRevA.96.062103}
\begin{equation}
F_Q = \frac{\omega^2 \, \mathrm{csch}^2\left(\frac{\omega}{2T}\right)}{4 T^4}.
\end{equation}
This QFI is depicted as the black dashed curve in Fig.~\ref{fig:sm_mph} and coincides exactly with the CFI evaluated at \( r = 0.01 \), confirming the consistency of our analysis.
\begin{figure}
    \centering
    \includegraphics[scale=0.77]{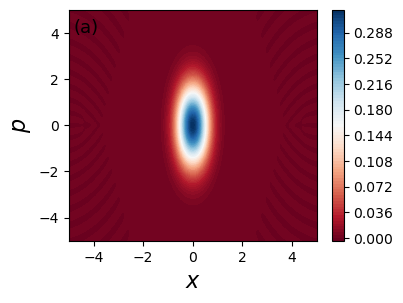}
       \includegraphics[scale=0.6]{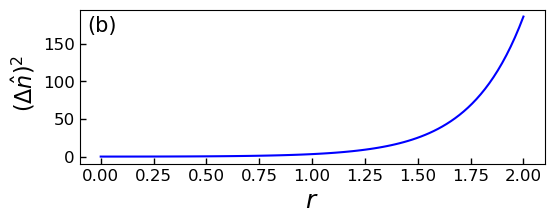}
    \caption{Plots of the Wigner function of single squeezed thermal state (Eq.~\eqref{sqTS}) and Variance $\hat{n}$ (Eq.~\eqref{var:sm}) as a function of squeezing strength $r$. The rest of the parameters are set to $\omega=1$, $T=0.1$, and $r=0.5$. }
    \label{fig:WF}
\end{figure}

We note that squeezing modifies the shape of the Wigner function, as seen in Fig.~\ref{fig:WF}(a) for a squeezed thermal state. In addition, it leads to an exponential growth in the photon number variance with the squeezing strength \( r \) (Fig.~\ref{fig:WF}(b)). This rapid increase in noise dominates the thermal sensitivity encoded in \(\langle \hat{n} \rangle\), reducing the CFI for temperature estimation. Specifically, the variance \(\mathrm{Var}(\hat{n})\) scales as \( e^{4r} \), while the thermal sensitivity \(\partial_T \langle \hat{n} \rangle\) grows only as \( e^{2r} \), causing the signal-to-noise ratio to degrade. Consequently, squeezing alone is counterproductive for thermometry when using photon-number measurements, as it renders \(\hat{n}\) an inefficient estimator for single-mode probes (see Appendix~\ref{quadrature:sm} for quadrature-based measurement). This can be circumvented using two-mode probes with squeezed thermal states or adaptive measurements.\\

\subsection{Squeezed thermal states for two-mode probes }
We consider the system described by Eq.~\eqref{diagHamil}, assumed to be in thermal equilibrium at temperature \( T \), and prepared in a global Gibbs state:
\begin{equation}
    \rho(T) = \frac{1}{\mathcal{Z}} e^{-\tilde{H}_2/T} = \rho_A \otimes \rho_B,
\end{equation}
where each mode independently occupies a thermal state, given by
\begin{equation}
\begin{aligned}
    \rho_A &= \frac{1}{\mathcal{Z}_A} e^{-\tilde{\omega}_+ \hat{A}^\dagger \hat{A} / T}, \quad
    \rho_B &= \frac{1}{\mathcal{Z}_B} e^{-\tilde{\omega}_- \hat{B}^\dagger \hat{B} / T}.
\end{aligned}
\end{equation}
Here, \( \tilde{\omega}_+ \) and \( \tilde{\omega}_- \) are the effective frequencies of the normal modes \( \hat{A} \) and \( \hat{B} \), respectively. The total partition function of the system is then
\begin{equation}
    \mathcal{Z} = \frac{1}{(1 - e^{-\tilde{\omega}_+/T})(1 - e^{-\tilde{\omega}_-/T})}.
\end{equation}
We now present our results based on the CFI evaluated for different measurement observables to estimate the temperature \( T \) in the next section. In particular, we consider: (i) quadrature-based measurements~\cite{RevModPhys.81.299}, (ii) optimal (energy) measurements that maximize the CFI~\cite{individualProbes,Mehboudi_2019}. The results for the population difference between the two modes are discussed in Appendix~\ref{population}. This comparison enables us to investigate various practical measurements for enhancing thermal sensitivity using two-mode squeezed thermal states.
\begin{figure}[t!]
    \centering
    \includegraphics[scale=0.82]{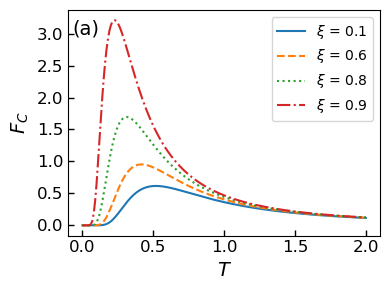}
        \includegraphics[scale=0.82]{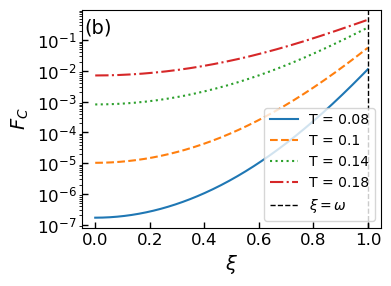}
    \caption{(a) CFI $F_C$ as a function of $T$ for different values of effective coupling constant $\xi$ using quadrature measurements. \textbf{(b)} Approximate low-temperature CFI \(F_C\) (Eq.~\eqref{approx:CFIX}) as a function of \(\xi\) for different temperatures \(T\). The CFI increases with \(\xi\) and exhibits a sharp peak as \(\xi \to \omega\). We fixed \(\omega = 1\) for all curves.}
    \label{fig:CFIX}
\end{figure}
\subsubsection{CFI based on quadrature observable}

We define the two-mode quadrature operator $\hat{X}_\theta$ as
\begin{equation}
    \hat{X}_\theta = \frac{1}{\sqrt{2}} \left( e^{-i\theta}(\hat{a} + \hat{b}) + e^{i\theta}(\hat{a}^\dagger + \hat{b}^\dagger) \right).
\end{equation}
We express \( \hat{a}, \hat{b} \) in terms of \( \hat{A}, \hat{B} \) and plug it in $\hat{X}_\theta$, the resulting expression becomes
\begin{equation}
    \hat{X}_\theta = \frac{1}{\sqrt{2}} \left[ u(\theta)(\hat{A} + \hat{B}) + u^*(\theta)(\hat{A}^\dagger + \hat{B}^\dagger) \right],
\end{equation}
where $ u(\theta) = \cosh r \, e^{-i\theta} - \sinh r \, e^{i\theta}$ is defined for simplicity. The mean \( \langle \hat{X}_\theta \rangle = 0 \) in this case, while the variance is given by

\begin{equation}
    (\Delta\hat{X}_\theta)^2 = \langle \hat{X}_\theta^2 \rangle = |u(\theta)|^2 \left( \langle \hat{A}^\dagger \hat{A}\rangle + \langle \hat{B}^\dagger \hat{B} \rangle + 1 \right).
\end{equation}

Using $\langle \hat{A}^\dagger \hat{A} \rangle =(e^{\tilde{\omega}_+ / T} - 1)^{-1}$ and $\langle \hat{B}^\dagger \hat{B} \rangle=(e^{\tilde{\omega}_- / T} - 1)^{-1}$, we get the variance, which is
\begin{equation}\label{quad}
    (\Delta\hat{X}_\theta)^2 = |u(\theta)|^2 \left( \coth\left( \frac{\tilde{\omega}_+}{2T} \right) + \coth\left( \frac{\tilde{\omega}_-}{2T} \right) \right).
\end{equation}
\begin{figure}[t!]
    \centering
    \includegraphics[scale=0.8]{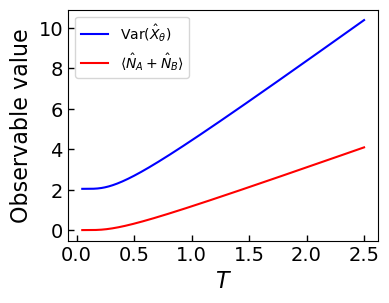}
    \caption{Quadrature variance $\text{Var}(\hat{X}_\theta)$ (blue curve) and the total population average $\langle\hat{N}_A+\hat{N}_B\rangle$ (red curve) as a function of $T$. The rest of the parameters are set to $\omega=1$, $\xi=0.1$, and $\theta=0$.}
    \label{fig:Quad+Total pop}
\end{figure}

Substituting the quadrature variance from Eq.~\eqref{quad} into Eq.~\eqref{CFI:var} yields the following explicit expression for the CFI:
\begin{equation}
    F_C = \frac{1}{2} \left( \frac{\partial_T \coth\left( \frac{\tilde{\omega}_+}{2T} \right) + \partial_T \coth\left( \frac{\tilde{\omega}_-}{2T} \right)}{\coth\left( \frac{\tilde{\omega}_+}{2T} \right) + \coth\left( \frac{\tilde{\omega}_-}{2T} \right)} \right)^2.
\end{equation}
In the degenerate stimulated PDC case, we have $\tilde{\omega}_+=\tilde{\omega}_-=\tilde{\omega}=\sqrt{\omega^2-\xi^2}$, the CFI simplifies to 
\begin{equation}
   F_C = \frac{1}{2} \left( \frac{\partial_T \coth\left( \frac{\tilde{\omega}}{2T} \right) }{\coth\left( \frac{\tilde{\omega}}{2T} \right) } \right)^2.  
\end{equation}
Here, the effective coupling constant $\xi$ quantifies the squeezing and is related to the squeezing parameter $r$ as given in Eq.~\eqref{eq:squeezing}.
We can analyze the CFI in the low-temperature regime, assuming the temperature \( T \) is much smaller than the characteristic frequencies \( \tilde{\omega} \), i.e., \( T \ll \tilde{\omega} \). In this limit, the hyperbolic cotangent function and its temperature derivative can be approximated by their leading exponential terms as
\begin{align}
    \coth\left( \frac{\tilde{\omega}_\pm}{2T} \right) &\approx 1 + 2e^{-\tilde{\omega} / T}, \\
    \partial_T \coth\left( \frac{\tilde{\omega}}{2T} \right) &\approx \frac{2\tilde{\omega}}{T^2} e^{-\tilde{\omega} / T}.
\end{align}
Using these approximations, the CFI simplifies to
\begin{equation}
    F_C \approx \frac{1}{2} \left( \frac{2\tilde{\omega} T^{-2} e^{-\tilde{\omega} / T} }{1 + 2e^{-\tilde{\omega} / T} } \right)^2.
\end{equation}
To leading order in the low-temperature limit $T \to 0$, $e^{-\tilde{\omega}/T}\ll1$ the dominant contribution to the CFI is given by
\begin{equation}
    F_C \sim \frac{\tilde{\omega}^2}{2T^4} \, e^{-2\tilde{\omega}/T}.
\end{equation}
 Next, we analyze the low-temperature behavior of $F_C$ as a function of the effective coupling constant $\xi$. At low temperatures, the dominant contribution to the CFI arises from the lowest accessible excitation energy of the system, which controls the leading exponential suppression. 
For small coupling ($\xi \ll \omega$), this minimal excitation energy can be approximated as
\begin{equation}
\tilde{\omega}_{\min} \approx \omega - \frac{\xi^2}{2\omega},
\end{equation}
and thus the CFI in the low-temperature limit scales as
\begin{equation}\label{approx:CFIX}
     F_C \sim \frac{\big(\omega - \frac{\xi^2}{2\omega}\big)^2}{2T^4} 
    \exp\!\Big[-\frac{2\big(\omega - \frac{\xi^2}{2\omega}\big)}{T}\Big].
\end{equation}
This result reveals that, for \( \omega > \xi \), the precision of temperature estimation improves exponentially as the coupling \( \xi \) increases. To validate the approximation~\eqref{approx:CFIX}, Fig.~\ref{fig:CFIX}(b) shows that \(F_C\) increases with \(\xi\) and diverges sharply as $\xi$ approaches \(\omega\). This marks the boundary where the approximation remains valid and squeezing is beneficial. Additionally, \(F_C\) increases with increasing temperature, consistent with the inverse \(T^4\) scaling and the exponential suppression in Eq.~\eqref{approx:CFIX}.

\begin{figure}[t!]
    \centering
    \includegraphics[scale=0.82]{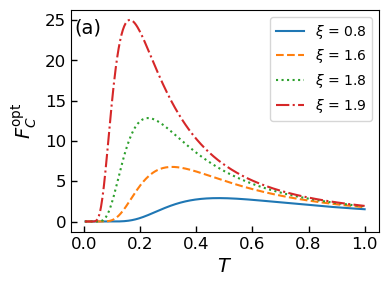}
        \includegraphics[scale=0.82]{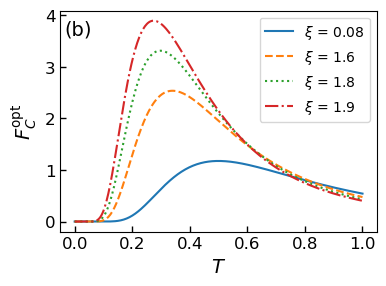}
    \caption{ (a) Optimal CFI \(F_C^{\mathrm{opt}}\) as a function of temperature \(T\) computed using the exact formula with optimal measurements. (b) Low-temperature approximation of the optimal CFI \(F_C^{\mathrm{opt}}\) versus \(T\). Different curves represent various values of effective coupling \(\xi\). In both plots, the base frequency is fixed at \(\omega = 2\). The results illustrate how increasing the coupling $\xi$ affects the temperature sensitivity, especially in the low-temperature regime.}
    \label{fig:optimalCFI}
\end{figure}
\begin{table*}[t!]\label{tab:summary}
\centering
\caption{Summary of temperature estimation performance across regimes, probe types, and observables.
Abbreviations: NE = non-equilibrium, E = equilibrium, SM = single-mode, TM = two-mode,
QFI = quantum Fisher information, CFI = classical Fisher information.}
\label{tab:summary}
\renewcommand{\arraystretch}{1.3}
\begin{tabular}{|p{1.1cm}|p{1.1cm}|p{2.5cm}|p{2cm}|p{1.3cm}|p{2.4cm}|p{4.4cm}|}
\hline
Regime & Probe & Input State & Observable & Speed & Precision & Notes \\
\hline
NE & SM & Vacuum / Coherent & Energy & Slow & Low & Baseline Gaussian case; used as reference for comparison \\
\hline
NE & SM & Squeezed Vacuum & Energy & Moderate & Moderate & Precision improves at higher $T$; performs poorly at low $T$ due to excess noise \\
\hline
NE & SM & Fock / Cat / GKP (non-Gaussian) & Energy & Fast & Moderate to High & Enables early-time sensing; GKP performs best, followed by Cat and Fock \\
\hline
NE & TM & Squeezed Vacuum & Energy & Fastest & High & Improves both speed and QFI over SM probes; accessible via SPDC \\
\hline
NE & TM & NOON / Entangled Cat (non-Gaussian) & Energy & Faster & Moderate to High & Non-Gaussian entangled probes offer faster and precise estimation at short times \\
\hline
E & SM & Squeezed Thermal & $\langle \hat{n} \rangle$, $\hat{x}$ & N/A & Low to Moderate & Squeezing increases noise in photon number; quadrature not sensitive to squeezing \\
\hline
E & TM & Squeezed Thermal & $\langle \hat{H}_2 \rangle$, $\hat{D}$, $\hat{X}_\theta$ & N/A & Moderate to High & Precision grows with squeezing; energy and population difference observables near optimal \\
\hline
\end{tabular}
\end{table*}
Figure~\ref{fig:CFIX}(a) shows the CFI \(F_C\) from quadrature measurement \(\hat{X}_\theta^2\) versus temperature \(T\) for various coupling constant \(\xi\). Increasing \(\xi\) enhances the CFI, improving temperature sensitivity, unlike the single-mode squeezed thermal state, where higher $r$ (squeezing strength) reduces the CFI. This improvement stems from mode correlations in two-mode squeezed thermal states, suggesting that quadrature measurements in two-mode setups outperform single-mode probes, especially at low temperatures. Figure~\ref{fig:Quad+Total pop} shows the quadrature variance \(\mathrm{Var}(\hat{X}_\theta)\) and total population \(\langle \hat{N}_A + \hat{N}_B \rangle\) as functions of temperature \(T\). Although quadrature variance is suboptimal for temperature estimation compared to population measurements, its near-linear dependence on \(T\) makes it a practical and accessible observable. The total population also varies approximately linearly with \(T\) and relates closely to the optimal measurement via the heat capacity.

Quadrature measurements and mean photon number provide experimentally accessible alternatives to optimal POVMs. In particular, quadrature variance can capture temperature information in dissipative regimes~\cite{PhysRevA.96.062103}. It is worth mentioning that quadratures of trapped particles are either directly
measurable~\cite{Bastin_2006} or accessible via state tomography~\cite{PhysRevA.54.R25,PhysRevA.53.R1966}.
\subsubsection{Optimal CFI from partition function and dependence on squeezing}

We recall that the thermal state is factorized, and one can easily calculate the average energy using this state. The average energy is given by
\begin{equation}
    \langle \tilde{H}_2 \rangle = \frac{\tilde{\omega}_+}{e^{\tilde{\omega}_+ / T} - 1} + \frac{\tilde{\omega}_-}{e^{\tilde{\omega}_- / T} - 1}.
\end{equation}
The specific heat $C(T)$ is defined as the temperature derivative of the average energy, such that
\begin{equation}
    C(T) = \frac{d\langle \tilde{H}_2 \rangle}{dT} = \frac{\tilde{\omega}_+^2 e^{\tilde{\omega}_+ / T}}{T^2 (e^{\tilde{\omega}_+ / T} - 1)^2} + \frac{\tilde{\omega}_-^2 e^{\tilde{\omega}_- / T}}{T^2 (e^{\tilde{\omega}_- / T} - 1)^2}.
\end{equation}

Hence, the optimal CFI based on heat capacity becomes
\begin{equation}
    F_C^{\text{opt}} = \frac{C(T)}{T^2} = \frac{\tilde{\omega}_+^2 e^{\tilde{\omega}_+ / T}}{T^4 (e^{\tilde{\omega}_+ / T} - 1)^2} + \frac{\tilde{\omega}_-^2 e^{\tilde{\omega}_- / T}}{T^4 (e^{\tilde{\omega}_- / T} - 1)^2}.
\end{equation}
In the degenerate stimulated PDC case, where \(\tilde{\omega}_+ = \tilde{\omega}_- = \tilde{\omega} = \sqrt{\omega^2 - \xi^2}\), these expressions reduce to
\begin{equation}\label{optimal:CFI}    
\begin{aligned}
    \langle \tilde{H}_2 \rangle &= \frac{2 \tilde{\omega}}{e^{\tilde{\omega} / T} - 1}, \\
    C(T) &= \frac{2 \tilde{\omega}^2 e^{\tilde{\omega} / T}}{T^2 (e^{\tilde{\omega} / T} - 1)^2}, \\
    F_C^{\mathrm{opt}} &= \frac{2 \tilde{\omega}^2 e^{\tilde{\omega} / T}}{T^4 (e^{\tilde{\omega} / T} - 1)^2}.
\end{aligned}
\end{equation}
We plot the optimal CFI, given in Eq.~\eqref{optimal:CFI} as a function of temperature \( T \) for various values of $\xi$ in Fig.~\ref{fig:optimalCFI}(a). The CFI obtained through optimal measurements significantly outperforms both the single-mode CFI and the CFI based on quadrature measurements in the two-mode scenario. Notably, the estimation precision improves markedly in the low-temperature regime, with the CFI exhibiting substantial enhancement for stronger coupling constant $\xi$ values, particularly for \( \xi = 1.9 \).

We now focus on the low-temperature regime \(T \ll \omega\), where the dominant contribution to the optimal CFI is approximated by
\begin{equation}\label{approx:CFI}
    F_C^{\mathrm{opt}} \sim \frac{\left(\omega - \frac{\xi^2}{2\omega}\right)^2}{T^4} \exp\left[-\frac{\omega - \frac{\xi^2}{2\omega}}{T}\right].
\end{equation}
This expression indicates that, in the extreme low-temperature limit, increasing the value of \(\xi\) exponentially enhances the CFI, improving the thermal sensitivity of optimal energy measurements~\cite{Glatthard2023energymeasurements}. Figure~\ref{fig:optimalCFI}(b) compares this approximation with exact numerical results for various values of \(\xi\), showing that the CFI is very small compared to the CFI obtained using the exact expression.

\section{Summary of results}\label{summary}

In this section, we present a comparative summary of the main findings from our analysis of temperature estimation across different quantum probes, input states, and measurement observables. The aim is to highlight the relative advantages and limitations of each strategy, both in non-equilibrium and equilibrium regimes. Table~\ref{tab:summary} compiles the performance in terms of estimation speed, achievable precision, and qualitative insights, serving as a compact reference that complements the detailed results presented in earlier sections. This overview may also guide experimental implementation by identifying practically accessible observables and probe configurations that offer enhanced thermometric performance.

\section{Conclusion}\label{conc}
We investigated quantum thermometry, both in transient and equilibrium regimes, using single-mode probes evolving through a linear medium, used as a benchmark, and two-mode probes generated via a stimulated PDC process. In stimulated PDC, a nonlinear crystal at finite temperature is driven by a strong coherent pump, which produces correlated signal and idler photons seeded with non-vacuum inputs, enabling flexible probe engineering in both Gaussian and non-Gaussian regimes. These modes serve as a quantum probe to estimate the crystal's temperature. In the transient regime, we show that temperature sensing can be significantly accelerated by preparing the probes in suitable non-Gaussian states.

We found that single-mode non-Gaussian probe states, such as Fock, GKP, and odd cat states, enable faster access to temperature information compared to Gaussian probes like the squeezed vacuum state, particularly in the non-equilibrium regime. In the two-mode setting, entangled states—including the TMSVS, NOON state, and entangled cat state—allow temperature information to be extracted at even earlier times. This highlights the advantage of non-Gaussian and entangled probes for accelerating thermometry beyond the capabilities of conventional Gaussian strategies.

In the equilibrium regime, we evaluated the CFI under different practical measurement strategies. Our results indicate that single‑mode squeezed thermal states provide lower estimation precision than thermal states when employing mean‑photon‑number and quadrature‑based measurements. On the other hand, two-mode squeezed thermal states demonstrated enhanced precision in temperature estimation. For two-mode squeezed thermal states, quadrature-based measurements yield lower CFI compared to other approaches, while energy measurements provide the highest precision. Similarly, measurements based on the population difference achieve near-optimal precision, closely matching that of the energy measurement. This precision improvement stemmed from the ability of squeezing to suppress fluctuations in the population difference, directly enhancing the precision of temperature estimation.

In summary, our analysis highlights the complementary advantages of single- and two-mode quantum probes for temperature estimation. Non-Gaussian and entangled states enable faster temperature sensing in the transient regime, while equilibrium-based strategies can achieve higher precision through optimized measurements. These findings establish a trade-off between speed and precision in quantum thermometry, providing useful guidance for experimental implementations across different operating regimes. While our findings are model-specific, the PDC framework is widely applicable in quantum optics and offers a promising platform for quantum thermometry. The probe states analyzed here could be implemented in platforms such as integrated photonics~\cite{Kues2017,Wang2020} or superconducting circuits~\cite{,PhysRevLett.117.110501, PhysRevLett.109.183901}, where PDC-like interactions are accessible.

\section*{Acknowledgment}
This work is supported by the Scientific and Technological Research Council (TÜBİTAK) of T\"urkiye under Project Grant No. 123F150.

\appendix
\section{QFI for an Initial Fock State}\label{QFI:Fock}

The master Eq.~\eqref{sm:master_equation} for a single-mode coupled to a thermal bath admits the following closed-form solution~\cite{Bracken_2013, grochowski2025optimal}
\begin{align}
\rho(t) &= \frac{e^{\frac{\Delta t}{2}}}{F(t)} 
\sum_{n=0}^{\infty} \frac{G(t)^n}{n!} (\hat{a}^\dagger)^n
\left[ e^{-i\omega t} F(t)^{-\hat{N}} \right. \notag\\
&\quad \left. \times \left( 
    \sum_{m=0}^{\infty} \frac{E(t)^m}{m!} \hat{a}^m \rho(0) (\hat{a}^\dagger)^m 
\right)
F(t)^{-\hat{N}} e^{i\omega t} \right] \hat{a}^n.
\label{eq:rho_solution}
\end{align}
where \( \hat{N} = \hat{a}^\dagger \hat{a} \) and
\begin{equation}\label{eqs:coeff}
\begin{aligned}
E(t) &= \frac{2a}{a - b} \frac{\sinh\left( \frac{a - b}{2} t \right)}{F(t)},  \\
F(t) &= \cosh\left( \frac{a - b}{2} t \right) + \frac{a + b}{a - b} \sinh\left( \frac{a - b}{2} t \right),  \\
G(t) &= \frac{2b}{a - b} \frac{\sinh\left( \frac{a - b}{2} t \right)}{F(t)},
\end{aligned}
\end{equation}
with \( \Delta = a - b = \gamma \), $a = \gamma (\bar{n} +1)$, and $b = \gamma \bar{n}$. Assume the initial state is \( \rho(0) = |n_0\rangle \langle n_0| \), and define the populations,
\begin{equation}
p_r(t) = \langle r | \rho(t) | r \rangle. \label{eq:pop_def}
\end{equation}
Substituting Eq.~\eqref{eq:pop_def} into Eq.~\eqref{eq:rho_solution} gives
\begin{equation}
\begin{aligned}
p_r(t) &= \frac{e^{\Delta t/2}}{F(t)} \sum_{n,m \geq 0} \frac{G(t)^n}{n!} \frac{E(t)^m}{m!} \\
&\quad\times\langle r | (\hat{a}^\dagger)^n F(t)^{-\hat{N}} \hat{a}^m | n_0 \rangle \langle n_0 | (\hat{a}^\dagger)^m F(t)^{-\hat{N}} \hat{a}^n | r \rangle. \label{eq:pop_expanded}
\end{aligned}
\end{equation}
Using the identities:
\begin{align}
\hat{a}^m |k\rangle &= \sqrt{\frac{k!}{(k - m)!}} |k - m\rangle, \\
(\hat{a}^\dagger)^n |k\rangle &= \sqrt{\frac{(k + n)!}{k!}} |k + n\rangle, \\
F^{-\hat{N}} |k\rangle &= F^{-k} |k\rangle,
\end{align}
we find that the non-zero contributions in Eq.~\eqref{eq:pop_expanded} occur only if \( m = n + n_0 - r \). This implies the summation range:
\begin{equation}
n = \max(0, r - n_0), \ldots, r. \label{eq:summation_range}
\end{equation}
After simplification, the matrix elements become:
\begin{equation}
\begin{aligned}
&\langle r | (\hat{a}^\dagger)^n F^{-\hat{N}} \hat{a}^m | n_0 \rangle \langle n_0 | (\hat{a}^\dagger)^m F^{-\hat{N}} \hat{a}^n | r \rangle\\
&= \frac{n_0! \, r!}{[(r - n)!]^2} F(t)^{-2(r - n)}. 
\label{eq:matrix_result}
\end{aligned}
\end{equation}
Collecting all terms, the final expression for the population reads
\begin{equation}
\begin{aligned}
p_r(T;t) &= \frac{e^{\Delta t / 2}}{F(t)} \sum_{n = \max(0, r - n_0)}^r 
\frac{G(t)^n}{n!}
\binom{n_0}{n_0 + n - r} \\
&\quad\times E(t)^{n_0 + n - r}
F(t)^{-2(r - n)} \cdot \frac{r!}{(r - n)!}. \label{eq:pop_final}
\end{aligned}
\end{equation}
As $\rho(t)$ is diagonal in the number basis, the symmetric logarithmic derivative $L_T$ is diagonal as well; hence, we can use the following expression to calculate the QFI
\begin{equation}\label{CFI:pop}
    F_Q(T; t) = \sum_{r=0}^{\infty} \frac{[\partial_T p_r(T; t)]^2}{p_r(T; t)}.
\end{equation}
To compute the QFI for temperature estimation, we need the temperature derivatives of the population \( p_r(t) \). These derivatives come from the temperature dependence of \( a \), \( b \), and hence \( \Delta = a - b \), as well as \( F(t) \), \( E(t) \), and \( G(t) \). 
The derivative of the Bose–Einstein distribution is
\begin{equation}
\frac{dn_{\text{th}}}{dT}= n'_{th}= -\frac{\omega}{T^2} n_{\text{th}} (n_{\text{th}} + 1). \label{eq:dT_nth}
\end{equation}
We can define the logarithmic derivatives
\begin{equation}
\frac{F'}{F} = \frac{2S}{F} n'_{\text{th}}, \quad
\frac{E'}{E} = \frac{n'_{\text{th}}}{n_{\text{th}} + 1} - \frac{F'}{F}, \quad
\frac{G'}{G} = \frac{n'_{\text{th}}}{n_{\text{th}}} - \frac{F'}{F}.
\end{equation}
Rewriting Eq.~\eqref{eq:pop_final} such as  
\begin{equation}\label{eq:pop:simp}
    p_r = \frac{e^{\gamma t/2}}{F}\sum_r, 
\end{equation}
where $\Sigma_r=\sum_{r.n}S_{r.n}$ with $S_{r,n}$ being the $n$-th sum in Eq.~\eqref{eq:pop_final}. A direct differentiation of population $p_r$ yields 
\begin{equation}\label{Dp}
    \partial_T p_r = p_r \left[ -\frac{F'}{F} + \Theta_r \right],
\end{equation}
where
\begin{equation}\label{theta}
    \Theta_r = \sum_n w_{r,n} \left( n \frac{G'}{G} + (n_0 + n - r) \frac{E'}{E} - 2(r - n) \frac{F'}{F} \right),
\end{equation}
with $w_{r,n} = S_{r,n} / \Sigma_r$ is a normalized weight.

\begin{figure}
    \centering
    \includegraphics[scale=0.65]{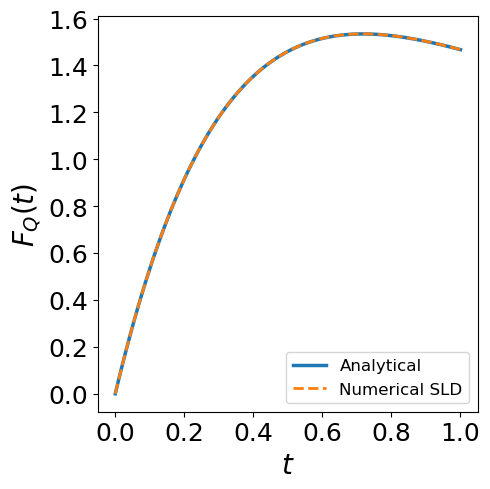}
\caption{QFI $F_Q(t)$ as a function of time $t$ obtained by numerically solving Eq.~\eqref{sm:master_equation} (orange dashed curve) and using an exact expression Eq.~\eqref{exact:QFI} (solid blue curve). The initial state in both cases is a Fock state $\ket{6}$. The other parameters are set to $\omega=1$, $T=0.4$, and $\gamma=0.2$}
    \label{fig:comparison}
\end{figure}

Substituting Eqs.~\eqref{eq:pop:simp} and \eqref{Dp} into Eq.~\eqref{CFI:pop} and expanding the square reproduces the compact formula for QFI, that is
\begin{equation}\label{exact:QFI}
    F_Q(T; t) = \sum_{r=0}^{\infty} \sum_{n,n' = n_{\min}}^r
    \frac{e^{\gamma t} S_{r,n} S_{r,n'}}{F(T; t)^2 \, p_r(T; t)} \, 
    \Theta_n^{(r)} \, \Theta_{n'}^{(r)},
\end{equation}
with $\Theta_n^{(r)}$ is given in Eq.~\eqref{theta}. The exact analytical expression for the QFI is plotted in Fig.~\ref{fig:comparison} and compared with the numerical results obtained by solving the master equation~\eqref{sm:master_equation}, which gives the same results.\\
We now perform some consistency checks to verify the correctness of our analytical results. For an initial time $t \to 0$: $E = G = 0$, $F \to 1$, only $r = n_0$ contributes, hence $F_Q(0) = 0$. In a long time limit, such that when $t \to \infty$: $p_r$ converges to the thermal distribution and Eq.~\eqref{CFI:pop} reduces to the known thermal-state QFI, given as
\begin{equation}
    F_Q^{\text{th}}(T) = \frac{\omega^2}{4 T^4} \, \text{csch}^2\left( \frac{\omega}{2T} \right).
\end{equation}

\subsection{\quad Short--time expansion $t \ll \gamma^{-1}$}
Define the dimensionless time
\begin{equation}
    x \equiv \frac{\gamma t}{2} \quad (x \ll 1).
\end{equation}
And recall that
\begin{equation}
\begin{aligned}
    F(x) &= 1 + (2n_{\text{th}} + 1)x + \frac{x^2}{2} + \mathcal{O}(x^3), \\
    E(x) &= 2(n_{\text{th}} + 1)x + \mathcal{O}(x^2), \quad G(x) = 2n_{\text{th}}x + \mathcal{O}(x^2). 
\end{aligned}
\end{equation}
Inserting the above equations into the exact population formula~\eqref{eq:pop_final}. To linear order in $x$ only the
central level $r = n_0$ and the nearest sidebands $r = n_0 \pm 1$ survive:
\begin{equation}\label{asyp:pop}
\begin{aligned}
    p_{n_0}(t) &= 1 + \mathcal{O}(x),  \\
    p_{n_0 - 1}(t) &= 2n_0(n_{\text{th}} + 1)x + \mathcal{O}(x^2),  \\
    p_{n_0 + 1}(t) &= 2(n_0 + 1)n_{\text{th}}x + \mathcal{O}(x^2).
\end{aligned}
\end{equation}
Using the log--derivative identities, we obtain
\begin{equation}\label{pop:der}
\begin{aligned}
    \partial_T p_{n_0} &= -2(2n_0 + 1)x n'_{\text{th}} + \mathcal{O}(x^2), \\
    \partial_T p_{n_0 - 1} &= 2n_0 x n'_{\text{th}} + \mathcal{O}(x^2), \\
    \partial_T p_{n_0 + 1} &= 2(n_0 + 1)x n'_{\text{th}} + \mathcal{O}(x^2)]. 
\end{aligned}
\end{equation}
Inserting Eqs.~\eqref{asyp:pop} and \eqref{pop:der} into the CFI expression~\eqref{CFI:pop} and keeping the lowest non-vanishing order in $x$ yields the following result
\begin{equation}\label{shorttime}
    F_Q(T, t) = \gamma t \cdot n_{\text{th}}'^2 \left[ \frac{n_0}{n_{\text{th}} + 1} + \frac{n_0 + 1}{n_{\text{th}}} \right] + \mathcal{O}(t^2).
\end{equation}
At very early times, the QFI grows linearly with $t$; its slope is exponentially suppressed at low $T$ through $n_{\text{th}}' \propto e^{-\omega/T}$ but enhanced by the factor $n_0 + 1$ if the initial state is excited.

\begin{figure}
    \centering
    \includegraphics[scale=0.67]{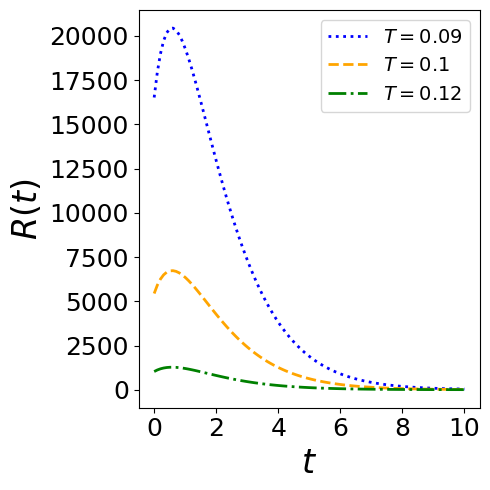}
    \caption{QFI ratio $R(t)$ as a function of time $t$ for different values of temperatures. The other parameters are set to $\omega=1$, $\gamma=0.2$, and $n_0=4$. The two states have fixed energy for which $r=\sinh^{-1}{(\sqrt{n_0})}$.}
    \label{fig:R_long}
\end{figure}
\section{QFI for initial SVS and ratio $R$ in short-time limit}\label{R_short}

A closed expression of the QFI for the SVS under dissipative evolution can be written compactly as (see Eq.~\eqref{eq:QFI})
\begin{equation}
    F_Q^{\text{SVS}}(r, t) = \frac{(1 - e^{-\gamma t})^2 (\partial_T \nu)^2}{2} \cdot \frac{A^2 + B^2 + 2}{(AB)^2 - 1}, \label{eq:SVS_QFI}
\end{equation}
where
\begin{align}
    A &= e^{-\gamma t} r + (1 - e^{-\gamma t}) \nu, \label{eq:A_def} \\
    B &= e^{-\gamma t} r^{-1} + (1 - e^{-\gamma t}) \nu. \label{eq:B_def}
\end{align}
Consider the short-time regime for which we consider $x \equiv \gamma t \ll 1$ and after using the Taylor series expansions for the quantities \(A\) and \(B\), we find
\begin{align}
    A &\approx r + x (\nu - r) + \frac{x^2}{2}(r - \nu)+\mathcal{O}(x^3), \\
    B &\approx r^{-1} + x (\nu - r^{-1}) + \frac{x^2}{2}(r^{-1} - \nu)+\mathcal{O}(x^3),
\end{align}
where \( r \) is the squeezing parameter, and \( \nu = 2\bar{n} + 1 \) with \( \bar{n} \) the thermal occupancy. The product \(AB\) to linear order is
\begin{align}
    AB &\approx 1 + x \left[ \nu(r + r^{-1}) - 2 \right]+\mathcal{O}(x^2).
\end{align}
Defining $C \equiv \nu(r + r^{-1}) - 2,$ such that $AB \approx 1 + x C+\mathcal{O}(x^2)$.
The denominator term is
\begin{equation}
    (AB)^2 - 1 \approx 2x C+\mathcal{O}(x^2).
\end{equation}
For the numerator, at zeroth order in \( x \):
\begin{equation}
    A^2 + B^2 + 2 \approx (r + r^{-1})^2 \equiv u^2,
\end{equation}
where \( u = r + r^{-1} \).
The prefactor expands to
\begin{equation}
    (1 - e^{-\gamma t})^2=\big(x-\frac{x^2}{2}+\mathcal{O}(x^3)\big)^2 \approx x^2+\mathcal{O}(x^3).
\end{equation}
We assemble all the terms such that the QFI then simplifies to
\begin{align}
    F_Q^{\text{SVS}}(t) &\approx \frac{x^2 (\partial_T \nu)^2 u^2}{2 \times 2x C} ,\\
    &= \frac{x (\partial_T \nu)^2 u^2}{4C} + \mathcal{O}(x^2).
\end{align}
Substituting \( x = \gamma t \) and $C=\nu u-2$ back, we obtain
\begin{equation}
    F_Q^{\text{SVS}}(t) \approx \frac{\gamma t}{4} (\partial_T \nu)^2 \frac{u^2}{\nu u - 2}+\mathcal{O}(x^2).
\end{equation}
\begin{figure*}[t!]
\subfloat[]{
    \includegraphics[scale=0.63]{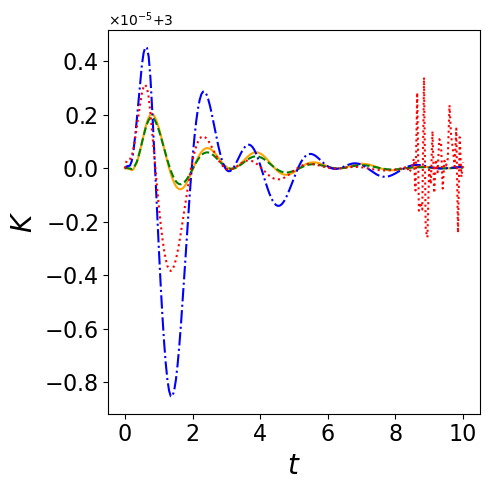}}
    \subfloat[]{
    \includegraphics[scale=0.63]{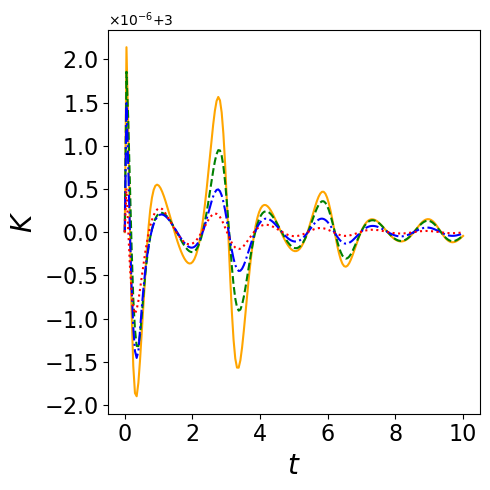}}\\
    \subfloat[]{
    \includegraphics[scale=0.63]{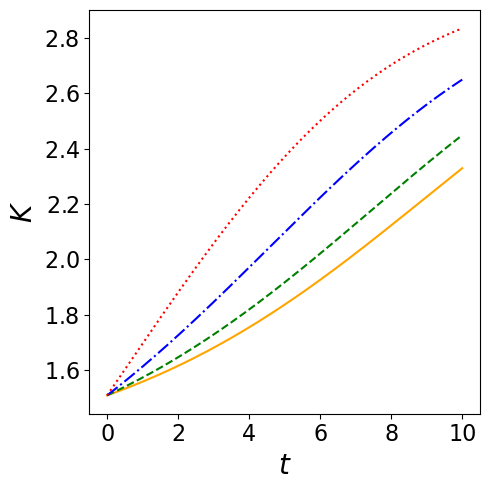}}
    \subfloat[]{
    \includegraphics[scale=0.63]{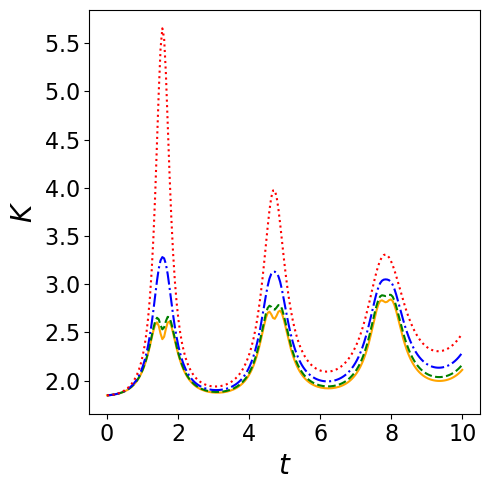}}
    \caption{Kurtosis \( K(t) \) as a function of time \( t \) for different initial single-mode states of the probe. The orange solid, green dashed, blue dot-dashed, and red dotted curves correspond to bath temperatures \( T = 0.1 \), \( 0.5 \), \( 1.0 \), and \( 2.0 \), respectively. Panels (a)--(d) display results for the following initial states: (a) coherent state with \( \alpha = 2 \), (b) squeezed vacuum state with \( r = 0.5 \), (c) Fock state \( |6\rangle \), and (d) GKP state with grid parameter \( \delta = 0.08 \). The oscillator frequency and coupling rate are fixed at \( \omega = 1 \) and \( \gamma = 0.2 \).}
    \label{fig:nG}
\end{figure*}
Expressing in terms of the initial mean photon number \( n_{sv} = \sinh^2 r \) and using \( u = 2n_{sv} + 1 \), the expression becomes
\begin{equation}\label{QFI:SVS_short_time}
    F_Q^{\text{SVS}}(t) \approx \frac{\gamma t}{2} (\partial_T \nu)^2 \frac{(2n_{sv} + 1)^2}{\nu (2n_{sv} + 1) - 1} + \mathcal{O}(x^2).
\end{equation}
The ratio $R$ of the short-time QFI for the Fock state to that of the squeezed vacuum state (SVS) when $n_0=n_{sv}=n$, after simplification, yields the compact expression:
\begin{equation}
R \approx \frac{\nu^2-\frac{1}{(2n+1)^2}}{\nu^2-1}.
\end{equation}
Since the temperature ensures $\nu>1$, the sign of $R-1$ depends on whether
\begin{equation}
    \frac{1}{(2n+1)^2}<1,
\end{equation}
which holds for all $n\ge0$ (equal only at $n=0$). Thus,
\begin{equation}
R(n; \nu) = \begin{cases}
1, & n = 0, \\
> 1, & n > 0,
\end{cases}
\end{equation}
showing that the short-time QFI of a Fock state exceeds that of an equal-energy squeezed vacuum state whenever $n>0$.  
\section{Non-Gaussian characteristics of single-mode probe states}\label{App:Kutosis}
To better understand the role of non-Gaussianity in single-mode quantum probes for thermometry, we analyze how key non-Gaussian features evolve under dissipative dynamics. We track the decay of non-Gaussianity using kurtosis as a witness, highlighting distinct dynamical behaviors across different initial states. This approach offers insight into how fast such probes lose their non-Gaussian character.

In this section, we characterize the probe states using two indicators of non-Gaussianity: statistical skewness and kurtosis.
These indicators capture, respectively, the asymmetry of the quadrature distribution, its peakedness or tails, and the overall departure from Gaussianity in the quantum statistical sense. We analyze these measures dynamically for different bath temperatures
and initial states to reveal how non-Gaussian features develop or decay, and how they correlate with the estimation precision of
temperature quantified earlier via the QFI. We consider the quadrature associated with the position operator, given by
\begin{equation}
    \hat{x} = \frac{1}{\sqrt{2}}(\hat{a} + \hat{a}^\dagger),
\end{equation}
where \(\hat{a}\) and \(\hat{a}^\dagger\) are the annihilation and creation operators of the mode. Let \(\mu = \langle \hat{x} \rangle\) and \(\sigma^2 = \langle (\hat{x} - \mu)^2 \rangle\) denote the mean and variance of the quadrature distribution. We consider two indicators of non-Gaussianity. The first is Skewness, which is the third
standardized moment $\tilde{\mu}$, and quantifies asymmetry,
\begin{equation}
    \gamma_1 := \tilde{\mu}=\frac{\langle (\hat{x} - \mu)^3 \rangle}{\sigma^3}.
\end{equation}
 For a symmetric Gaussian distribution, \(\gamma_1 = 0\). Deviations from zero signal asymmetry and hence non-Gaussianity. This means that the deviations from zero signal asymmetry show non-Gaussianity. The second measure is the Kurtosis, the fourth standardized moment, reflecting peakedness and tail behavior~\cite{PhysRevA.82.052341},
 \begin{equation}
    K := \frac{\langle (\hat{x} - \mu)^4 \rangle}{\sigma^4}.
\end{equation}
A Gaussian distribution has \(K = 3\), and deviations from this value indicate non-Gaussianity.

To investigate the role of initial quantum states and thermal effects on the non-Gaussian features of the probe’s dynamics, we compute the time evolution of the kurtosis \( K(t) \) for various initial single-mode states, as shown in Fig.~10. The four subplots correspond to different initial preparations: (a) a coherent state with amplitude \( \alpha = 2 \), (b) a squeezed vacuum state with squeezing parameter \( r = 0.5 \), (c) a Fock state with \( n = 6 \), and (d) a grid state approximating a GKP encoding with \( \delta = 0.08 \). For each case, we study the influence of the bath temperature by varying \( T \in \{0.1, 0.5, 1.0, 2.0\} \). We emphasize that the skewness \( \gamma_1 \) is zero for all states investigated in this study, indicating that this measure is not sensitive to the non-Gaussian features of the states. Therefore, we only discuss our results on the time-dependent kurtosis. 

In Fig.~\ref{fig:nG}(a), the coherent state exhibits extremely weak non-Gaussianity, with \( K(t) \) remaining within approximately \( 10^{-5} \) of the Gaussian value 3 throughout the evolution, and showing only minor temperature dependence. This reaffirms its quasi-classical character and limited thermometric sensitivity, as reflected in the QFI results (solid blue curve in Fig.~\ref{fig:Smode}). Figure~\ref{fig:nG}(b) shows similar Gaussian-like behavior for the squeezed vacuum state, though with oscillations and temperature sensitivity at short times due to its nonclassical squeezing. The value of \( K(t) \) remains close to 3. In contrast, Fig.~\ref{fig:nG}(c) reveals a distinct monotonic increase in kurtosis for the Fock state, with clearly separated curves for each temperature. This indicates enhanced sensitivity to thermal fluctuations and the development of significant non-Gaussian features over time, with \( K(t) \) saturating around 3. This behavior suggests that the system, starting from a non-Gaussian state, gradually thermalizes into a Gaussian state. Finally, Fig.~\ref{fig:nG}(d) shows that the GKP state exhibits sharp peaks and revival-like features in \( K(t) \), indicative of strong non-Gaussianity and quantum interference effects. These features become increasingly washed out at higher temperatures, highlighting the fragility of the GKP encoding under thermal noise.

Overall, these results demonstrate that kurtosis is a sensitive indicator of the probe’s non-Gaussianity over time. While the coherent and squeezed states remain near-Gaussian throughout thermalization, the Fock and GKP states display pronounced, temperature-dependent deviations, which can offer advantages in quantum thermometry protocols based on non-Gaussian probes. Notably, the temporal and thermal structure of \( K(t) \) mirrors the QFI-based sensitivity hierarchy shown in Fig.~\ref{fig:Smode}: the strong non-Gaussianity and thermal responsiveness of the Fock and GKP states correlate with their superior thermometric performance in the non-equilibrium regime, whereas the coherent and squeezed states exhibit delayed and weak kurtosis responses, consistent with their performance crossover in QFI sensitivity.
\begin{figure}[b!]
    \centering
    \includegraphics[scale=0.62]{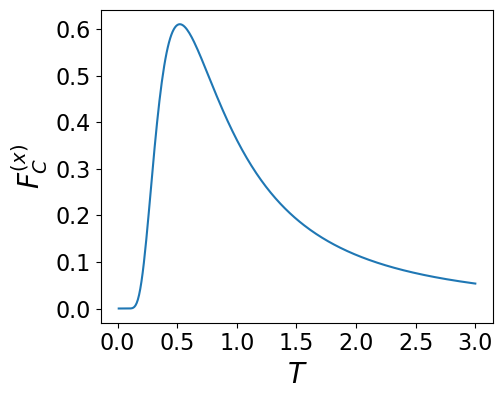}
    \caption{CFI as a function of temperature $T$ for a single-mode squeezed thermal state using quadrature $\hat{x}$ as an observable. Here, we set $\omega=1$.}
    \label{fig:smQ}
\end{figure}
\section{Quadrature-based measurements for a single-mode squeezed thermal states}\label{quadrature:sm}
We now examine the quadrature-based measurement for a single-mode squeezed thermal state~\cite{RevModPhys.81.299}. To this end, for this state, we have
\begin{equation}
    \langle\hat{x}\rangle=0, \quad \sigma_x^2=(\bar{n}+\frac{1}{2})e^{-2r}.
\end{equation}
Substituting the values of $\langle\hat{x}\rangle$ and \( \sigma_x^2 \) in Eq.~\eqref{CFI:var} and simplifying, we find the expression of CFI, given as
\begin{equation}
    F_C^{(x)} = 2 \left( \frac{\partial_T \bar{n}}{2\bar{n} + 1} \right)^2.
\end{equation}

Using the explicit form of the mean thermal photon number $\bar{n}$, we obtain
\begin{equation}\label{CFI:smX}
F_C^{(x)} = \frac{2 \omega^2 e^{2\omega/T}}{T^4 (e^{\omega/T} - 1)^2 (1 + e^{\omega/T})^2}.
\end{equation}
We observe that Eq.~\eqref{CFI:smX} is independent of the effective coupling $\xi$, and depends solely on the temperature $T$ and the mode frequency $\omega$. Figure~\ref{fig:smQ} shows the CFI based on the $x$-quadrature measurement as a function of $T$. The precision offered by this observable is lower than that obtained from measurements based on the mean photon number $\langle \hat{n} \rangle$. 

\section{CFI based on population difference}\label{population}
In this section, we consider the population difference between the two normal modes, defined as 
\begin{equation}
\hat{D} = \hat{N}_A - \hat{N}_B,
\end{equation}
as an experimentally accessible observable for temperature estimation~\cite{PhysRevLett.101.053601}. This quantity measures the difference in thermal populations between the two effective modes. The mean population difference is the difference of the thermal occupation numbers
$\bar{n}_+ = \langle \hat{A}^\dagger \hat{A} \rangle$ and $ \bar{n}_- = \langle \hat{B}^\dagger \hat{B} \rangle$,
which for thermal states follow Bose-Einstein distributions $\bar{n}_\pm =(e^{\tilde{\omega}_\pm / T} - 1)^{-1}$. Since the modes are independent, the mean and the variance of \(\hat{D}\) are the sum of the individual variances:
\begin{equation}
\begin{aligned}
    \langle \hat{D} \rangle &= \bar{n}_+ - \bar{n}_-, \\
    (\Delta \hat{D})^2 &= \bar{n}_+ (1 + \bar{n}_+) + \bar{n}_- (1 + \bar{n}_-).
\end{aligned}
\end{equation}
\begin{figure}[t!]
    \centering
    \includegraphics[scale=0.82]{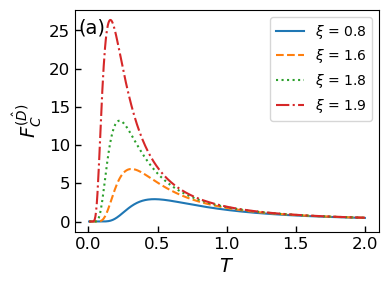}
    \caption{CFI $F_C^{(\hat{D})}$ (see Eq.~\eqref{CFI:popDiff}) associated with the population difference observable \( \hat{D} = \hat{N}_A - \hat{N}_B \), plotted as a function of temperature \( T \) for different values of effective coupling \( \xi \). Here, we consider a near-generate case for which we fixed $\omega_s=2$ and $\omega_i=1.99$.
    }
    \label{fig:PopDiffCFI}
\end{figure}
Applying the CFI expression~\eqref{CFI:Formula} to the population difference (observable), the corresponding CFI becomes
\begin{equation}\label{CFI:popDiff}
\begin{aligned}
F_C^{(\hat{D})} = \frac{1}{(\Delta \hat{D})^2}\left( \dfrac{d\bar{n}_+}{dT} - \dfrac{d\bar{n}_-}{dT} \right)^2.
\end{aligned}
\end{equation}
The derivatives of the occupation numbers are
\begin{equation}
    \frac{d\bar{n}_\pm}{dT} = \frac{\tilde{\omega}_\pm e^{\tilde{\omega}_\pm / T}}{T^2 \left(e^{\tilde{\omega}_\pm / T} - 1\right)^2}.
\end{equation}
In Fig.~\ref{fig:PopDiffCFI}, we plot \( F_C^{(\hat{D})} \) as a function of temperature for different values of effective coupling \( \xi \) for a near-degenerate case. The CFI based on \(\hat{D}\) shows qualitatively similar behavior to the Fisher information obtained from energy measurements (i.e., the heat capacity). Since population measurements are experimentally feasible (e.g., via photon counting or atom number measurements), this observable provides a practical approach to thermometry in coupled bosonic modes. This suggests the potential usefulness of energy-asymmetric observables like \(\hat{D}\) or total population in precision thermometry.

\bibliography{2modestates.bib}

\begin{thebibliography}{104}%
\makeatletter
\providecommand \@ifxundefined [1]{%
 \@ifx{#1\undefined}
}%
\providecommand \@ifnum [1]{%
 \ifnum #1\expandafter \@firstoftwo
 \else \expandafter \@secondoftwo
 \fi
}%
\providecommand \@ifx [1]{%
 \ifx #1\expandafter \@firstoftwo
 \else \expandafter \@secondoftwo
 \fi
}%
\providecommand \natexlab [1]{#1}%
\providecommand \enquote  [1]{``#1''}%
\providecommand \bibnamefont  [1]{#1}%
\providecommand \bibfnamefont [1]{#1}%
\providecommand \citenamefont [1]{#1}%
\providecommand \href@noop [0]{\@secondoftwo}%
\providecommand \href [0]{\begingroup \@sanitize@url \@href}%
\providecommand \@href[1]{\@@startlink{#1}\@@href}%
\providecommand \@@href[1]{\endgroup#1\@@endlink}%
\providecommand \@sanitize@url [0]{\catcode `\\12\catcode `\$12\catcode `\&12\catcode `\#12\catcode `\^12\catcode `\_12\catcode `\%12\relax}%
\providecommand \@@startlink[1]{}%
\providecommand \@@endlink[0]{}%
\providecommand \url  [0]{\begingroup\@sanitize@url \@url }%
\providecommand \@url [1]{\endgroup\@href {#1}{\urlprefix }}%
\providecommand \urlprefix  [0]{URL }%
\providecommand \Eprint [0]{\href }%
\providecommand \doibase [0]{http://dx.doi.org/}%
\providecommand \selectlanguage [0]{\@gobble}%
\providecommand \bibinfo  [0]{\@secondoftwo}%
\providecommand \bibfield  [0]{\@secondoftwo}%
\providecommand \translation [1]{[#1]}%
\providecommand \BibitemOpen [0]{}%
\providecommand \bibitemStop [0]{}%
\providecommand \bibitemNoStop [0]{.\EOS\space}%
\providecommand \EOS [0]{\spacefactor3000\relax}%
\providecommand \BibitemShut  [1]{\csname bibitem#1\endcsname}%
\let\auto@bib@innerbib\@empty
\bibitem [{\citenamefont {De~Pasquale}\ \emph {et~al.}(2016)\citenamefont {De~Pasquale}, \citenamefont {Rossini}, \citenamefont {Fazio},\ and\ \citenamefont {Giovannetti}}]{DePasquale2016}%
  \BibitemOpen
  \bibfield  {author} {\bibinfo {author} {\bibfnamefont {Antonella}\ \bibnamefont {De~Pasquale}}, \bibinfo {author} {\bibfnamefont {Davide}\ \bibnamefont {Rossini}}, \bibinfo {author} {\bibfnamefont {Rosario}\ \bibnamefont {Fazio}}, \ and\ \bibinfo {author} {\bibfnamefont {Vittorio}\ \bibnamefont {Giovannetti}},\ }\bibfield  {title} {\enquote {\bibinfo {title} {Local quantum thermal susceptibility},}\ }\href {\doibase 10.1038/ncomms12782} {\bibfield  {journal} {\bibinfo  {journal} {Nat. Commun.}\ }\textbf {\bibinfo {volume} {7}},\ \bibinfo {pages} {12782} (\bibinfo {year} {2016})}\BibitemShut {NoStop}%
\bibitem [{\citenamefont {Goold}\ \emph {et~al.}(2016)\citenamefont {Goold}, \citenamefont {Huber}, \citenamefont {Riera}, \citenamefont {Rio},\ and\ \citenamefont {Skrzypczyk}}]{Goold_2016}%
  \BibitemOpen
  \bibfield  {author} {\bibinfo {author} {\bibfnamefont {John}\ \bibnamefont {Goold}}, \bibinfo {author} {\bibfnamefont {Marcus}\ \bibnamefont {Huber}}, \bibinfo {author} {\bibfnamefont {Arnau}\ \bibnamefont {Riera}}, \bibinfo {author} {\bibfnamefont {Lídia~del}\ \bibnamefont {Rio}}, \ and\ \bibinfo {author} {\bibfnamefont {Paul}\ \bibnamefont {Skrzypczyk}},\ }\bibfield  {title} {\enquote {\bibinfo {title} {The role of quantum information in thermodynamics—a topical review},}\ }\href {\doibase 10.1088/1751-8113/49/14/143001} {\bibfield  {journal} {\bibinfo  {journal} {J. Phys. A: Math. Theor.}\ }\textbf {\bibinfo {volume} {49}},\ \bibinfo {pages} {143001} (\bibinfo {year} {2016})}\BibitemShut {NoStop}%
\bibitem [{\citenamefont {Levitin}\ \emph {et~al.}(2022)\citenamefont {Levitin}, \citenamefont {van~der Vliet}, \citenamefont {Theisen}, \citenamefont {Dimitriadis}, \citenamefont {Lucas}, \citenamefont {Corcoles}, \citenamefont {Ny{\'e}ki}, \citenamefont {Casey}, \citenamefont {Creeth}, \citenamefont {Farrer}, \citenamefont {Ritchie}, \citenamefont {Nicholls},\ and\ \citenamefont {Saunders}}]{Levitin2022}%
  \BibitemOpen
  \bibfield  {author} {\bibinfo {author} {\bibfnamefont {Lev~V.}\ \bibnamefont {Levitin}}, \bibinfo {author} {\bibfnamefont {Harriet}\ \bibnamefont {van~der Vliet}}, \bibinfo {author} {\bibfnamefont {Terje}\ \bibnamefont {Theisen}}, \bibinfo {author} {\bibfnamefont {Stefanos}\ \bibnamefont {Dimitriadis}}, \bibinfo {author} {\bibfnamefont {Marijn}\ \bibnamefont {Lucas}}, \bibinfo {author} {\bibfnamefont {Antonio~D.}\ \bibnamefont {Corcoles}}, \bibinfo {author} {\bibfnamefont {J{\'a}n}\ \bibnamefont {Ny{\'e}ki}}, \bibinfo {author} {\bibfnamefont {Andrew~J.}\ \bibnamefont {Casey}}, \bibinfo {author} {\bibfnamefont {Graham}\ \bibnamefont {Creeth}}, \bibinfo {author} {\bibfnamefont {Ian}\ \bibnamefont {Farrer}}, \bibinfo {author} {\bibfnamefont {David~A.}\ \bibnamefont {Ritchie}}, \bibinfo {author} {\bibfnamefont {James~T.}\ \bibnamefont {Nicholls}}, \ and\ \bibinfo {author} {\bibfnamefont {John}\ \bibnamefont {Saunders}},\ }\bibfield  {title} {\enquote {\bibinfo {title} {Cooling low-dimensional electron systems
  into the microkelvin regime},}\ }\href {\doibase 10.1038/s41467-022-28222-x} {\bibfield  {journal} {\bibinfo  {journal} {Nat. Commun.}\ }\textbf {\bibinfo {volume} {13}},\ \bibinfo {pages} {667} (\bibinfo {year} {2022})}\BibitemShut {NoStop}%
\bibitem [{\citenamefont {Sarsby}\ \emph {et~al.}(2020)\citenamefont {Sarsby}, \citenamefont {Yurttag{\"u}l},\ and\ \citenamefont {Geresdi}}]{Sarsby2020}%
  \BibitemOpen
  \bibfield  {author} {\bibinfo {author} {\bibfnamefont {Matthew}\ \bibnamefont {Sarsby}}, \bibinfo {author} {\bibfnamefont {Nikolai}\ \bibnamefont {Yurttag{\"u}l}}, \ and\ \bibinfo {author} {\bibfnamefont {Attila}\ \bibnamefont {Geresdi}},\ }\bibfield  {title} {\enquote {\bibinfo {title} {500 microkelvin nanoelectronics},}\ }\href {\doibase 10.1038/s41467-020-15201-3} {\bibfield  {journal} {\bibinfo  {journal} {Nat. Commun.}\ }\textbf {\bibinfo {volume} {11}},\ \bibinfo {pages} {1492} (\bibinfo {year} {2020})}\BibitemShut {NoStop}%
\bibitem [{\citenamefont {Correa}\ \emph {et~al.}(2015)\citenamefont {Correa}, \citenamefont {Mehboudi}, \citenamefont {Adesso},\ and\ \citenamefont {Sanpera}}]{individualProbes}%
  \BibitemOpen
  \bibfield  {author} {\bibinfo {author} {\bibfnamefont {Luis~A.}\ \bibnamefont {Correa}}, \bibinfo {author} {\bibfnamefont {Mohammad}\ \bibnamefont {Mehboudi}}, \bibinfo {author} {\bibfnamefont {Gerardo}\ \bibnamefont {Adesso}}, \ and\ \bibinfo {author} {\bibfnamefont {Anna}\ \bibnamefont {Sanpera}},\ }\bibfield  {title} {\enquote {\bibinfo {title} {Individual quantum probes for optimal thermometry},}\ }\href {\doibase 10.1103/PhysRevLett.114.220405} {\bibfield  {journal} {\bibinfo  {journal} {Phys. Rev. Lett.}\ }\textbf {\bibinfo {volume} {114}},\ \bibinfo {pages} {220405} (\bibinfo {year} {2015})}\BibitemShut {NoStop}%
\bibitem [{\citenamefont {De~Pasquale}\ and\ \citenamefont {Stace}(2018)}]{DePasquale2018}%
  \BibitemOpen
  \bibfield  {author} {\bibinfo {author} {\bibfnamefont {Antonella}\ \bibnamefont {De~Pasquale}}\ and\ \bibinfo {author} {\bibfnamefont {Thomas~M.}\ \bibnamefont {Stace}},\ }\bibfield  {title} {\enquote {\bibinfo {title} {Quantum thermometry},}\ }in\ \href {\doibase 10.1007/978-3-319-99046-0_21} {\emph {\bibinfo {booktitle} {Thermodynamics in the Quantum Regime: Fundamental Aspects and New Directions}}},\ \bibinfo {editor} {edited by\ \bibinfo {editor} {\bibfnamefont {Felix}\ \bibnamefont {Binder}}, \bibinfo {editor} {\bibfnamefont {Luis~A.}\ \bibnamefont {Correa}}, \bibinfo {editor} {\bibfnamefont {Christian}\ \bibnamefont {Gogolin}}, \bibinfo {editor} {\bibfnamefont {Janet}\ \bibnamefont {Anders}}, \ and\ \bibinfo {editor} {\bibfnamefont {Gerardo}\ \bibnamefont {Adesso}}}\ (\bibinfo  {publisher} {Springer},\ \bibinfo {address} {New York},\ \bibinfo {year} {2018})\ pp.\ \bibinfo {pages} {503--527}\BibitemShut {NoStop}%
\bibitem [{\citenamefont {Mehboudi}\ \emph {et~al.}(2019{\natexlab{a}})\citenamefont {Mehboudi}, \citenamefont {Sanpera},\ and\ \citenamefont {Correa}}]{Mehboudi_2019}%
  \BibitemOpen
  \bibfield  {author} {\bibinfo {author} {\bibfnamefont {Mohammad}\ \bibnamefont {Mehboudi}}, \bibinfo {author} {\bibfnamefont {Anna}\ \bibnamefont {Sanpera}}, \ and\ \bibinfo {author} {\bibfnamefont {Luis~A}\ \bibnamefont {Correa}},\ }\bibfield  {title} {\enquote {\bibinfo {title} {Thermometry in the quantum regime: recent theoretical progress},}\ }\href {\doibase 10.1088/1751-8121/ab2828} {\bibfield  {journal} {\bibinfo  {journal} {J. Phys. A: Math. Theor.}\ }\textbf {\bibinfo {volume} {52}},\ \bibinfo {pages} {303001} (\bibinfo {year} {2019}{\natexlab{a}})}\BibitemShut {NoStop}%
\bibitem [{\citenamefont {Montenegro}\ \emph {et~al.}(2025)\citenamefont {Montenegro}, \citenamefont {Mukhopadhyay}, \citenamefont {Yousefjani}, \citenamefont {Sarkar}, \citenamefont {Mishra}, \citenamefont {Paris},\ and\ \citenamefont {Bayat}}]{Montenegro2025}%
  \BibitemOpen
  \bibfield  {author} {\bibinfo {author} {\bibfnamefont {Victor}\ \bibnamefont {Montenegro}}, \bibinfo {author} {\bibfnamefont {Chiranjib}\ \bibnamefont {Mukhopadhyay}}, \bibinfo {author} {\bibfnamefont {Rozhin}\ \bibnamefont {Yousefjani}}, \bibinfo {author} {\bibfnamefont {Saubhik}\ \bibnamefont {Sarkar}}, \bibinfo {author} {\bibfnamefont {Utkarsh}\ \bibnamefont {Mishra}}, \bibinfo {author} {\bibfnamefont {Matteo~G.A.}\ \bibnamefont {Paris}}, \ and\ \bibinfo {author} {\bibfnamefont {Abolfazl}\ \bibnamefont {Bayat}},\ }\bibfield  {title} {\enquote {\bibinfo {title} {Review: Quantum metrology and sensing with many-body systems},}\ }\href {\doibase https://doi.org/10.1016/j.physrep.2025.05.005} {\bibfield  {journal} {\bibinfo  {journal} {Phys. Rep.}\ }\textbf {\bibinfo {volume} {1134}},\ \bibinfo {pages} {1--62} (\bibinfo {year} {2025})}\BibitemShut {NoStop}%
\bibitem [{\citenamefont {Jevtic}\ \emph {et~al.}(2015)\citenamefont {Jevtic}, \citenamefont {Newman}, \citenamefont {Rudolph},\ and\ \citenamefont {Stace}}]{PhysRevA.91.012331}%
  \BibitemOpen
  \bibfield  {author} {\bibinfo {author} {\bibfnamefont {Sania}\ \bibnamefont {Jevtic}}, \bibinfo {author} {\bibfnamefont {David}\ \bibnamefont {Newman}}, \bibinfo {author} {\bibfnamefont {Terry}\ \bibnamefont {Rudolph}}, \ and\ \bibinfo {author} {\bibfnamefont {T.~M.}\ \bibnamefont {Stace}},\ }\bibfield  {title} {\enquote {\bibinfo {title} {Single-qubit thermometry},}\ }\href {\doibase 10.1103/PhysRevA.91.012331} {\bibfield  {journal} {\bibinfo  {journal} {Phys. Rev. A}\ }\textbf {\bibinfo {volume} {91}},\ \bibinfo {pages} {012331} (\bibinfo {year} {2015})}\BibitemShut {NoStop}%
\bibitem [{\citenamefont {Ullah}\ \emph {et~al.}(2025{\natexlab{a}})\citenamefont {Ullah}, \citenamefont {Upadhyay},\ and\ \citenamefont {Müstecaplıoğlu}}]{UllahSpinchains}%
  \BibitemOpen
  \bibfield  {author} {\bibinfo {author} {\bibfnamefont {Asghar}\ \bibnamefont {Ullah}}, \bibinfo {author} {\bibfnamefont {Vipul}\ \bibnamefont {Upadhyay}}, \ and\ \bibinfo {author} {\bibfnamefont {Özgür~E.}\ \bibnamefont {Müstecaplıoğlu}},\ }\bibfield  {title} {\enquote {\bibinfo {title} {Quantum thermometry for ultra-low temperatures using probe and ancilla qubit chains},}\ }\href {\doibase 10.3390/e27020204} {\bibfield  {journal} {\bibinfo  {journal} {Entropy}\ }\textbf {\bibinfo {volume} {27}} (\bibinfo {year} {2025}{\natexlab{a}}),\ 10.3390/e27020204}\BibitemShut {NoStop}%
\bibitem [{\citenamefont {Gebbia}\ \emph {et~al.}(2020)\citenamefont {Gebbia}, \citenamefont {Benedetti}, \citenamefont {Benatti}, \citenamefont {Floreanini}, \citenamefont {Bina},\ and\ \citenamefont {Paris}}]{PhysRevA.101.032112}%
  \BibitemOpen
  \bibfield  {author} {\bibinfo {author} {\bibfnamefont {Francesca}\ \bibnamefont {Gebbia}}, \bibinfo {author} {\bibfnamefont {Claudia}\ \bibnamefont {Benedetti}}, \bibinfo {author} {\bibfnamefont {Fabio}\ \bibnamefont {Benatti}}, \bibinfo {author} {\bibfnamefont {Roberto}\ \bibnamefont {Floreanini}}, \bibinfo {author} {\bibfnamefont {Matteo}\ \bibnamefont {Bina}}, \ and\ \bibinfo {author} {\bibfnamefont {Matteo G.~A.}\ \bibnamefont {Paris}},\ }\bibfield  {title} {\enquote {\bibinfo {title} {Two-qubit quantum probes for the temperature of an ohmic environment},}\ }\href {\doibase 10.1103/PhysRevA.101.032112} {\bibfield  {journal} {\bibinfo  {journal} {Phys. Rev. A}\ }\textbf {\bibinfo {volume} {101}},\ \bibinfo {pages} {032112} (\bibinfo {year} {2020})}\BibitemShut {NoStop}%
\bibitem [{\citenamefont {Aiache}\ \emph {et~al.}(2024{\natexlab{a}})\citenamefont {Aiache}, \citenamefont {El~Allati},\ and\ \citenamefont {El~Anouz}}]{PhysRevA.110.032605}%
  \BibitemOpen
  \bibfield  {author} {\bibinfo {author} {\bibfnamefont {Y.}~\bibnamefont {Aiache}}, \bibinfo {author} {\bibfnamefont {A.}~\bibnamefont {El~Allati}}, \ and\ \bibinfo {author} {\bibfnamefont {K.}~\bibnamefont {El~Anouz}},\ }\bibfield  {title} {\enquote {\bibinfo {title} {Harnessing coherence generation for precision single- and two-qubit quantum thermometry},}\ }\href {\doibase 10.1103/PhysRevA.110.032605} {\bibfield  {journal} {\bibinfo  {journal} {Phys. Rev. A}\ }\textbf {\bibinfo {volume} {110}},\ \bibinfo {pages} {032605} (\bibinfo {year} {2024}{\natexlab{a}})}\BibitemShut {NoStop}%
\bibitem [{\citenamefont {Brunelli}\ \emph {et~al.}(2011)\citenamefont {Brunelli}, \citenamefont {Olivares},\ and\ \citenamefont {Paris}}]{PhysRevA.84.032105}%
  \BibitemOpen
  \bibfield  {author} {\bibinfo {author} {\bibfnamefont {Matteo}\ \bibnamefont {Brunelli}}, \bibinfo {author} {\bibfnamefont {Stefano}\ \bibnamefont {Olivares}}, \ and\ \bibinfo {author} {\bibfnamefont {Matteo G.~A.}\ \bibnamefont {Paris}},\ }\bibfield  {title} {\enquote {\bibinfo {title} {Qubit thermometry for micromechanical resonators},}\ }\href {\doibase 10.1103/PhysRevA.84.032105} {\bibfield  {journal} {\bibinfo  {journal} {Phys. Rev. A}\ }\textbf {\bibinfo {volume} {84}},\ \bibinfo {pages} {032105} (\bibinfo {year} {2011})}\BibitemShut {NoStop}%
\bibitem [{\citenamefont {Ullah}\ \emph {et~al.}(2023)\citenamefont {Ullah}, \citenamefont {Naseem},\ and\ \citenamefont {M\"ustecapl\ifmmode \imath \else \i \fi{}o\ifmmode~\breve{g}\else \u{g}\fi{}lu}}]{PhysRevResearch.5.043184}%
  \BibitemOpen
  \bibfield  {author} {\bibinfo {author} {\bibfnamefont {Asghar}\ \bibnamefont {Ullah}}, \bibinfo {author} {\bibfnamefont {M.~Tahir}\ \bibnamefont {Naseem}}, \ and\ \bibinfo {author} {\bibfnamefont {\"Ozg\"ur~E.}\ \bibnamefont {M\"ustecapl\ifmmode \imath \else \i \fi{}o\ifmmode~\breve{g}\else \u{g}\fi{}lu}},\ }\bibfield  {title} {\enquote {\bibinfo {title} {Low-temperature quantum thermometry boosted by coherence generation},}\ }\href {\doibase 10.1103/PhysRevResearch.5.043184} {\bibfield  {journal} {\bibinfo  {journal} {Phys. Rev. Res.}\ }\textbf {\bibinfo {volume} {5}},\ \bibinfo {pages} {043184} (\bibinfo {year} {2023})}\BibitemShut {NoStop}%
\bibitem [{\citenamefont {Brunelli}\ \emph {et~al.}(2012)\citenamefont {Brunelli}, \citenamefont {Olivares}, \citenamefont {Paternostro},\ and\ \citenamefont {Paris}}]{PhysRevA.86.012125}%
  \BibitemOpen
  \bibfield  {author} {\bibinfo {author} {\bibfnamefont {Matteo}\ \bibnamefont {Brunelli}}, \bibinfo {author} {\bibfnamefont {Stefano}\ \bibnamefont {Olivares}}, \bibinfo {author} {\bibfnamefont {Mauro}\ \bibnamefont {Paternostro}}, \ and\ \bibinfo {author} {\bibfnamefont {Matteo G.~A.}\ \bibnamefont {Paris}},\ }\bibfield  {title} {\enquote {\bibinfo {title} {Qubit-assisted thermometry of a quantum harmonic oscillator},}\ }\href {\doibase 10.1103/PhysRevA.86.012125} {\bibfield  {journal} {\bibinfo  {journal} {Phys. Rev. A}\ }\textbf {\bibinfo {volume} {86}},\ \bibinfo {pages} {012125} (\bibinfo {year} {2012})}\BibitemShut {NoStop}%
\bibitem [{\citenamefont {Razavian}\ \emph {et~al.}(2019)\citenamefont {Razavian}, \citenamefont {Benedetti}, \citenamefont {Bina}, \citenamefont {Akbari-Kourbolagh},\ and\ \citenamefont {Paris}}]{Razavian2019}%
  \BibitemOpen
  \bibfield  {author} {\bibinfo {author} {\bibfnamefont {Sholeh}\ \bibnamefont {Razavian}}, \bibinfo {author} {\bibfnamefont {Claudia}\ \bibnamefont {Benedetti}}, \bibinfo {author} {\bibfnamefont {Matteo}\ \bibnamefont {Bina}}, \bibinfo {author} {\bibfnamefont {Yahya}\ \bibnamefont {Akbari-Kourbolagh}}, \ and\ \bibinfo {author} {\bibfnamefont {Matteo G.~A.}\ \bibnamefont {Paris}},\ }\bibfield  {title} {\enquote {\bibinfo {title} {Quantum thermometry by single-qubit dephasing},}\ }\href {\doibase 10.1140/epjp/i2019-12708-9} {\bibfield  {journal} {\bibinfo  {journal} {Eur. Phys. J. Plus}\ }\textbf {\bibinfo {volume} {134}},\ \bibinfo {pages} {284} (\bibinfo {year} {2019})}\BibitemShut {NoStop}%
\bibitem [{\citenamefont {Feyles}\ \emph {et~al.}(2019)\citenamefont {Feyles}, \citenamefont {Mancino}, \citenamefont {Sbroscia}, \citenamefont {Gianani},\ and\ \citenamefont {Barbieri}}]{PhysRevA.99.062114}%
  \BibitemOpen
  \bibfield  {author} {\bibinfo {author} {\bibfnamefont {Michele~M.}\ \bibnamefont {Feyles}}, \bibinfo {author} {\bibfnamefont {Luca}\ \bibnamefont {Mancino}}, \bibinfo {author} {\bibfnamefont {Marco}\ \bibnamefont {Sbroscia}}, \bibinfo {author} {\bibfnamefont {Ilaria}\ \bibnamefont {Gianani}}, \ and\ \bibinfo {author} {\bibfnamefont {Marco}\ \bibnamefont {Barbieri}},\ }\bibfield  {title} {\enquote {\bibinfo {title} {Dynamical role of quantum signatures in quantum thermometry},}\ }\href {\doibase 10.1103/PhysRevA.99.062114} {\bibfield  {journal} {\bibinfo  {journal} {Phys. Rev. A}\ }\textbf {\bibinfo {volume} {99}},\ \bibinfo {pages} {062114} (\bibinfo {year} {2019})}\BibitemShut {NoStop}%
\bibitem [{\citenamefont {Ullah}\ \emph {et~al.}(2025{\natexlab{b}})\citenamefont {Ullah}, \citenamefont {Cattaneo},\ and\ \citenamefont {M\"ustecapl\ifmmode \imath \else \i \fi{}o\ifmmode~\breve{g}\else \u{g}\fi{}lu}}]{Singlequbit}%
  \BibitemOpen
  \bibfield  {author} {\bibinfo {author} {\bibfnamefont {Asghar}\ \bibnamefont {Ullah}}, \bibinfo {author} {\bibfnamefont {Marco}\ \bibnamefont {Cattaneo}}, \ and\ \bibinfo {author} {\bibfnamefont {\"Ozg\"ur~E.}\ \bibnamefont {M\"ustecapl\ifmmode \imath \else \i \fi{}o\ifmmode~\breve{g}\else \u{g}\fi{}lu}},\ }\bibfield  {title} {\enquote {\bibinfo {title} {Single-qubit probes for temperature estimation in the presence of collective baths},}\ }\href {\doibase 10.1103/PhysRevA.111.062201} {\bibfield  {journal} {\bibinfo  {journal} {Phys. Rev. A}\ }\textbf {\bibinfo {volume} {111}},\ \bibinfo {pages} {062201} (\bibinfo {year} {2025}{\natexlab{b}})}\BibitemShut {NoStop}%
\bibitem [{\citenamefont {Abiuso}\ \emph {et~al.}(2024)\citenamefont {Abiuso}, \citenamefont {Andrea~Erdman}, \citenamefont {Ronen}, \citenamefont {Noé}, \citenamefont {Haack},\ and\ \citenamefont {Perarnau-Llobet}}]{Abiuso_2024}%
  \BibitemOpen
  \bibfield  {author} {\bibinfo {author} {\bibfnamefont {Paolo}\ \bibnamefont {Abiuso}}, \bibinfo {author} {\bibfnamefont {Paolo}\ \bibnamefont {Andrea~Erdman}}, \bibinfo {author} {\bibfnamefont {Michael}\ \bibnamefont {Ronen}}, \bibinfo {author} {\bibfnamefont {Frank}\ \bibnamefont {Noé}}, \bibinfo {author} {\bibfnamefont {Géraldine}\ \bibnamefont {Haack}}, \ and\ \bibinfo {author} {\bibfnamefont {Martí}\ \bibnamefont {Perarnau-Llobet}},\ }\bibfield  {title} {\enquote {\bibinfo {title} {Optimal thermometers with spin networks},}\ }\href {\doibase 10.1088/2058-9565/ad37d3} {\bibfield  {journal} {\bibinfo  {journal} {Quantum Sci. Technol.}\ }\textbf {\bibinfo {volume} {9}},\ \bibinfo {pages} {035008} (\bibinfo {year} {2024})}\BibitemShut {NoStop}%
\bibitem [{\citenamefont {Aybar}\ \emph {et~al.}(2022)\citenamefont {Aybar}, \citenamefont {Niezgoda}, \citenamefont {Mirkhalaf}, \citenamefont {Mitchell}, \citenamefont {Benedicto~Orenes},\ and\ \citenamefont {Witkowska}}]{Aybar2022criticalquantum}%
  \BibitemOpen
  \bibfield  {author} {\bibinfo {author} {\bibfnamefont {Enes}\ \bibnamefont {Aybar}}, \bibinfo {author} {\bibfnamefont {Artur}\ \bibnamefont {Niezgoda}}, \bibinfo {author} {\bibfnamefont {Safoura~S.}\ \bibnamefont {Mirkhalaf}}, \bibinfo {author} {\bibfnamefont {Morgan~W.}\ \bibnamefont {Mitchell}}, \bibinfo {author} {\bibfnamefont {Daniel}\ \bibnamefont {Benedicto~Orenes}}, \ and\ \bibinfo {author} {\bibfnamefont {Emilia}\ \bibnamefont {Witkowska}},\ }\bibfield  {title} {\enquote {\bibinfo {title} {Critical quantum thermometry and its feasibility in spin systems},}\ }\href {\doibase 10.22331/q-2022-09-19-808} {\bibfield  {journal} {\bibinfo  {journal} {{Quantum}}\ }\textbf {\bibinfo {volume} {6}},\ \bibinfo {pages} {808} (\bibinfo {year} {2022})}\BibitemShut {NoStop}%
\bibitem [{\citenamefont {Guo}\ \emph {et~al.}(2015)\citenamefont {Guo}, \citenamefont {Xu}, \citenamefont {Zou},\ and\ \citenamefont {Shao}}]{RingQuant}%
  \BibitemOpen
  \bibfield  {author} {\bibinfo {author} {\bibfnamefont {Li-Sha}\ \bibnamefont {Guo}}, \bibinfo {author} {\bibfnamefont {Bao-Ming}\ \bibnamefont {Xu}}, \bibinfo {author} {\bibfnamefont {Jian}\ \bibnamefont {Zou}}, \ and\ \bibinfo {author} {\bibfnamefont {Bin}\ \bibnamefont {Shao}},\ }\bibfield  {title} {\enquote {\bibinfo {title} {Improved thermometry of low-temperature quantum systems by a ring-structure probe},}\ }\href {\doibase 10.1103/PhysRevA.92.052112} {\bibfield  {journal} {\bibinfo  {journal} {Phys. Rev. A}\ }\textbf {\bibinfo {volume} {92}},\ \bibinfo {pages} {052112} (\bibinfo {year} {2015})}\BibitemShut {NoStop}%
\bibitem [{\citenamefont {Srivastava}\ \emph {et~al.}(2025)\citenamefont {Srivastava}, \citenamefont {Bhattacharya}, \citenamefont {Lewenstein},\ and\ \citenamefont {P\l{}odzie\ifmmode~\acute{n}\else \'{n}\fi{}}}]{Topological2025}%
  \BibitemOpen
  \bibfield  {author} {\bibinfo {author} {\bibfnamefont {Anubhav~Kumar}\ \bibnamefont {Srivastava}}, \bibinfo {author} {\bibfnamefont {Utso}\ \bibnamefont {Bhattacharya}}, \bibinfo {author} {\bibfnamefont {Maciej}\ \bibnamefont {Lewenstein}}, \ and\ \bibinfo {author} {\bibfnamefont {Marcin}\ \bibnamefont {P\l{}odzie\ifmmode~\acute{n}\else \'{n}\fi{}}},\ }\bibfield  {title} {\enquote {\bibinfo {title} {Topological quantum thermometry},}\ }\href {\doibase 10.1103/PhysRevA.111.052216} {\bibfield  {journal} {\bibinfo  {journal} {Phys. Rev. A}\ }\textbf {\bibinfo {volume} {111}},\ \bibinfo {pages} {052216} (\bibinfo {year} {2025})}\BibitemShut {NoStop}%
\bibitem [{\citenamefont {Ullah}\ \emph {et~al.}(2025{\natexlab{c}})\citenamefont {Ullah}, \citenamefont {Özgür E.~Müstecaplıoğlu},\ and\ \citenamefont {Paris}}]{ullah2025configuration}%
  \BibitemOpen
  \bibfield  {author} {\bibinfo {author} {\bibfnamefont {Asghar}\ \bibnamefont {Ullah}}, \bibinfo {author} {\bibnamefont {Özgür E.~Müstecaplıoğlu}}, \ and\ \bibinfo {author} {\bibfnamefont {Matteo G.~A.}\ \bibnamefont {Paris}},\ }\href {https://arxiv.org/abs/2505.22395} {\enquote {\bibinfo {title} {Configuration-dependent precision in magnetometry and thermometry using multi-qubit quantum sensors},}\ } (\bibinfo {year} {2025}{\natexlab{c}}),\ \Eprint {http://arxiv.org/abs/2505.22395} {arXiv:2505.22395} \BibitemShut {NoStop}%
\bibitem [{\citenamefont {Campbell}\ \emph {et~al.}(2018)\citenamefont {Campbell}, \citenamefont {Genoni},\ and\ \citenamefont {Deffner}}]{Campbell_2018}%
  \BibitemOpen
  \bibfield  {author} {\bibinfo {author} {\bibfnamefont {Steve}\ \bibnamefont {Campbell}}, \bibinfo {author} {\bibfnamefont {Marco~G}\ \bibnamefont {Genoni}}, \ and\ \bibinfo {author} {\bibfnamefont {Sebastian}\ \bibnamefont {Deffner}},\ }\bibfield  {title} {\enquote {\bibinfo {title} {Precision thermometry and the quantum speed limit},}\ }\href {\doibase 10.1088/2058-9565/aaa641} {\bibfield  {journal} {\bibinfo  {journal} {Quantum Sci. Technol.}\ }\textbf {\bibinfo {volume} {3}},\ \bibinfo {pages} {025002} (\bibinfo {year} {2018})}\BibitemShut {NoStop}%
\bibitem [{\citenamefont {Correa}\ \emph {et~al.}(2017)\citenamefont {Correa}, \citenamefont {Perarnau-Llobet}, \citenamefont {Hovhannisyan}, \citenamefont {Hern\'andez-Santana}, \citenamefont {Mehboudi},\ and\ \citenamefont {Sanpera}}]{PhysRevA.96.062103}%
  \BibitemOpen
  \bibfield  {author} {\bibinfo {author} {\bibfnamefont {Luis~A.}\ \bibnamefont {Correa}}, \bibinfo {author} {\bibfnamefont {Mart\'{\i}}\ \bibnamefont {Perarnau-Llobet}}, \bibinfo {author} {\bibfnamefont {Karen~V.}\ \bibnamefont {Hovhannisyan}}, \bibinfo {author} {\bibfnamefont {Senaida}\ \bibnamefont {Hern\'andez-Santana}}, \bibinfo {author} {\bibfnamefont {Mohammad}\ \bibnamefont {Mehboudi}}, \ and\ \bibinfo {author} {\bibfnamefont {Anna}\ \bibnamefont {Sanpera}},\ }\bibfield  {title} {\enquote {\bibinfo {title} {Enhancement of low-temperature thermometry by strong coupling},}\ }\href {\doibase 10.1103/PhysRevA.96.062103} {\bibfield  {journal} {\bibinfo  {journal} {Phys. Rev. A}\ }\textbf {\bibinfo {volume} {96}},\ \bibinfo {pages} {062103} (\bibinfo {year} {2017})}\BibitemShut {NoStop}%
\bibitem [{\citenamefont {Campbell}\ \emph {et~al.}(2017)\citenamefont {Campbell}, \citenamefont {Mehboudi}, \citenamefont {Chiara},\ and\ \citenamefont {Paternostro}}]{Campbell_2017}%
  \BibitemOpen
  \bibfield  {author} {\bibinfo {author} {\bibfnamefont {Steve}\ \bibnamefont {Campbell}}, \bibinfo {author} {\bibfnamefont {Mohammad}\ \bibnamefont {Mehboudi}}, \bibinfo {author} {\bibfnamefont {Gabriele~De}\ \bibnamefont {Chiara}}, \ and\ \bibinfo {author} {\bibfnamefont {Mauro}\ \bibnamefont {Paternostro}},\ }\bibfield  {title} {\enquote {\bibinfo {title} {Global and local thermometry schemes in coupled quantum systems},}\ }\href {\doibase 10.1088/1367-2630/aa7fac} {\bibfield  {journal} {\bibinfo  {journal} {New J. Phys.}\ }\textbf {\bibinfo {volume} {19}},\ \bibinfo {pages} {103003} (\bibinfo {year} {2017})}\BibitemShut {NoStop}%
\bibitem [{\citenamefont {Ullah}\ \emph {et~al.}(2024)\citenamefont {Ullah}, \citenamefont {Tahir~Naseem},\ and\ \citenamefont {Müstecaplıoğlu}}]{Ullah_2025MTCSs}%
  \BibitemOpen
  \bibfield  {author} {\bibinfo {author} {\bibfnamefont {Asghar}\ \bibnamefont {Ullah}}, \bibinfo {author} {\bibfnamefont {M}~\bibnamefont {Tahir~Naseem}}, \ and\ \bibinfo {author} {\bibfnamefont {Özgür~E}\ \bibnamefont {Müstecaplıoğlu}},\ }\bibfield  {title} {\enquote {\bibinfo {title} {Mixing thermal coherent states for precision and range enhancement in quantum thermometry},}\ }\href {\doibase 10.1088/2058-9565/ad994a} {\bibfield  {journal} {\bibinfo  {journal} {Quantum Sci. Technol.}\ }\textbf {\bibinfo {volume} {10}},\ \bibinfo {pages} {015044} (\bibinfo {year} {2024})}\BibitemShut {NoStop}%
\bibitem [{\citenamefont {Yang}\ \emph {et~al.}(2024)\citenamefont {Yang}, \citenamefont {Montenegro},\ and\ \citenamefont {Bayat}}]{PhysRevApplied.22.024069}%
  \BibitemOpen
  \bibfield  {author} {\bibinfo {author} {\bibfnamefont {Yaoling}\ \bibnamefont {Yang}}, \bibinfo {author} {\bibfnamefont {Victor}\ \bibnamefont {Montenegro}}, \ and\ \bibinfo {author} {\bibfnamefont {Abolfazl}\ \bibnamefont {Bayat}},\ }\bibfield  {title} {\enquote {\bibinfo {title} {Sequential-measurement thermometry with quantum many-body probes},}\ }\href {\doibase 10.1103/PhysRevApplied.22.024069} {\bibfield  {journal} {\bibinfo  {journal} {Phys. Rev. Appl.}\ }\textbf {\bibinfo {volume} {22}},\ \bibinfo {pages} {024069} (\bibinfo {year} {2024})}\BibitemShut {NoStop}%
\bibitem [{\citenamefont {Mok}\ \emph {et~al.}(2021)\citenamefont {Mok}, \citenamefont {Bharti}, \citenamefont {Kwek},\ and\ \citenamefont {Bayat}}]{Mok2021}%
  \BibitemOpen
  \bibfield  {author} {\bibinfo {author} {\bibfnamefont {Wai-Keong}\ \bibnamefont {Mok}}, \bibinfo {author} {\bibfnamefont {Kishor}\ \bibnamefont {Bharti}}, \bibinfo {author} {\bibfnamefont {Leong-Chuan}\ \bibnamefont {Kwek}}, \ and\ \bibinfo {author} {\bibfnamefont {Abolfazl}\ \bibnamefont {Bayat}},\ }\bibfield  {title} {\enquote {\bibinfo {title} {Optimal probes for global quantum thermometry},}\ }\href {\doibase 10.1038/s42005-021-00572-w} {\bibfield  {journal} {\bibinfo  {journal} {Commun. Phys.}\ }\textbf {\bibinfo {volume} {4}},\ \bibinfo {pages} {62} (\bibinfo {year} {2021})}\BibitemShut {NoStop}%
\bibitem [{\citenamefont {Hovhannisyan}\ and\ \citenamefont {Correa}(2018)}]{PhysRevB.98.045101}%
  \BibitemOpen
  \bibfield  {author} {\bibinfo {author} {\bibfnamefont {Karen~V.}\ \bibnamefont {Hovhannisyan}}\ and\ \bibinfo {author} {\bibfnamefont {Luis~A.}\ \bibnamefont {Correa}},\ }\bibfield  {title} {\enquote {\bibinfo {title} {Measuring the temperature of cold many-body quantum systems},}\ }\href {\doibase 10.1103/PhysRevB.98.045101} {\bibfield  {journal} {\bibinfo  {journal} {Phys. Rev. B}\ }\textbf {\bibinfo {volume} {98}},\ \bibinfo {pages} {045101} (\bibinfo {year} {2018})}\BibitemShut {NoStop}%
\bibitem [{\citenamefont {Baak}\ and\ \citenamefont {Fischer}(2024)}]{PhysRevLett.132.240803}%
  \BibitemOpen
  \bibfield  {author} {\bibinfo {author} {\bibfnamefont {Jae-Gyun}\ \bibnamefont {Baak}}\ and\ \bibinfo {author} {\bibfnamefont {Uwe~R.}\ \bibnamefont {Fischer}},\ }\bibfield  {title} {\enquote {\bibinfo {title} {Self-consistent many-body metrology},}\ }\href {\doibase 10.1103/PhysRevLett.132.240803} {\bibfield  {journal} {\bibinfo  {journal} {Phys. Rev. Lett.}\ }\textbf {\bibinfo {volume} {132}},\ \bibinfo {pages} {240803} (\bibinfo {year} {2024})}\BibitemShut {NoStop}%
\bibitem [{\citenamefont {Seah}\ \emph {et~al.}(2019)\citenamefont {Seah}, \citenamefont {Nimmrichter}, \citenamefont {Grimmer}, \citenamefont {Santos}, \citenamefont {Scarani},\ and\ \citenamefont {Landi}}]{PhysRevLett.123.180602}%
  \BibitemOpen
  \bibfield  {author} {\bibinfo {author} {\bibfnamefont {Stella}\ \bibnamefont {Seah}}, \bibinfo {author} {\bibfnamefont {Stefan}\ \bibnamefont {Nimmrichter}}, \bibinfo {author} {\bibfnamefont {Daniel}\ \bibnamefont {Grimmer}}, \bibinfo {author} {\bibfnamefont {Jader~P.}\ \bibnamefont {Santos}}, \bibinfo {author} {\bibfnamefont {Valerio}\ \bibnamefont {Scarani}}, \ and\ \bibinfo {author} {\bibfnamefont {Gabriel~T.}\ \bibnamefont {Landi}},\ }\bibfield  {title} {\enquote {\bibinfo {title} {Collisional quantum thermometry},}\ }\href {\doibase 10.1103/PhysRevLett.123.180602} {\bibfield  {journal} {\bibinfo  {journal} {Phys. Rev. Lett.}\ }\textbf {\bibinfo {volume} {123}},\ \bibinfo {pages} {180602} (\bibinfo {year} {2019})}\BibitemShut {NoStop}%
\bibitem [{\citenamefont {Purdy}\ \emph {et~al.}(2017)\citenamefont {Purdy}, \citenamefont {Grutter}, \citenamefont {Srinivasan},\ and\ \citenamefont {Taylor}}]{Purdy}%
  \BibitemOpen
  \bibfield  {author} {\bibinfo {author} {\bibfnamefont {T.~P.}\ \bibnamefont {Purdy}}, \bibinfo {author} {\bibfnamefont {K.~E.}\ \bibnamefont {Grutter}}, \bibinfo {author} {\bibfnamefont {K.}~\bibnamefont {Srinivasan}}, \ and\ \bibinfo {author} {\bibfnamefont {J.~M.}\ \bibnamefont {Taylor}},\ }\bibfield  {title} {\enquote {\bibinfo {title} {Quantum correlations from a room-temperature optomechanical cavity},}\ }\href {\doibase 10.1126/science.aag1407} {\bibfield  {journal} {\bibinfo  {journal} {Science}\ }\textbf {\bibinfo {volume} {356}},\ \bibinfo {pages} {1265--1268} (\bibinfo {year} {2017})}\BibitemShut {NoStop}%
\bibitem [{\citenamefont {Mehboudi}\ \emph {et~al.}(2019{\natexlab{b}})\citenamefont {Mehboudi}, \citenamefont {Lampo}, \citenamefont {Charalambous}, \citenamefont {Correa}, \citenamefont {Garc\'{\i}a-March},\ and\ \citenamefont {Lewenstein}}]{PhysRevLett.122.030403}%
  \BibitemOpen
  \bibfield  {author} {\bibinfo {author} {\bibfnamefont {Mohammad}\ \bibnamefont {Mehboudi}}, \bibinfo {author} {\bibfnamefont {Aniello}\ \bibnamefont {Lampo}}, \bibinfo {author} {\bibfnamefont {Christos}\ \bibnamefont {Charalambous}}, \bibinfo {author} {\bibfnamefont {Luis~A.}\ \bibnamefont {Correa}}, \bibinfo {author} {\bibfnamefont {Miguel~\'Angel}\ \bibnamefont {Garc\'{\i}a-March}}, \ and\ \bibinfo {author} {\bibfnamefont {Maciej}\ \bibnamefont {Lewenstein}},\ }\bibfield  {title} {\enquote {\bibinfo {title} {Using polarons for sub-nk quantum nondemolition thermometry in a bose-einstein condensate},}\ }\href {\doibase 10.1103/PhysRevLett.122.030403} {\bibfield  {journal} {\bibinfo  {journal} {Phys. Rev. Lett.}\ }\textbf {\bibinfo {volume} {122}},\ \bibinfo {pages} {030403} (\bibinfo {year} {2019}{\natexlab{b}})}\BibitemShut {NoStop}%
\bibitem [{\citenamefont {Stace}(2010)}]{PhysRevA.82.011611}%
  \BibitemOpen
  \bibfield  {author} {\bibinfo {author} {\bibfnamefont {Thomas~M.}\ \bibnamefont {Stace}},\ }\bibfield  {title} {\enquote {\bibinfo {title} {Quantum limits of thermometry},}\ }\href {\doibase 10.1103/PhysRevA.82.011611} {\bibfield  {journal} {\bibinfo  {journal} {Phys. Rev. A}\ }\textbf {\bibinfo {volume} {82}},\ \bibinfo {pages} {011611} (\bibinfo {year} {2010})}\BibitemShut {NoStop}%
\bibitem [{\citenamefont {P\l{}odzie\ifmmode~\acute{n}\else \'{n}\fi{}}\ \emph {et~al.}(2018)\citenamefont {P\l{}odzie\ifmmode~\acute{n}\else \'{n}\fi{}}, \citenamefont {Demkowicz-Dobrza\ifmmode~\acute{n}\else \'{n}\fi{}ski},\ and\ \citenamefont {Sowi\ifmmode~\acute{n}\else \'{n}\fi{}ski}}]{PhysRevA.97.063619}%
  \BibitemOpen
  \bibfield  {author} {\bibinfo {author} {\bibfnamefont {Marcin}\ \bibnamefont {P\l{}odzie\ifmmode~\acute{n}\else \'{n}\fi{}}}, \bibinfo {author} {\bibfnamefont {Rafa\l{}}\ \bibnamefont {Demkowicz-Dobrza\ifmmode~\acute{n}\else \'{n}\fi{}ski}}, \ and\ \bibinfo {author} {\bibfnamefont {Tomasz}\ \bibnamefont {Sowi\ifmmode~\acute{n}\else \'{n}\fi{}ski}},\ }\bibfield  {title} {\enquote {\bibinfo {title} {Few-fermion thermometry},}\ }\href {\doibase 10.1103/PhysRevA.97.063619} {\bibfield  {journal} {\bibinfo  {journal} {Phys. Rev. A}\ }\textbf {\bibinfo {volume} {97}},\ \bibinfo {pages} {063619} (\bibinfo {year} {2018})}\BibitemShut {NoStop}%
\bibitem [{\citenamefont {Carlos}\ and\ \citenamefont {Palacio}(2015)}]{Carlos}%
  \BibitemOpen
  \bibfield  {author} {\bibinfo {author} {\bibfnamefont {Luís~Dias}\ \bibnamefont {Carlos}}\ and\ \bibinfo {author} {\bibfnamefont {Fernando}\ \bibnamefont {Palacio}},\ }\href {\doibase 10.1039/9781782622031} {\emph {\bibinfo {title} {Thermometry at the Nanoscale: Techniques and Selected Applications}}}\ (\bibinfo  {publisher} {The Royal Society of Chemistry, London},\ \bibinfo {year} {2015})\BibitemShut {NoStop}%
\bibitem [{\citenamefont {Nguyen}\ \emph {et~al.}(2018)\citenamefont {Nguyen}, \citenamefont {Evans}, \citenamefont {Sipahigil}, \citenamefont {Bhaskar}, \citenamefont {Sukachev}, \citenamefont {Agafonov}, \citenamefont {Davydov}, \citenamefont {Kulikova}, \citenamefont {Jelezko},\ and\ \citenamefont {Lukin}}]{Nguyen}%
  \BibitemOpen
  \bibfield  {author} {\bibinfo {author} {\bibfnamefont {Christian~T.}\ \bibnamefont {Nguyen}}, \bibinfo {author} {\bibfnamefont {Ruffin~E.}\ \bibnamefont {Evans}}, \bibinfo {author} {\bibfnamefont {Alp}\ \bibnamefont {Sipahigil}}, \bibinfo {author} {\bibfnamefont {Mihir~K.}\ \bibnamefont {Bhaskar}}, \bibinfo {author} {\bibfnamefont {Denis~D.}\ \bibnamefont {Sukachev}}, \bibinfo {author} {\bibfnamefont {Viatcheslav~N.}\ \bibnamefont {Agafonov}}, \bibinfo {author} {\bibfnamefont {Valery~A.}\ \bibnamefont {Davydov}}, \bibinfo {author} {\bibfnamefont {Liudmila~F.}\ \bibnamefont {Kulikova}}, \bibinfo {author} {\bibfnamefont {Fedor}\ \bibnamefont {Jelezko}}, \ and\ \bibinfo {author} {\bibfnamefont {Mikhail~D.}\ \bibnamefont {Lukin}},\ }\bibfield  {title} {\enquote {\bibinfo {title} {All-optical nanoscale thermometry with silicon-vacancy centers in diamond},}\ }\href {\doibase 10.1063/1.5029904} {\bibfield  {journal} {\bibinfo  {journal} {Appl. Phys. Lett}\ }\textbf {\bibinfo {volume} {112}},\ \bibinfo {pages}
  {203102} (\bibinfo {year} {2018})}\BibitemShut {NoStop}%
\bibitem [{\citenamefont {Cavina}\ \emph {et~al.}(2018)\citenamefont {Cavina}, \citenamefont {Mancino}, \citenamefont {De~Pasquale}, \citenamefont {Gianani}, \citenamefont {Sbroscia}, \citenamefont {Booth}, \citenamefont {Roccia}, \citenamefont {Raimondi}, \citenamefont {Giovannetti},\ and\ \citenamefont {Barbieri}}]{PhysRevA.98.050101}%
  \BibitemOpen
  \bibfield  {author} {\bibinfo {author} {\bibfnamefont {Vasco}\ \bibnamefont {Cavina}}, \bibinfo {author} {\bibfnamefont {Luca}\ \bibnamefont {Mancino}}, \bibinfo {author} {\bibfnamefont {Antonella}\ \bibnamefont {De~Pasquale}}, \bibinfo {author} {\bibfnamefont {Ilaria}\ \bibnamefont {Gianani}}, \bibinfo {author} {\bibfnamefont {Marco}\ \bibnamefont {Sbroscia}}, \bibinfo {author} {\bibfnamefont {Robert~I.}\ \bibnamefont {Booth}}, \bibinfo {author} {\bibfnamefont {Emanuele}\ \bibnamefont {Roccia}}, \bibinfo {author} {\bibfnamefont {Roberto}\ \bibnamefont {Raimondi}}, \bibinfo {author} {\bibfnamefont {Vittorio}\ \bibnamefont {Giovannetti}}, \ and\ \bibinfo {author} {\bibfnamefont {Marco}\ \bibnamefont {Barbieri}},\ }\bibfield  {title} {\enquote {\bibinfo {title} {Bridging thermodynamics and metrology in nonequilibrium quantum thermometry},}\ }\href {\doibase 10.1103/PhysRevA.98.050101} {\bibfield  {journal} {\bibinfo  {journal} {Phys. Rev. A}\ }\textbf {\bibinfo {volume} {98}},\ \bibinfo {pages} {050101}
  (\bibinfo {year} {2018})}\BibitemShut {NoStop}%
\bibitem [{\citenamefont {Ravell~Rodríguez}\ \emph {et~al.}(2024)\citenamefont {Ravell~Rodríguez}, \citenamefont {Mehboudi}, \citenamefont {Horodecki},\ and\ \citenamefont {Perarnau-Llobet}}]{Ravell}%
  \BibitemOpen
  \bibfield  {author} {\bibinfo {author} {\bibfnamefont {Ricard}\ \bibnamefont {Ravell~Rodríguez}}, \bibinfo {author} {\bibfnamefont {Mohammad}\ \bibnamefont {Mehboudi}}, \bibinfo {author} {\bibfnamefont {Michał}\ \bibnamefont {Horodecki}}, \ and\ \bibinfo {author} {\bibfnamefont {Martí}\ \bibnamefont {Perarnau-Llobet}},\ }\bibfield  {title} {\enquote {\bibinfo {title} {Strongly coupled fermionic probe for nonequilibrium thermometry},}\ }\href {\doibase 10.1088/1367-2630/ad1d75} {\bibfield  {journal} {\bibinfo  {journal} {New J. Phys.}\ }\textbf {\bibinfo {volume} {26}},\ \bibinfo {pages} {013046} (\bibinfo {year} {2024})}\BibitemShut {NoStop}%
\bibitem [{\citenamefont {Brattegard}\ and\ \citenamefont {Mitchison}(2024)}]{PhysRevA.109.023309}%
  \BibitemOpen
  \bibfield  {author} {\bibinfo {author} {\bibfnamefont {Sindre}\ \bibnamefont {Brattegard}}\ and\ \bibinfo {author} {\bibfnamefont {Mark~T.}\ \bibnamefont {Mitchison}},\ }\bibfield  {title} {\enquote {\bibinfo {title} {Thermometry by correlated dephasing of impurities in a one-dimensional fermi gas},}\ }\href {\doibase 10.1103/PhysRevA.109.023309} {\bibfield  {journal} {\bibinfo  {journal} {Phys. Rev. A}\ }\textbf {\bibinfo {volume} {109}},\ \bibinfo {pages} {023309} (\bibinfo {year} {2024})}\BibitemShut {NoStop}%
\bibitem [{\citenamefont {Aiache}\ \emph {et~al.}(2024{\natexlab{b}})\citenamefont {Aiache}, \citenamefont {Seida}, \citenamefont {El~Anouz},\ and\ \citenamefont {El~Allati}}]{PhysRevE.110.024132}%
  \BibitemOpen
  \bibfield  {author} {\bibinfo {author} {\bibfnamefont {Y.}~\bibnamefont {Aiache}}, \bibinfo {author} {\bibfnamefont {C.}~\bibnamefont {Seida}}, \bibinfo {author} {\bibfnamefont {K.}~\bibnamefont {El~Anouz}}, \ and\ \bibinfo {author} {\bibfnamefont {A.}~\bibnamefont {El~Allati}},\ }\bibfield  {title} {\enquote {\bibinfo {title} {Non-markovian enhancement of nonequilibrium quantum thermometry},}\ }\href {\doibase 10.1103/PhysRevE.110.024132} {\bibfield  {journal} {\bibinfo  {journal} {Phys. Rev. E}\ }\textbf {\bibinfo {volume} {110}},\ \bibinfo {pages} {024132} (\bibinfo {year} {2024}{\natexlab{b}})}\BibitemShut {NoStop}%
\bibitem [{\citenamefont {Anto-Sztrikacs}\ \emph {et~al.}(2024)\citenamefont {Anto-Sztrikacs}, \citenamefont {Miller}, \citenamefont {Nazir},\ and\ \citenamefont {Segal}}]{PhysRevA.109.L060201}%
  \BibitemOpen
  \bibfield  {author} {\bibinfo {author} {\bibfnamefont {Nicholas}\ \bibnamefont {Anto-Sztrikacs}}, \bibinfo {author} {\bibfnamefont {Harry J.~D.}\ \bibnamefont {Miller}}, \bibinfo {author} {\bibfnamefont {Ahsan}\ \bibnamefont {Nazir}}, \ and\ \bibinfo {author} {\bibfnamefont {Dvira}\ \bibnamefont {Segal}},\ }\bibfield  {title} {\enquote {\bibinfo {title} {Bypassing thermalization timescales in temperature estimation using prethermal probes},}\ }\href {\doibase 10.1103/PhysRevA.109.L060201} {\bibfield  {journal} {\bibinfo  {journal} {Phys. Rev. A}\ }\textbf {\bibinfo {volume} {109}},\ \bibinfo {pages} {L060201} (\bibinfo {year} {2024})}\BibitemShut {NoStop}%
\bibitem [{\citenamefont {Mirkhalaf}\ \emph {et~al.}(2024)\citenamefont {Mirkhalaf}, \citenamefont {Mehboudi}, \citenamefont {Nafari~Qaleh},\ and\ \citenamefont {Rahimi-Keshari}}]{Mirkhalaf_2024}%
  \BibitemOpen
  \bibfield  {author} {\bibinfo {author} {\bibfnamefont {Safoura}\ \bibnamefont {Mirkhalaf}}, \bibinfo {author} {\bibfnamefont {Mohammad}\ \bibnamefont {Mehboudi}}, \bibinfo {author} {\bibfnamefont {Zohre}\ \bibnamefont {Nafari~Qaleh}}, \ and\ \bibinfo {author} {\bibfnamefont {Saleh}\ \bibnamefont {Rahimi-Keshari}},\ }\bibfield  {title} {\enquote {\bibinfo {title} {Operational significance of nonclassicality in nonequilibrium gaussian quantum thermometry},}\ }\href {\doibase 10.1088/1367-2630/ad23a1} {\bibfield  {journal} {\bibinfo  {journal} {New J. Phys.}\ }\textbf {\bibinfo {volume} {26}},\ \bibinfo {pages} {023046} (\bibinfo {year} {2024})}\BibitemShut {NoStop}%
\bibitem [{\citenamefont {Weedbrook}\ \emph {et~al.}(2012)\citenamefont {Weedbrook}, \citenamefont {Pirandola}, \citenamefont {Garc\'{\i}a-Patr\'on}, \citenamefont {Cerf}, \citenamefont {Ralph}, \citenamefont {Shapiro},\ and\ \citenamefont {Lloyd}}]{RevModPhys.84.621}%
  \BibitemOpen
  \bibfield  {author} {\bibinfo {author} {\bibfnamefont {Christian}\ \bibnamefont {Weedbrook}}, \bibinfo {author} {\bibfnamefont {Stefano}\ \bibnamefont {Pirandola}}, \bibinfo {author} {\bibfnamefont {Ra\'ul}\ \bibnamefont {Garc\'{\i}a-Patr\'on}}, \bibinfo {author} {\bibfnamefont {Nicolas~J.}\ \bibnamefont {Cerf}}, \bibinfo {author} {\bibfnamefont {Timothy~C.}\ \bibnamefont {Ralph}}, \bibinfo {author} {\bibfnamefont {Jeffrey~H.}\ \bibnamefont {Shapiro}}, \ and\ \bibinfo {author} {\bibfnamefont {Seth}\ \bibnamefont {Lloyd}},\ }\bibfield  {title} {\enquote {\bibinfo {title} {Gaussian quantum information},}\ }\href {\doibase 10.1103/RevModPhys.84.621} {\bibfield  {journal} {\bibinfo  {journal} {Rev. Mod. Phys.}\ }\textbf {\bibinfo {volume} {84}},\ \bibinfo {pages} {621--669} (\bibinfo {year} {2012})}\BibitemShut {NoStop}%
\bibitem [{\citenamefont {Cenni}\ \emph {et~al.}(2022)\citenamefont {Cenni}, \citenamefont {Lami}, \citenamefont {Ac{\'{i}}n},\ and\ \citenamefont {Mehboudi}}]{Cenni2022thermometryof}%
  \BibitemOpen
  \bibfield  {author} {\bibinfo {author} {\bibfnamefont {Marina~F.B.}\ \bibnamefont {Cenni}}, \bibinfo {author} {\bibfnamefont {Ludovico}\ \bibnamefont {Lami}}, \bibinfo {author} {\bibfnamefont {Antonio}\ \bibnamefont {Ac{\'{i}}n}}, \ and\ \bibinfo {author} {\bibfnamefont {Mohammad}\ \bibnamefont {Mehboudi}},\ }\bibfield  {title} {\enquote {\bibinfo {title} {Thermometry of {G}aussian quantum systems using {G}aussian measurements},}\ }\href {\doibase 10.22331/q-2022-06-23-743} {\bibfield  {journal} {\bibinfo  {journal} {{Quantum}}\ }\textbf {\bibinfo {volume} {6}},\ \bibinfo {pages} {743} (\bibinfo {year} {2022})}\BibitemShut {NoStop}%
\bibitem [{\citenamefont {Alves}\ \emph {et~al.}(2024)\citenamefont {Alves}, \citenamefont {Santos},\ and\ \citenamefont {Landi}}]{PhysRevA.110.052421}%
  \BibitemOpen
  \bibfield  {author} {\bibinfo {author} {\bibfnamefont {Gabriel~O.}\ \bibnamefont {Alves}}, \bibinfo {author} {\bibfnamefont {Marcelo A.~F.}\ \bibnamefont {Santos}}, \ and\ \bibinfo {author} {\bibfnamefont {Gabriel~T.}\ \bibnamefont {Landi}},\ }\bibfield  {title} {\enquote {\bibinfo {title} {Collisional thermometry for gaussian systems},}\ }\href {\doibase 10.1103/PhysRevA.110.052421} {\bibfield  {journal} {\bibinfo  {journal} {Phys. Rev. A}\ }\textbf {\bibinfo {volume} {110}},\ \bibinfo {pages} {052421} (\bibinfo {year} {2024})}\BibitemShut {NoStop}%
\bibitem [{\citenamefont {Mukhopadhyay}\ \emph {et~al.}(2025)\citenamefont {Mukhopadhyay}, \citenamefont {Paris},\ and\ \citenamefont {Bayat}}]{5wn8-d9ks}%
  \BibitemOpen
  \bibfield  {author} {\bibinfo {author} {\bibfnamefont {Chiranjib}\ \bibnamefont {Mukhopadhyay}}, \bibinfo {author} {\bibfnamefont {Matteo~G.A.}\ \bibnamefont {Paris}}, \ and\ \bibinfo {author} {\bibfnamefont {Abolfazl}\ \bibnamefont {Bayat}},\ }\bibfield  {title} {\enquote {\bibinfo {title} {Saturable global quantum sensing},}\ }\href {\doibase 10.1103/5wn8-d9ks} {\bibfield  {journal} {\bibinfo  {journal} {Phys. Rev. Appl.}\ }\textbf {\bibinfo {volume} {24}},\ \bibinfo {pages} {014012} (\bibinfo {year} {2025})}\BibitemShut {NoStop}%
\bibitem [{\citenamefont {Frigerio}\ \emph {et~al.}(2022)\citenamefont {Frigerio}, \citenamefont {Olivares},\ and\ \citenamefont {Paris}}]{Frigerio_2022}%
  \BibitemOpen
  \bibfield  {author} {\bibinfo {author} {\bibfnamefont {Massimo}\ \bibnamefont {Frigerio}}, \bibinfo {author} {\bibfnamefont {Stefano}\ \bibnamefont {Olivares}}, \ and\ \bibinfo {author} {\bibfnamefont {Matteo G~A}\ \bibnamefont {Paris}},\ }\bibfield  {title} {\enquote {\bibinfo {title} {Cost-effective estimation of single-mode thermal states by probabilistic quantum metrology},}\ }\href {\doibase 10.1088/2058-9565/ac6dfe} {\bibfield  {journal} {\bibinfo  {journal} {Quantum Sci. Technol.}\ }\textbf {\bibinfo {volume} {7}},\ \bibinfo {pages} {035011} (\bibinfo {year} {2022})}\BibitemShut {NoStop}%
\bibitem [{\citenamefont {Deng}\ \emph {et~al.}(2024)\citenamefont {Deng}, \citenamefont {Li}, \citenamefont {Chen}, \citenamefont {Ni}, \citenamefont {Cai}, \citenamefont {Mai}, \citenamefont {Zhang}, \citenamefont {Zheng}, \citenamefont {Yu}, \citenamefont {Zou}, \citenamefont {Liu}, \citenamefont {Yan}, \citenamefont {Xu},\ and\ \citenamefont {Yu}}]{Deng2024}%
  \BibitemOpen
  \bibfield  {author} {\bibinfo {author} {\bibfnamefont {Xiaowei}\ \bibnamefont {Deng}}, \bibinfo {author} {\bibfnamefont {Sai}\ \bibnamefont {Li}}, \bibinfo {author} {\bibfnamefont {Zi-Jie}\ \bibnamefont {Chen}}, \bibinfo {author} {\bibfnamefont {Zhongchu}\ \bibnamefont {Ni}}, \bibinfo {author} {\bibfnamefont {Yanyan}\ \bibnamefont {Cai}}, \bibinfo {author} {\bibfnamefont {Jiasheng}\ \bibnamefont {Mai}}, \bibinfo {author} {\bibfnamefont {Libo}\ \bibnamefont {Zhang}}, \bibinfo {author} {\bibfnamefont {Pan}\ \bibnamefont {Zheng}}, \bibinfo {author} {\bibfnamefont {Haifeng}\ \bibnamefont {Yu}}, \bibinfo {author} {\bibfnamefont {Chang-Ling}\ \bibnamefont {Zou}}, \bibinfo {author} {\bibfnamefont {Song}\ \bibnamefont {Liu}}, \bibinfo {author} {\bibfnamefont {Fei}\ \bibnamefont {Yan}}, \bibinfo {author} {\bibfnamefont {Yuan}\ \bibnamefont {Xu}}, \ and\ \bibinfo {author} {\bibfnamefont {Dapeng}\ \bibnamefont {Yu}},\ }\bibfield  {title} {\enquote {\bibinfo {title} {Quantum-enhanced metrology with large fock
  states},}\ }\href {\doibase 10.1038/s41567-024-02619-5} {\bibfield  {journal} {\bibinfo  {journal} {Nat. Phys.}\ }\textbf {\bibinfo {volume} {20}},\ \bibinfo {pages} {1874--1880} (\bibinfo {year} {2024})}\BibitemShut {NoStop}%
\bibitem [{\citenamefont {Oh}\ \emph {et~al.}(2020)\citenamefont {Oh}, \citenamefont {Park}, \citenamefont {Filip}, \citenamefont {Jeong},\ and\ \citenamefont {Marek}}]{Oh_2020}%
  \BibitemOpen
  \bibfield  {author} {\bibinfo {author} {\bibfnamefont {Changhun}\ \bibnamefont {Oh}}, \bibinfo {author} {\bibfnamefont {Kimin}\ \bibnamefont {Park}}, \bibinfo {author} {\bibfnamefont {Radim}\ \bibnamefont {Filip}}, \bibinfo {author} {\bibfnamefont {Hyunseok}\ \bibnamefont {Jeong}}, \ and\ \bibinfo {author} {\bibfnamefont {Petr}\ \bibnamefont {Marek}},\ }\bibfield  {title} {\enquote {\bibinfo {title} {Optical estimation of unitary gaussian processes without phase reference using fock states},}\ }\href {\doibase 10.1088/1367-2630/abd0b8} {\bibfield  {journal} {\bibinfo  {journal} {New J. Phys.}\ }\textbf {\bibinfo {volume} {22}},\ \bibinfo {pages} {123039} (\bibinfo {year} {2020})}\BibitemShut {NoStop}%
\bibitem [{\citenamefont {Tatsuta}\ \emph {et~al.}(2019)\citenamefont {Tatsuta}, \citenamefont {Matsuzaki},\ and\ \citenamefont {Shimizu}}]{PhysRevA.100.032318}%
  \BibitemOpen
  \bibfield  {author} {\bibinfo {author} {\bibfnamefont {Mamiko}\ \bibnamefont {Tatsuta}}, \bibinfo {author} {\bibfnamefont {Yuichiro}\ \bibnamefont {Matsuzaki}}, \ and\ \bibinfo {author} {\bibfnamefont {Akira}\ \bibnamefont {Shimizu}},\ }\bibfield  {title} {\enquote {\bibinfo {title} {Quantum metrology with generalized cat states},}\ }\href {\doibase 10.1103/PhysRevA.100.032318} {\bibfield  {journal} {\bibinfo  {journal} {Phys. Rev. A}\ }\textbf {\bibinfo {volume} {100}},\ \bibinfo {pages} {032318} (\bibinfo {year} {2019})}\BibitemShut {NoStop}%
\bibitem [{\citenamefont {Rahman}\ \emph {et~al.}(2025)\citenamefont {Rahman}, \citenamefont {Kladari\ifmmode~\acute{c}\else \'{c}\fi{}}, \citenamefont {Kern}, \citenamefont {Lachman}, \citenamefont {Chu}, \citenamefont {Filip},\ and\ \citenamefont {Fadel}}]{PhysRevLett.134.180801}%
  \BibitemOpen
  \bibfield  {author} {\bibinfo {author} {\bibfnamefont {Q.~Rumman}\ \bibnamefont {Rahman}}, \bibinfo {author} {\bibfnamefont {Igor}\ \bibnamefont {Kladari\ifmmode~\acute{c}\else \'{c}\fi{}}}, \bibinfo {author} {\bibfnamefont {Max-Emanuel}\ \bibnamefont {Kern}}, \bibinfo {author} {\bibfnamefont {Luk\'a\ifmmode \check{s}\else~\v{s}\fi{}}\ \bibnamefont {Lachman}}, \bibinfo {author} {\bibfnamefont {Yiwen}\ \bibnamefont {Chu}}, \bibinfo {author} {\bibfnamefont {Radim}\ \bibnamefont {Filip}}, \ and\ \bibinfo {author} {\bibfnamefont {Matteo}\ \bibnamefont {Fadel}},\ }\bibfield  {title} {\enquote {\bibinfo {title} {Genuine quantum non-gaussianity and metrological sensitivity of fock states prepared in a mechanical resonator},}\ }\href {\doibase 10.1103/PhysRevLett.134.180801} {\bibfield  {journal} {\bibinfo  {journal} {Phys. Rev. Lett.}\ }\textbf {\bibinfo {volume} {134}},\ \bibinfo {pages} {180801} (\bibinfo {year} {2025})}\BibitemShut {NoStop}%
\bibitem [{\citenamefont {Santos}(2025)}]{santos2025}%
  \BibitemOpen
  \bibfield  {author} {\bibinfo {author} {\bibfnamefont {Jonas F.~G.}\ \bibnamefont {Santos}},\ }\href {https://arxiv.org/abs/2505.14870} {\enquote {\bibinfo {title} {Enhanced frequency estimation by non-gaussianity of fock states},}\ } (\bibinfo {year} {2025}),\ \Eprint {http://arxiv.org/abs/2505.14870} {arXiv:2505.14870 [quant-ph]} \BibitemShut {NoStop}%
\bibitem [{\citenamefont {Monras}\ and\ \citenamefont {Paris}(2007)}]{PhysRevLett.98.160401}%
  \BibitemOpen
  \bibfield  {author} {\bibinfo {author} {\bibfnamefont {Alex}\ \bibnamefont {Monras}}\ and\ \bibinfo {author} {\bibfnamefont {Matteo G.~A.}\ \bibnamefont {Paris}},\ }\bibfield  {title} {\enquote {\bibinfo {title} {Optimal quantum estimation of loss in bosonic channels},}\ }\href {\doibase 10.1103/PhysRevLett.98.160401} {\bibfield  {journal} {\bibinfo  {journal} {Phys. Rev. Lett.}\ }\textbf {\bibinfo {volume} {98}},\ \bibinfo {pages} {160401} (\bibinfo {year} {2007})}\BibitemShut {NoStop}%
\bibitem [{\citenamefont {Adesso}\ \emph {et~al.}(2009)\citenamefont {Adesso}, \citenamefont {Dell'Anno}, \citenamefont {De~Siena}, \citenamefont {Illuminati},\ and\ \citenamefont {Souza}}]{PhysRevA.79.040305}%
  \BibitemOpen
  \bibfield  {author} {\bibinfo {author} {\bibfnamefont {G.}~\bibnamefont {Adesso}}, \bibinfo {author} {\bibfnamefont {F.}~\bibnamefont {Dell'Anno}}, \bibinfo {author} {\bibfnamefont {S.}~\bibnamefont {De~Siena}}, \bibinfo {author} {\bibfnamefont {F.}~\bibnamefont {Illuminati}}, \ and\ \bibinfo {author} {\bibfnamefont {L.~A.~M.}\ \bibnamefont {Souza}},\ }\bibfield  {title} {\enquote {\bibinfo {title} {Optimal estimation of losses at the ultimate quantum limit with non-gaussian states},}\ }\href {\doibase 10.1103/PhysRevA.79.040305} {\bibfield  {journal} {\bibinfo  {journal} {Phys. Rev. A}\ }\textbf {\bibinfo {volume} {79}},\ \bibinfo {pages} {040305} (\bibinfo {year} {2009})}\BibitemShut {NoStop}%
\bibitem [{\citenamefont {Birrittella}\ \emph {et~al.}(2015)\citenamefont {Birrittella}, \citenamefont {Gura},\ and\ \citenamefont {Gerry}}]{PhysRevA.91.053801}%
  \BibitemOpen
  \bibfield  {author} {\bibinfo {author} {\bibfnamefont {Richard}\ \bibnamefont {Birrittella}}, \bibinfo {author} {\bibfnamefont {Anna}\ \bibnamefont {Gura}}, \ and\ \bibinfo {author} {\bibfnamefont {Christopher~C.}\ \bibnamefont {Gerry}},\ }\bibfield  {title} {\enquote {\bibinfo {title} {Coherently stimulated parametric down-conversion, phase effects, and quantum-optical interferometry},}\ }\href {\doibase 10.1103/PhysRevA.91.053801} {\bibfield  {journal} {\bibinfo  {journal} {Phys. Rev. A}\ }\textbf {\bibinfo {volume} {91}},\ \bibinfo {pages} {053801} (\bibinfo {year} {2015})}\BibitemShut {NoStop}%
\bibitem [{\citenamefont {Kolkiran}\ and\ \citenamefont {Agarwal}(2008)}]{Kolkiran2008}%
  \BibitemOpen
  \bibfield  {author} {\bibinfo {author} {\bibfnamefont {Aziz}\ \bibnamefont {Kolkiran}}\ and\ \bibinfo {author} {\bibfnamefont {G.~S.}\ \bibnamefont {Agarwal}},\ }\bibfield  {title} {\enquote {\bibinfo {title} {Quantum interferometry using coherent beam stimulated parametric down-conversion},}\ }\href {\doibase 10.1364/OE.16.006479} {\bibfield  {journal} {\bibinfo  {journal} {Opt. Express}\ }\textbf {\bibinfo {volume} {16}},\ \bibinfo {pages} {6479--6485} (\bibinfo {year} {2008})}\BibitemShut {NoStop}%
\bibitem [{\citenamefont {Roux}(2021)}]{Roux2021}%
  \BibitemOpen
  \bibfield  {author} {\bibinfo {author} {\bibfnamefont {Filippus~S.}\ \bibnamefont {Roux}},\ }\bibfield  {title} {\enquote {\bibinfo {title} {Stimulated parametric down-conversion for spatiotemporal metrology},}\ }\href {\doibase 10.1103/PhysRevA.104.043514} {\bibfield  {journal} {\bibinfo  {journal} {Phys. Rev. A}\ }\textbf {\bibinfo {volume} {104}},\ \bibinfo {pages} {043514} (\bibinfo {year} {2021})}\BibitemShut {NoStop}%
\bibitem [{\citenamefont {Andersen}\ \emph {et~al.}(2016)\citenamefont {Andersen}, \citenamefont {Gehring}, \citenamefont {Marquardt},\ and\ \citenamefont {Leuchs}}]{Andersen_2016}%
  \BibitemOpen
  \bibfield  {author} {\bibinfo {author} {\bibfnamefont {Ulrik~L}\ \bibnamefont {Andersen}}, \bibinfo {author} {\bibfnamefont {Tobias}\ \bibnamefont {Gehring}}, \bibinfo {author} {\bibfnamefont {Christoph}\ \bibnamefont {Marquardt}}, \ and\ \bibinfo {author} {\bibfnamefont {Gerd}\ \bibnamefont {Leuchs}},\ }\bibfield  {title} {\enquote {\bibinfo {title} {30 years of squeezed light generation},}\ }\href {\doibase 10.1088/0031-8949/91/5/053001} {\bibfield  {journal} {\bibinfo  {journal} {Phys. Scr.}\ }\textbf {\bibinfo {volume} {91}},\ \bibinfo {pages} {053001} (\bibinfo {year} {2016})}\BibitemShut {NoStop}%
\bibitem [{\citenamefont {Takeno}\ \emph {et~al.}(2007)\citenamefont {Takeno}, \citenamefont {Yukawa}, \citenamefont {Yonezawa},\ and\ \citenamefont {Furusawa}}]{Takeno:07}%
  \BibitemOpen
  \bibfield  {author} {\bibinfo {author} {\bibfnamefont {Yuishi}\ \bibnamefont {Takeno}}, \bibinfo {author} {\bibfnamefont {Mitsuyoshi}\ \bibnamefont {Yukawa}}, \bibinfo {author} {\bibfnamefont {Hidehiro}\ \bibnamefont {Yonezawa}}, \ and\ \bibinfo {author} {\bibfnamefont {Akira}\ \bibnamefont {Furusawa}},\ }\bibfield  {title} {\enquote {\bibinfo {title} {Observation of -9 db quadrature squeezing with improvement of phase stability in homodyne measurement},}\ }\href {\doibase 10.1364/OE.15.004321} {\bibfield  {journal} {\bibinfo  {journal} {Opt. Express}\ }\textbf {\bibinfo {volume} {15}},\ \bibinfo {pages} {4321--4327} (\bibinfo {year} {2007})}\BibitemShut {NoStop}%
\bibitem [{\citenamefont {Vahlbruch}\ \emph {et~al.}(2016)\citenamefont {Vahlbruch}, \citenamefont {Mehmet}, \citenamefont {Danzmann},\ and\ \citenamefont {Schnabel}}]{PhysRevLett.117.110801}%
  \BibitemOpen
  \bibfield  {author} {\bibinfo {author} {\bibfnamefont {Henning}\ \bibnamefont {Vahlbruch}}, \bibinfo {author} {\bibfnamefont {Moritz}\ \bibnamefont {Mehmet}}, \bibinfo {author} {\bibfnamefont {Karsten}\ \bibnamefont {Danzmann}}, \ and\ \bibinfo {author} {\bibfnamefont {Roman}\ \bibnamefont {Schnabel}},\ }\bibfield  {title} {\enquote {\bibinfo {title} {Detection of 15 db squeezed states of light and their application for the absolute calibration of photoelectric quantum efficiency},}\ }\href {\doibase 10.1103/PhysRevLett.117.110801} {\bibfield  {journal} {\bibinfo  {journal} {Phys. Rev. Lett.}\ }\textbf {\bibinfo {volume} {117}},\ \bibinfo {pages} {110801} (\bibinfo {year} {2016})}\BibitemShut {NoStop}%
\bibitem [{\citenamefont {Eberle}\ \emph {et~al.}(2010)\citenamefont {Eberle}, \citenamefont {Steinlechner}, \citenamefont {Bauchrowitz}, \citenamefont {H\"andchen}, \citenamefont {Vahlbruch}, \citenamefont {Mehmet}, \citenamefont {M\"uller-Ebhardt},\ and\ \citenamefont {Schnabel}}]{PhysRevLett.104.251102}%
  \BibitemOpen
  \bibfield  {author} {\bibinfo {author} {\bibfnamefont {Tobias}\ \bibnamefont {Eberle}}, \bibinfo {author} {\bibfnamefont {Sebastian}\ \bibnamefont {Steinlechner}}, \bibinfo {author} {\bibfnamefont {J\"oran}\ \bibnamefont {Bauchrowitz}}, \bibinfo {author} {\bibfnamefont {Vitus}\ \bibnamefont {H\"andchen}}, \bibinfo {author} {\bibfnamefont {Henning}\ \bibnamefont {Vahlbruch}}, \bibinfo {author} {\bibfnamefont {Moritz}\ \bibnamefont {Mehmet}}, \bibinfo {author} {\bibfnamefont {Helge}\ \bibnamefont {M\"uller-Ebhardt}}, \ and\ \bibinfo {author} {\bibfnamefont {Roman}\ \bibnamefont {Schnabel}},\ }\bibfield  {title} {\enquote {\bibinfo {title} {Quantum enhancement of the zero-area sagnac interferometer topology for gravitational wave detection},}\ }\href {\doibase 10.1103/PhysRevLett.104.251102} {\bibfield  {journal} {\bibinfo  {journal} {Phys. Rev. Lett.}\ }\textbf {\bibinfo {volume} {104}},\ \bibinfo {pages} {251102} (\bibinfo {year} {2010})}\BibitemShut {NoStop}%
\bibitem [{\citenamefont {Helstrom}(1969)}]{Helstrom1969}%
  \BibitemOpen
  \bibfield  {author} {\bibinfo {author} {\bibfnamefont {Carl~W.}\ \bibnamefont {Helstrom}},\ }\bibfield  {title} {\enquote {\bibinfo {title} {Quantum detection and estimation theory},}\ }\href {\doibase 10.1007/BF01007479} {\bibfield  {journal} {\bibinfo  {journal} {J. Stat. Phys.}\ }\textbf {\bibinfo {volume} {1}},\ \bibinfo {pages} {231--252} (\bibinfo {year} {1969})}\BibitemShut {NoStop}%
\bibitem [{\citenamefont {Paris}(2009)}]{Paris2009}%
  \BibitemOpen
  \bibfield  {author} {\bibinfo {author} {\bibfnamefont {Matteo G.~A.}\ \bibnamefont {Paris}},\ }\bibfield  {title} {\enquote {\bibinfo {title} {Quantum estimation for quantum technology},}\ }\href {\doibase 10.1142/S0219749909004839} {\bibfield  {journal} {\bibinfo  {journal} {Int. J. Quantum Inf}\ }\textbf {\bibinfo {volume} {07}},\ \bibinfo {pages} {125--137} (\bibinfo {year} {2009})}\BibitemShut {NoStop}%
\bibitem [{\citenamefont {Agarwal}(2012)}]{Agarwal_2012}%
  \BibitemOpen
  \bibfield  {author} {\bibinfo {author} {\bibfnamefont {Girish~S.}\ \bibnamefont {Agarwal}},\ }\href@noop {} {\emph {\bibinfo {title} {Quantum Optics}}}\ (\bibinfo  {publisher} {Cambridge University Press},\ \bibinfo {year} {2012})\BibitemShut {NoStop}%
\bibitem [{\citenamefont {Scully}\ and\ \citenamefont {Zubairy}(1997)}]{Scully_Zubairy_1997}%
  \BibitemOpen
  \bibfield  {author} {\bibinfo {author} {\bibfnamefont {Marlan~O.}\ \bibnamefont {Scully}}\ and\ \bibinfo {author} {\bibfnamefont {M.~Suhail}\ \bibnamefont {Zubairy}},\ }\href@noop {} {\emph {\bibinfo {title} {Quantum Optics}}}\ (\bibinfo  {publisher} {Cambridge University Press},\ \bibinfo {year} {1997})\BibitemShut {NoStop}%
\bibitem [{\citenamefont {Holevo}\ and\ \citenamefont {Werner}(2001)}]{PhysRevA.63.032312}%
  \BibitemOpen
  \bibfield  {author} {\bibinfo {author} {\bibfnamefont {A.~S.}\ \bibnamefont {Holevo}}\ and\ \bibinfo {author} {\bibfnamefont {R.~F.}\ \bibnamefont {Werner}},\ }\bibfield  {title} {\enquote {\bibinfo {title} {Evaluating capacities of bosonic gaussian channels},}\ }\href {\doibase 10.1103/PhysRevA.63.032312} {\bibfield  {journal} {\bibinfo  {journal} {Phys. Rev. A}\ }\textbf {\bibinfo {volume} {63}},\ \bibinfo {pages} {032312} (\bibinfo {year} {2001})}\BibitemShut {NoStop}%
\bibitem [{\citenamefont {Lindblad}(2000)}]{Göran}%
  \BibitemOpen
  \bibfield  {author} {\bibinfo {author} {\bibfnamefont {Göran}\ \bibnamefont {Lindblad}},\ }\bibfield  {title} {\enquote {\bibinfo {title} {Cloning the quantum oscillator},}\ }\href {\doibase 10.1088/0305-4470/33/28/310} {\bibfield  {journal} {\bibinfo  {journal} {J. Phys. A: Math. Gen.}\ }\textbf {\bibinfo {volume} {33}},\ \bibinfo {pages} {5059} (\bibinfo {year} {2000})}\BibitemShut {NoStop}%
\bibitem [{\citenamefont {Pinel}\ \emph {et~al.}(2013)\citenamefont {Pinel}, \citenamefont {Jian}, \citenamefont {Treps}, \citenamefont {Fabre},\ and\ \citenamefont {Braun}}]{PhysRevA.88.040102}%
  \BibitemOpen
  \bibfield  {author} {\bibinfo {author} {\bibfnamefont {O.}~\bibnamefont {Pinel}}, \bibinfo {author} {\bibfnamefont {P.}~\bibnamefont {Jian}}, \bibinfo {author} {\bibfnamefont {N.}~\bibnamefont {Treps}}, \bibinfo {author} {\bibfnamefont {C.}~\bibnamefont {Fabre}}, \ and\ \bibinfo {author} {\bibfnamefont {D.}~\bibnamefont {Braun}},\ }\bibfield  {title} {\enquote {\bibinfo {title} {Quantum parameter estimation using general single-mode gaussian states},}\ }\href {\doibase 10.1103/PhysRevA.88.040102} {\bibfield  {journal} {\bibinfo  {journal} {Phys. Rev. A}\ }\textbf {\bibinfo {volume} {88}},\ \bibinfo {pages} {040102} (\bibinfo {year} {2013})}\BibitemShut {NoStop}%
\bibitem [{\citenamefont {Breuer}\ and\ \citenamefont {Petruccione}(2007)}]{Breuer}%
  \BibitemOpen
  \bibfield  {author} {\bibinfo {author} {\bibfnamefont {Heinz-Peter}\ \bibnamefont {Breuer}}\ and\ \bibinfo {author} {\bibfnamefont {Francesco}\ \bibnamefont {Petruccione}},\ }\href {\doibase 10.1093/acprof:oso/9780199213900.001.0001} {\emph {\bibinfo {title} {The Theory of Open Quantum Systems}}}\ (\bibinfo  {publisher} {Oxford University Press, Oxford},\ \bibinfo {year} {2007})\BibitemShut {NoStop}%
\bibitem [{\citenamefont {Seifoory}\ \emph {et~al.}(2017)\citenamefont {Seifoory}, \citenamefont {Doutre}, \citenamefont {Dignam},\ and\ \citenamefont {Sipe}}]{Seifoory:17}%
  \BibitemOpen
  \bibfield  {author} {\bibinfo {author} {\bibfnamefont {Hossein}\ \bibnamefont {Seifoory}}, \bibinfo {author} {\bibfnamefont {Sean}\ \bibnamefont {Doutre}}, \bibinfo {author} {\bibfnamefont {Marc.~M.}\ \bibnamefont {Dignam}}, \ and\ \bibinfo {author} {\bibfnamefont {J.~E.}\ \bibnamefont {Sipe}},\ }\bibfield  {title} {\enquote {\bibinfo {title} {Squeezed thermal states: the result of parametric down conversion in lossy cavities},}\ }\href {\doibase 10.1364/JOSAB.34.001587} {\bibfield  {journal} {\bibinfo  {journal} {J. Opt. Soc. Am. B}\ }\textbf {\bibinfo {volume} {34}},\ \bibinfo {pages} {1587--1596} (\bibinfo {year} {2017})}\BibitemShut {NoStop}%
\bibitem [{\citenamefont {Arzani}\ \emph {et~al.}(2017)\citenamefont {Arzani}, \citenamefont {Treps},\ and\ \citenamefont {Ferrini}}]{PhysRevA.95.052352}%
  \BibitemOpen
  \bibfield  {author} {\bibinfo {author} {\bibfnamefont {Francesco}\ \bibnamefont {Arzani}}, \bibinfo {author} {\bibfnamefont {Nicolas}\ \bibnamefont {Treps}}, \ and\ \bibinfo {author} {\bibfnamefont {Giulia}\ \bibnamefont {Ferrini}},\ }\bibfield  {title} {\enquote {\bibinfo {title} {Polynomial approximation of non-gaussian unitaries by counting one photon at a time},}\ }\href {\doibase 10.1103/PhysRevA.95.052352} {\bibfield  {journal} {\bibinfo  {journal} {Phys. Rev. A}\ }\textbf {\bibinfo {volume} {95}},\ \bibinfo {pages} {052352} (\bibinfo {year} {2017})}\BibitemShut {NoStop}%
\bibitem [{\citenamefont {Walschaers}(2021)}]{PRXQuantum.2.030204}%
  \BibitemOpen
  \bibfield  {author} {\bibinfo {author} {\bibfnamefont {Mattia}\ \bibnamefont {Walschaers}},\ }\bibfield  {title} {\enquote {\bibinfo {title} {Non-gaussian quantum states and where to find them},}\ }\href {\doibase 10.1103/PRXQuantum.2.030204} {\bibfield  {journal} {\bibinfo  {journal} {PRX Quantum}\ }\textbf {\bibinfo {volume} {2}},\ \bibinfo {pages} {030204} (\bibinfo {year} {2021})}\BibitemShut {NoStop}%
\bibitem [{\citenamefont {Gottesman}\ \emph {et~al.}(2001)\citenamefont {Gottesman}, \citenamefont {Kitaev},\ and\ \citenamefont {Preskill}}]{PhysRevA.64.012310}%
  \BibitemOpen
  \bibfield  {author} {\bibinfo {author} {\bibfnamefont {Daniel}\ \bibnamefont {Gottesman}}, \bibinfo {author} {\bibfnamefont {Alexei}\ \bibnamefont {Kitaev}}, \ and\ \bibinfo {author} {\bibfnamefont {John}\ \bibnamefont {Preskill}},\ }\bibfield  {title} {\enquote {\bibinfo {title} {Encoding a qubit in an oscillator},}\ }\href {\doibase 10.1103/PhysRevA.64.012310} {\bibfield  {journal} {\bibinfo  {journal} {Phys. Rev. A}\ }\textbf {\bibinfo {volume} {64}},\ \bibinfo {pages} {012310} (\bibinfo {year} {2001})}\BibitemShut {NoStop}%
\bibitem [{\citenamefont {Monroe}\ \emph {et~al.}(1996)\citenamefont {Monroe}, \citenamefont {Meekhof}, \citenamefont {King},\ and\ \citenamefont {Wineland}}]{science.272.5265.1131}%
  \BibitemOpen
  \bibfield  {author} {\bibinfo {author} {\bibfnamefont {C.}~\bibnamefont {Monroe}}, \bibinfo {author} {\bibfnamefont {D.~M.}\ \bibnamefont {Meekhof}}, \bibinfo {author} {\bibfnamefont {B.~E.}\ \bibnamefont {King}}, \ and\ \bibinfo {author} {\bibfnamefont {D.~J.}\ \bibnamefont {Wineland}},\ }\bibfield  {title} {\enquote {\bibinfo {title} {A “schrödinger cat” superposition state of an atom},}\ }\href {\doibase 10.1126/science.272.5265.1131} {\bibfield  {journal} {\bibinfo  {journal} {Science}\ }\textbf {\bibinfo {volume} {272}},\ \bibinfo {pages} {1131--1136} (\bibinfo {year} {1996})}\BibitemShut {NoStop}%
\bibitem [{\citenamefont {Brune}\ \emph {et~al.}(1996)\citenamefont {Brune}, \citenamefont {Hagley}, \citenamefont {Dreyer}, \citenamefont {Ma\^{\i}tre}, \citenamefont {Maali}, \citenamefont {Wunderlich}, \citenamefont {Raimond},\ and\ \citenamefont {Haroche}}]{PhysRevLett.77.4887}%
  \BibitemOpen
  \bibfield  {author} {\bibinfo {author} {\bibfnamefont {M.}~\bibnamefont {Brune}}, \bibinfo {author} {\bibfnamefont {E.}~\bibnamefont {Hagley}}, \bibinfo {author} {\bibfnamefont {J.}~\bibnamefont {Dreyer}}, \bibinfo {author} {\bibfnamefont {X.}~\bibnamefont {Ma\^{\i}tre}}, \bibinfo {author} {\bibfnamefont {A.}~\bibnamefont {Maali}}, \bibinfo {author} {\bibfnamefont {C.}~\bibnamefont {Wunderlich}}, \bibinfo {author} {\bibfnamefont {J.~M.}\ \bibnamefont {Raimond}}, \ and\ \bibinfo {author} {\bibfnamefont {S.}~\bibnamefont {Haroche}},\ }\bibfield  {title} {\enquote {\bibinfo {title} {Observing the progressive decoherence of the ``meter'' in a quantum measurement},}\ }\href {\doibase 10.1103/PhysRevLett.77.4887} {\bibfield  {journal} {\bibinfo  {journal} {Phys. Rev. Lett.}\ }\textbf {\bibinfo {volume} {77}},\ \bibinfo {pages} {4887--4890} (\bibinfo {year} {1996})}\BibitemShut {NoStop}%
\bibitem [{\citenamefont {Gao}\ \emph {et~al.}(2010)\citenamefont {Gao}, \citenamefont {Lu}, \citenamefont {Yao}, \citenamefont {Xu}, \citenamefont {G{\"u}hne}, \citenamefont {Goebel}, \citenamefont {Chen}, \citenamefont {Peng}, \citenamefont {Chen},\ and\ \citenamefont {Pan}}]{Gao2010}%
  \BibitemOpen
  \bibfield  {author} {\bibinfo {author} {\bibfnamefont {Wei-Bo}\ \bibnamefont {Gao}}, \bibinfo {author} {\bibfnamefont {Chao-Yang}\ \bibnamefont {Lu}}, \bibinfo {author} {\bibfnamefont {Xing-Can}\ \bibnamefont {Yao}}, \bibinfo {author} {\bibfnamefont {Ping}\ \bibnamefont {Xu}}, \bibinfo {author} {\bibfnamefont {Otfried}\ \bibnamefont {G{\"u}hne}}, \bibinfo {author} {\bibfnamefont {Alexander}\ \bibnamefont {Goebel}}, \bibinfo {author} {\bibfnamefont {Yu-Ao}\ \bibnamefont {Chen}}, \bibinfo {author} {\bibfnamefont {Cheng-Zhi}\ \bibnamefont {Peng}}, \bibinfo {author} {\bibfnamefont {Zeng-Bing}\ \bibnamefont {Chen}}, \ and\ \bibinfo {author} {\bibfnamefont {Jian-Wei}\ \bibnamefont {Pan}},\ }\bibfield  {title} {\enquote {\bibinfo {title} {Experimental demonstration of a hyper-entangled ten-qubit schr{\"o}dinger cat state},}\ }\href {\doibase 10.1038/nphys1603} {\bibfield  {journal} {\bibinfo  {journal} {Nat. Phys.}\ }\textbf {\bibinfo {volume} {6}},\ \bibinfo {pages} {331--335} (\bibinfo {year}
  {2010})}\BibitemShut {NoStop}%
\bibitem [{\citenamefont {Bild}\ \emph {et~al.}(2023)\citenamefont {Bild}, \citenamefont {Fadel}, \citenamefont {Yang}, \citenamefont {von Lüpke}, \citenamefont {Martin}, \citenamefont {Bruno},\ and\ \citenamefont {Chu}}]{science.adf7553}%
  \BibitemOpen
  \bibfield  {author} {\bibinfo {author} {\bibfnamefont {Marius}\ \bibnamefont {Bild}}, \bibinfo {author} {\bibfnamefont {Matteo}\ \bibnamefont {Fadel}}, \bibinfo {author} {\bibfnamefont {Yu}~\bibnamefont {Yang}}, \bibinfo {author} {\bibfnamefont {Uwe}\ \bibnamefont {von Lüpke}}, \bibinfo {author} {\bibfnamefont {Phillip}\ \bibnamefont {Martin}}, \bibinfo {author} {\bibfnamefont {Alessandro}\ \bibnamefont {Bruno}}, \ and\ \bibinfo {author} {\bibfnamefont {Yiwen}\ \bibnamefont {Chu}},\ }\bibfield  {title} {\enquote {\bibinfo {title} {Schrödinger cat states of a 16-microgram mechanical oscillator},}\ }\href {\doibase 10.1126/science.adf7553} {\bibfield  {journal} {\bibinfo  {journal} {Science}\ }\textbf {\bibinfo {volume} {380}},\ \bibinfo {pages} {274--278} (\bibinfo {year} {2023})}\BibitemShut {NoStop}%
\bibitem [{\citenamefont {Zhang}\ \emph {et~al.}(2022)\citenamefont {Zhang}, \citenamefont {Yu}, \citenamefont {Yuan}, \citenamefont {Wang}, \citenamefont {Demkowicz-Dobrza\ifmmode~\acute{n}\else \'{n}\fi{}ski},\ and\ \citenamefont {Liu}}]{PhysRevResearch.4.043057}%
  \BibitemOpen
  \bibfield  {author} {\bibinfo {author} {\bibfnamefont {Mao}\ \bibnamefont {Zhang}}, \bibinfo {author} {\bibfnamefont {Huai-Ming}\ \bibnamefont {Yu}}, \bibinfo {author} {\bibfnamefont {Haidong}\ \bibnamefont {Yuan}}, \bibinfo {author} {\bibfnamefont {Xiaoguang}\ \bibnamefont {Wang}}, \bibinfo {author} {\bibfnamefont {Rafa\l{}}\ \bibnamefont {Demkowicz-Dobrza\ifmmode~\acute{n}\else \'{n}\fi{}ski}}, \ and\ \bibinfo {author} {\bibfnamefont {Jing}\ \bibnamefont {Liu}},\ }\bibfield  {title} {\enquote {\bibinfo {title} {Quanestimation: An open-source toolkit for quantum parameter estimation},}\ }\href {\doibase 10.1103/PhysRevResearch.4.043057} {\bibfield  {journal} {\bibinfo  {journal} {Phys. Rev. Res.}\ }\textbf {\bibinfo {volume} {4}},\ \bibinfo {pages} {043057} (\bibinfo {year} {2022})}\BibitemShut {NoStop}%
\bibitem [{\citenamefont {Braunstein}\ and\ \citenamefont {van Loock}(2005)}]{RevModPhys.77.513}%
  \BibitemOpen
  \bibfield  {author} {\bibinfo {author} {\bibfnamefont {Samuel~L.}\ \bibnamefont {Braunstein}}\ and\ \bibinfo {author} {\bibfnamefont {Peter}\ \bibnamefont {van Loock}},\ }\bibfield  {title} {\enquote {\bibinfo {title} {Quantum information with continuous variables},}\ }\href {\doibase 10.1103/RevModPhys.77.513} {\bibfield  {journal} {\bibinfo  {journal} {Rev. Mod. Phys.}\ }\textbf {\bibinfo {volume} {77}},\ \bibinfo {pages} {513--577} (\bibinfo {year} {2005})}\BibitemShut {NoStop}%
\bibitem [{\citenamefont {Lvovsky}\ and\ \citenamefont {Raymer}(2009)}]{RevModPhys.81.299}%
  \BibitemOpen
  \bibfield  {author} {\bibinfo {author} {\bibfnamefont {A.~I.}\ \bibnamefont {Lvovsky}}\ and\ \bibinfo {author} {\bibfnamefont {M.~G.}\ \bibnamefont {Raymer}},\ }\bibfield  {title} {\enquote {\bibinfo {title} {Continuous-variable optical quantum-state tomography},}\ }\href {\doibase 10.1103/RevModPhys.81.299} {\bibfield  {journal} {\bibinfo  {journal} {Rev. Mod. Phys.}\ }\textbf {\bibinfo {volume} {81}},\ \bibinfo {pages} {299--332} (\bibinfo {year} {2009})}\BibitemShut {NoStop}%
\bibitem [{\citenamefont {Munro}\ \emph {et~al.}(2002)\citenamefont {Munro}, \citenamefont {Nemoto}, \citenamefont {Milburn},\ and\ \citenamefont {Braunstein}}]{PhysRevA.66.023819}%
  \BibitemOpen
  \bibfield  {author} {\bibinfo {author} {\bibfnamefont {W.~J.}\ \bibnamefont {Munro}}, \bibinfo {author} {\bibfnamefont {K.}~\bibnamefont {Nemoto}}, \bibinfo {author} {\bibfnamefont {G.~J.}\ \bibnamefont {Milburn}}, \ and\ \bibinfo {author} {\bibfnamefont {S.~L.}\ \bibnamefont {Braunstein}},\ }\bibfield  {title} {\enquote {\bibinfo {title} {Weak-force detection with superposed coherent states},}\ }\href {\doibase 10.1103/PhysRevA.66.023819} {\bibfield  {journal} {\bibinfo  {journal} {Phys. Rev. A}\ }\textbf {\bibinfo {volume} {66}},\ \bibinfo {pages} {023819} (\bibinfo {year} {2002})}\BibitemShut {NoStop}%
\bibitem [{\citenamefont {Sanders}(2012)}]{Sanders_2012}%
  \BibitemOpen
  \bibfield  {author} {\bibinfo {author} {\bibfnamefont {Barry~C}\ \bibnamefont {Sanders}},\ }\bibfield  {title} {\enquote {\bibinfo {title} {Review of entangled coherent states},}\ }\href {\doibase 10.1088/1751-8113/45/24/244002} {\bibfield  {journal} {\bibinfo  {journal} {J. Phys. A: Math. Theor.}\ }\textbf {\bibinfo {volume} {45}},\ \bibinfo {pages} {244002} (\bibinfo {year} {2012})}\BibitemShut {NoStop}%
\bibitem [{\citenamefont {Ourjoumtsev}\ \emph {et~al.}(2007)\citenamefont {Ourjoumtsev}, \citenamefont {Jeong}, \citenamefont {Tualle-Brouri},\ and\ \citenamefont {Grangier}}]{Ourjoumtsev2007}%
  \BibitemOpen
  \bibfield  {author} {\bibinfo {author} {\bibfnamefont {Alexei}\ \bibnamefont {Ourjoumtsev}}, \bibinfo {author} {\bibfnamefont {Hyunseok}\ \bibnamefont {Jeong}}, \bibinfo {author} {\bibfnamefont {Rosa}\ \bibnamefont {Tualle-Brouri}}, \ and\ \bibinfo {author} {\bibfnamefont {Philippe}\ \bibnamefont {Grangier}},\ }\bibfield  {title} {\enquote {\bibinfo {title} {Generation of optical `schr{\"o}dinger cats' from photon number states},}\ }\href {\doibase 10.1038/nature06054} {\bibfield  {journal} {\bibinfo  {journal} {Nature}\ }\textbf {\bibinfo {volume} {448}},\ \bibinfo {pages} {784--786} (\bibinfo {year} {2007})}\BibitemShut {NoStop}%
\bibitem [{\citenamefont {Dowling}(2008)}]{Dowling2008}%
  \BibitemOpen
  \bibfield  {author} {\bibinfo {author} {\bibfnamefont {Jonathan~P.}\ \bibnamefont {Dowling}},\ }\bibfield  {title} {\enquote {\bibinfo {title} {Quantum optical metrology: The lowdown on high-n00n states},}\ }\href {\doibase 10.1080/00107510802091298} {\bibfield  {journal} {\bibinfo  {journal} {Contemp. Phys.}\ }\textbf {\bibinfo {volume} {49}},\ \bibinfo {pages} {125--143} (\bibinfo {year} {2008})}\BibitemShut {NoStop}%
\bibitem [{\citenamefont {Ono}\ \emph {et~al.}(2013)\citenamefont {Ono}, \citenamefont {Okamoto},\ and\ \citenamefont {Takeuchi}}]{Ono2013}%
  \BibitemOpen
  \bibfield  {author} {\bibinfo {author} {\bibfnamefont {Takafumi}\ \bibnamefont {Ono}}, \bibinfo {author} {\bibfnamefont {Ryo}\ \bibnamefont {Okamoto}}, \ and\ \bibinfo {author} {\bibfnamefont {Shigeki}\ \bibnamefont {Takeuchi}},\ }\bibfield  {title} {\enquote {\bibinfo {title} {An entanglement-enhanced microscope},}\ }\href {\doibase 10.1038/ncomms3426} {\bibfield  {journal} {\bibinfo  {journal} {Nat. Commun.}\ }\textbf {\bibinfo {volume} {4}},\ \bibinfo {pages} {2426} (\bibinfo {year} {2013})}\BibitemShut {NoStop}%
\bibitem [{\citenamefont {Wolfgramm}\ \emph {et~al.}(2013)\citenamefont {Wolfgramm}, \citenamefont {Vitelli}, \citenamefont {Beduini}, \citenamefont {Godbout},\ and\ \citenamefont {Mitchell}}]{Wolfgramm2013}%
  \BibitemOpen
  \bibfield  {author} {\bibinfo {author} {\bibfnamefont {Florian}\ \bibnamefont {Wolfgramm}}, \bibinfo {author} {\bibfnamefont {Chiara}\ \bibnamefont {Vitelli}}, \bibinfo {author} {\bibfnamefont {Federica~A.}\ \bibnamefont {Beduini}}, \bibinfo {author} {\bibfnamefont {Nicolas}\ \bibnamefont {Godbout}}, \ and\ \bibinfo {author} {\bibfnamefont {Morgan~W.}\ \bibnamefont {Mitchell}},\ }\bibfield  {title} {\enquote {\bibinfo {title} {Entanglement-enhanced probing of a delicate material system},}\ }\href {\doibase 10.1038/nphoton.2012.300} {\bibfield  {journal} {\bibinfo  {journal} {Nat. Photonics}\ }\textbf {\bibinfo {volume} {7}},\ \bibinfo {pages} {28--32} (\bibinfo {year} {2013})}\BibitemShut {NoStop}%
\bibitem [{\citenamefont {D'Ambrosio}\ \emph {et~al.}(2013)\citenamefont {D'Ambrosio}, \citenamefont {Spagnolo}, \citenamefont {Del~Re}, \citenamefont {Slussarenko}, \citenamefont {Li}, \citenamefont {Kwek}, \citenamefont {Marrucci}, \citenamefont {Walborn}, \citenamefont {Aolita},\ and\ \citenamefont {Sciarrino}}]{D'Ambrosio2013}%
  \BibitemOpen
  \bibfield  {author} {\bibinfo {author} {\bibfnamefont {Vincenzo}\ \bibnamefont {D'Ambrosio}}, \bibinfo {author} {\bibfnamefont {Nicol{\`o}}\ \bibnamefont {Spagnolo}}, \bibinfo {author} {\bibfnamefont {Lorenzo}\ \bibnamefont {Del~Re}}, \bibinfo {author} {\bibfnamefont {Sergei}\ \bibnamefont {Slussarenko}}, \bibinfo {author} {\bibfnamefont {Ying}\ \bibnamefont {Li}}, \bibinfo {author} {\bibfnamefont {Leong~Chuan}\ \bibnamefont {Kwek}}, \bibinfo {author} {\bibfnamefont {Lorenzo}\ \bibnamefont {Marrucci}}, \bibinfo {author} {\bibfnamefont {Stephen~P.}\ \bibnamefont {Walborn}}, \bibinfo {author} {\bibfnamefont {Leandro}\ \bibnamefont {Aolita}}, \ and\ \bibinfo {author} {\bibfnamefont {Fabio}\ \bibnamefont {Sciarrino}},\ }\bibfield  {title} {\enquote {\bibinfo {title} {Photonic polarization gears for ultra-sensitive angular measurements},}\ }\href {\doibase 10.1038/ncomms3432} {\bibfield  {journal} {\bibinfo  {journal} {Nat. Commun.}\ }\textbf {\bibinfo {volume} {4}},\ \bibinfo {pages} {2432} (\bibinfo {year}
  {2013})}\BibitemShut {NoStop}%
\bibitem [{\citenamefont {Kolkiran}(2019)}]{Kolkiran2019}%
  \BibitemOpen
  \bibfield  {author} {\bibinfo {author} {\bibfnamefont {Aziz}\ \bibnamefont {Kolkiran}},\ }\bibfield  {title} {\enquote {\bibinfo {title} {High-noon states with high flux of photons using coherent beam stimulated noncollinear parametric down conversion},}\ }\href {\doibase https://doi.org/10.1155/2019/6871979} {\bibfield  {journal} {\bibinfo  {journal} {Int. J. Opt.}\ }\textbf {\bibinfo {volume} {2019}},\ \bibinfo {pages} {6871979} (\bibinfo {year} {2019})}\BibitemShut {NoStop}%
\bibitem [{\citenamefont {Mitchison}\ \emph {et~al.}(2020)\citenamefont {Mitchison}, \citenamefont {Fogarty}, \citenamefont {Guarnieri}, \citenamefont {Campbell}, \citenamefont {Busch},\ and\ \citenamefont {Goold}}]{PhysRevLett.125.080402}%
  \BibitemOpen
  \bibfield  {author} {\bibinfo {author} {\bibfnamefont {Mark~T.}\ \bibnamefont {Mitchison}}, \bibinfo {author} {\bibfnamefont {Thom\'as}\ \bibnamefont {Fogarty}}, \bibinfo {author} {\bibfnamefont {Giacomo}\ \bibnamefont {Guarnieri}}, \bibinfo {author} {\bibfnamefont {Steve}\ \bibnamefont {Campbell}}, \bibinfo {author} {\bibfnamefont {Thomas}\ \bibnamefont {Busch}}, \ and\ \bibinfo {author} {\bibfnamefont {John}\ \bibnamefont {Goold}},\ }\bibfield  {title} {\enquote {\bibinfo {title} {In situ thermometry of a cold fermi gas via dephasing impurities},}\ }\href {\doibase 10.1103/PhysRevLett.125.080402} {\bibfield  {journal} {\bibinfo  {journal} {Phys. Rev. Lett.}\ }\textbf {\bibinfo {volume} {125}},\ \bibinfo {pages} {080402} (\bibinfo {year} {2020})}\BibitemShut {NoStop}%
\bibitem [{\citenamefont {Khan}\ \emph {et~al.}(2022)\citenamefont {Khan}, \citenamefont {Mehboudi}, \citenamefont {Ter\ifmmode~\mbox{\c{c}}\else \c{c}\fi{}as}, \citenamefont {Lewenstein},\ and\ \citenamefont {Garcia-March}}]{PhysRevResearch.4.023191}%
  \BibitemOpen
  \bibfield  {author} {\bibinfo {author} {\bibfnamefont {Muhammad~Miskeen}\ \bibnamefont {Khan}}, \bibinfo {author} {\bibfnamefont {Mohammad}\ \bibnamefont {Mehboudi}}, \bibinfo {author} {\bibfnamefont {Hugo}\ \bibnamefont {Ter\ifmmode~\mbox{\c{c}}\else \c{c}\fi{}as}}, \bibinfo {author} {\bibfnamefont {Maciej}\ \bibnamefont {Lewenstein}}, \ and\ \bibinfo {author} {\bibfnamefont {Miguel~Angel}\ \bibnamefont {Garcia-March}},\ }\bibfield  {title} {\enquote {\bibinfo {title} {Subnanokelvin thermometry of an interacting $d$-dimensional homogeneous bose gas},}\ }\href {\doibase 10.1103/PhysRevResearch.4.023191} {\bibfield  {journal} {\bibinfo  {journal} {Phys. Rev. Res.}\ }\textbf {\bibinfo {volume} {4}},\ \bibinfo {pages} {023191} (\bibinfo {year} {2022})}\BibitemShut {NoStop}%
\bibitem [{\citenamefont {Bastin}\ \emph {et~al.}(2006)\citenamefont {Bastin}, \citenamefont {von Zanthier},\ and\ \citenamefont {Solano}}]{Bastin_2006}%
  \BibitemOpen
  \bibfield  {author} {\bibinfo {author} {\bibfnamefont {T}~\bibnamefont {Bastin}}, \bibinfo {author} {\bibfnamefont {J}~\bibnamefont {von Zanthier}}, \ and\ \bibinfo {author} {\bibfnamefont {E}~\bibnamefont {Solano}},\ }\bibfield  {title} {\enquote {\bibinfo {title} {Measure of phonon-number moments and motional quadratures through infinitesimal-time probing of trapped ions},}\ }\href {\doibase 10.1088/0953-4075/39/3/020} {\bibfield  {journal} {\bibinfo  {journal} {J. Phys. B: At. Mol. Opt. Phys.}\ }\textbf {\bibinfo {volume} {39}},\ \bibinfo {pages} {685} (\bibinfo {year} {2006})}\BibitemShut {NoStop}%
\bibitem [{\citenamefont {D'Helon}\ and\ \citenamefont {Milburn}(1996)}]{PhysRevA.54.R25}%
  \BibitemOpen
  \bibfield  {author} {\bibinfo {author} {\bibfnamefont {C.}~\bibnamefont {D'Helon}}\ and\ \bibinfo {author} {\bibfnamefont {G.~J.}\ \bibnamefont {Milburn}},\ }\bibfield  {title} {\enquote {\bibinfo {title} {Reconstructing the vibrational state of a trapped ion},}\ }\href {\doibase 10.1103/PhysRevA.54.R25} {\bibfield  {journal} {\bibinfo  {journal} {Phys. Rev. A}\ }\textbf {\bibinfo {volume} {54}},\ \bibinfo {pages} {R25--R28} (\bibinfo {year} {1996})}\BibitemShut {NoStop}%
\bibitem [{\citenamefont {Poyatos}\ \emph {et~al.}(1996)\citenamefont {Poyatos}, \citenamefont {Walser}, \citenamefont {Cirac}, \citenamefont {Zoller},\ and\ \citenamefont {Blatt}}]{PhysRevA.53.R1966}%
  \BibitemOpen
  \bibfield  {author} {\bibinfo {author} {\bibfnamefont {J.~F.}\ \bibnamefont {Poyatos}}, \bibinfo {author} {\bibfnamefont {R.}~\bibnamefont {Walser}}, \bibinfo {author} {\bibfnamefont {J.~I.}\ \bibnamefont {Cirac}}, \bibinfo {author} {\bibfnamefont {P.}~\bibnamefont {Zoller}}, \ and\ \bibinfo {author} {\bibfnamefont {R.}~\bibnamefont {Blatt}},\ }\bibfield  {title} {\enquote {\bibinfo {title} {Motion tomography of a single trapped ion},}\ }\href {\doibase 10.1103/PhysRevA.53.R1966} {\bibfield  {journal} {\bibinfo  {journal} {Phys. Rev. A}\ }\textbf {\bibinfo {volume} {53}},\ \bibinfo {pages} {R1966--R1969} (\bibinfo {year} {1996})}\BibitemShut {NoStop}%
\bibitem [{\citenamefont {Glatthard}\ \emph {et~al.}(2023)\citenamefont {Glatthard}, \citenamefont {Hovhannisyan}, \citenamefont {Perarnau-Llobet}, \citenamefont {Correa},\ and\ \citenamefont {Miller}}]{Glatthard2023energymeasurements}%
  \BibitemOpen
  \bibfield  {author} {\bibinfo {author} {\bibfnamefont {Jonas}\ \bibnamefont {Glatthard}}, \bibinfo {author} {\bibfnamefont {Karen~V.}\ \bibnamefont {Hovhannisyan}}, \bibinfo {author} {\bibfnamefont {Mart{\'{i}}}\ \bibnamefont {Perarnau-Llobet}}, \bibinfo {author} {\bibfnamefont {Luis~A.}\ \bibnamefont {Correa}}, \ and\ \bibinfo {author} {\bibfnamefont {Harry J.~D.}\ \bibnamefont {Miller}},\ }\bibfield  {title} {\enquote {\bibinfo {title} {Energy measurements remain thermometrically optimal beyond weak coupling},}\ }\href {\doibase 10.22331/q-2023-11-28-1190} {\bibfield  {journal} {\bibinfo  {journal} {{Quantum}}\ }\textbf {\bibinfo {volume} {7}},\ \bibinfo {pages} {1190} (\bibinfo {year} {2023})}\BibitemShut {NoStop}%
\bibitem [{\citenamefont {Kues}\ \emph {et~al.}(2017)\citenamefont {Kues}, \citenamefont {Reimer}, \citenamefont {Roztocki}, \citenamefont {Cort{\'e}s}, \citenamefont {Sciara}, \citenamefont {Wetzel}, \citenamefont {Zhang}, \citenamefont {Cino}, \citenamefont {Chu}, \citenamefont {Little}, \citenamefont {Moss}, \citenamefont {Caspani}, \citenamefont {Aza{\~{n}}a},\ and\ \citenamefont {Morandotti}}]{Kues2017}%
  \BibitemOpen
  \bibfield  {author} {\bibinfo {author} {\bibfnamefont {Michael}\ \bibnamefont {Kues}}, \bibinfo {author} {\bibfnamefont {Christian}\ \bibnamefont {Reimer}}, \bibinfo {author} {\bibfnamefont {Piotr}\ \bibnamefont {Roztocki}}, \bibinfo {author} {\bibfnamefont {Luis~Romero}\ \bibnamefont {Cort{\'e}s}}, \bibinfo {author} {\bibfnamefont {Stefania}\ \bibnamefont {Sciara}}, \bibinfo {author} {\bibfnamefont {Benjamin}\ \bibnamefont {Wetzel}}, \bibinfo {author} {\bibfnamefont {Yanbing}\ \bibnamefont {Zhang}}, \bibinfo {author} {\bibfnamefont {Alfonso}\ \bibnamefont {Cino}}, \bibinfo {author} {\bibfnamefont {Sai~T.}\ \bibnamefont {Chu}}, \bibinfo {author} {\bibfnamefont {Brent~E.}\ \bibnamefont {Little}}, \bibinfo {author} {\bibfnamefont {David~J.}\ \bibnamefont {Moss}}, \bibinfo {author} {\bibfnamefont {Lucia}\ \bibnamefont {Caspani}}, \bibinfo {author} {\bibfnamefont {Jos{\'e}}\ \bibnamefont {Aza{\~{n}}a}}, \ and\ \bibinfo {author} {\bibfnamefont {Roberto}\ \bibnamefont {Morandotti}},\ }\bibfield  {title} {\enquote
  {\bibinfo {title} {On-chip generation of high-dimensional entangled quantum states and their coherent control},}\ }\href {\doibase 10.1038/nature22986} {\bibfield  {journal} {\bibinfo  {journal} {Nature}\ }\textbf {\bibinfo {volume} {546}},\ \bibinfo {pages} {622--626} (\bibinfo {year} {2017})}\BibitemShut {NoStop}%
\bibitem [{\citenamefont {Wang}\ \emph {et~al.}(2020)\citenamefont {Wang}, \citenamefont {Sciarrino}, \citenamefont {Laing},\ and\ \citenamefont {Thompson}}]{Wang2020}%
  \BibitemOpen
  \bibfield  {author} {\bibinfo {author} {\bibfnamefont {Jianwei}\ \bibnamefont {Wang}}, \bibinfo {author} {\bibfnamefont {Fabio}\ \bibnamefont {Sciarrino}}, \bibinfo {author} {\bibfnamefont {Anthony}\ \bibnamefont {Laing}}, \ and\ \bibinfo {author} {\bibfnamefont {Mark~G.}\ \bibnamefont {Thompson}},\ }\bibfield  {title} {\enquote {\bibinfo {title} {Integrated photonic quantum technologies},}\ }\href {\doibase 10.1038/s41566-019-0532-1} {\bibfield  {journal} {\bibinfo  {journal} {Nat. Photonics}\ }\textbf {\bibinfo {volume} {14}},\ \bibinfo {pages} {273--284} (\bibinfo {year} {2020})}\BibitemShut {NoStop}%
\bibitem [{\citenamefont {Zhong}\ \emph {et~al.}(2016)\citenamefont {Zhong}, \citenamefont {Xu}, \citenamefont {Wang}, \citenamefont {Song}, \citenamefont {Guo}, \citenamefont {Liu}, \citenamefont {Xu}, \citenamefont {Xia}, \citenamefont {Lu}, \citenamefont {Han}, \citenamefont {Pan},\ and\ \citenamefont {Wang}}]{PhysRevLett.117.110501}%
  \BibitemOpen
  \bibfield  {author} {\bibinfo {author} {\bibfnamefont {Y.~P.}\ \bibnamefont {Zhong}}, \bibinfo {author} {\bibfnamefont {D.}~\bibnamefont {Xu}}, \bibinfo {author} {\bibfnamefont {P.}~\bibnamefont {Wang}}, \bibinfo {author} {\bibfnamefont {C.}~\bibnamefont {Song}}, \bibinfo {author} {\bibfnamefont {Q.~J.}\ \bibnamefont {Guo}}, \bibinfo {author} {\bibfnamefont {W.~X.}\ \bibnamefont {Liu}}, \bibinfo {author} {\bibfnamefont {K.}~\bibnamefont {Xu}}, \bibinfo {author} {\bibfnamefont {B.~X.}\ \bibnamefont {Xia}}, \bibinfo {author} {\bibfnamefont {C.-Y.}\ \bibnamefont {Lu}}, \bibinfo {author} {\bibfnamefont {Siyuan}\ \bibnamefont {Han}}, \bibinfo {author} {\bibfnamefont {Jian-Wei}\ \bibnamefont {Pan}}, \ and\ \bibinfo {author} {\bibfnamefont {H.}~\bibnamefont {Wang}},\ }\bibfield  {title} {\enquote {\bibinfo {title} {Emulating anyonic fractional statistical behavior in a superconducting quantum circuit},}\ }\href {\doibase 10.1103/PhysRevLett.117.110501} {\bibfield  {journal} {\bibinfo  {journal} {Phys. Rev. Lett.}\
  }\textbf {\bibinfo {volume} {117}},\ \bibinfo {pages} {110501} (\bibinfo {year} {2016})}\BibitemShut {NoStop}%
\bibitem [{\citenamefont {Flurin}\ \emph {et~al.}(2012)\citenamefont {Flurin}, \citenamefont {Roch}, \citenamefont {Mallet}, \citenamefont {Devoret},\ and\ \citenamefont {Huard}}]{PhysRevLett.109.183901}%
  \BibitemOpen
  \bibfield  {author} {\bibinfo {author} {\bibfnamefont {E.}~\bibnamefont {Flurin}}, \bibinfo {author} {\bibfnamefont {N.}~\bibnamefont {Roch}}, \bibinfo {author} {\bibfnamefont {F.}~\bibnamefont {Mallet}}, \bibinfo {author} {\bibfnamefont {M.~H.}\ \bibnamefont {Devoret}}, \ and\ \bibinfo {author} {\bibfnamefont {B.}~\bibnamefont {Huard}},\ }\bibfield  {title} {\enquote {\bibinfo {title} {Generating entangled microwave radiation over two transmission lines},}\ }\href {\doibase 10.1103/PhysRevLett.109.183901} {\bibfield  {journal} {\bibinfo  {journal} {Phys. Rev. Lett.}\ }\textbf {\bibinfo {volume} {109}},\ \bibinfo {pages} {183901} (\bibinfo {year} {2012})}\BibitemShut {NoStop}%
\bibitem [{\citenamefont {Bracken}(2013)}]{Bracken_2013}%
  \BibitemOpen
  \bibfield  {author} {\bibinfo {author} {\bibfnamefont {Paul}\ \bibnamefont {Bracken}},\ }\href {\doibase 10.5772/50232} {\emph {\bibinfo {title} {Advances in Quantum Mechanics}}}\ (\bibinfo  {publisher} {IntechOpen},\ \bibinfo {address} {Rijeka},\ \bibinfo {year} {2013})\BibitemShut {NoStop}%
\bibitem [{\citenamefont {Grochowski}\ and\ \citenamefont {Filip}(2025)}]{grochowski2025optimal}%
  \BibitemOpen
  \bibfield  {author} {\bibinfo {author} {\bibfnamefont {Piotr~T.}\ \bibnamefont {Grochowski}}\ and\ \bibinfo {author} {\bibfnamefont {Radim}\ \bibnamefont {Filip}},\ }\href {https://arxiv.org/abs/2505.20832} {\enquote {\bibinfo {title} {Optimal phase-insensitive force sensing with non-gaussian states},}\ } (\bibinfo {year} {2025}),\ \Eprint {http://arxiv.org/abs/2505.20832} {arXiv:2505.20832} \BibitemShut {NoStop}%
\bibitem [{\citenamefont {Genoni}\ and\ \citenamefont {Paris}(2010)}]{PhysRevA.82.052341}%
  \BibitemOpen
  \bibfield  {author} {\bibinfo {author} {\bibfnamefont {Marco~G.}\ \bibnamefont {Genoni}}\ and\ \bibinfo {author} {\bibfnamefont {Matteo G.~A.}\ \bibnamefont {Paris}},\ }\bibfield  {title} {\enquote {\bibinfo {title} {Quantifying non-gaussianity for quantum information},}\ }\href {\doibase 10.1103/PhysRevA.82.052341} {\bibfield  {journal} {\bibinfo  {journal} {Phys. Rev. A}\ }\textbf {\bibinfo {volume} {82}},\ \bibinfo {pages} {052341} (\bibinfo {year} {2010})}\BibitemShut {NoStop}%
\bibitem [{\citenamefont {Avenhaus}\ \emph {et~al.}(2008)\citenamefont {Avenhaus}, \citenamefont {Coldenstrodt-Ronge}, \citenamefont {Laiho}, \citenamefont {Mauerer}, \citenamefont {Walmsley},\ and\ \citenamefont {Silberhorn}}]{PhysRevLett.101.053601}%
  \BibitemOpen
  \bibfield  {author} {\bibinfo {author} {\bibfnamefont {M.}~\bibnamefont {Avenhaus}}, \bibinfo {author} {\bibfnamefont {H.~B.}\ \bibnamefont {Coldenstrodt-Ronge}}, \bibinfo {author} {\bibfnamefont {K.}~\bibnamefont {Laiho}}, \bibinfo {author} {\bibfnamefont {W.}~\bibnamefont {Mauerer}}, \bibinfo {author} {\bibfnamefont {I.~A.}\ \bibnamefont {Walmsley}}, \ and\ \bibinfo {author} {\bibfnamefont {C.}~\bibnamefont {Silberhorn}},\ }\bibfield  {title} {\enquote {\bibinfo {title} {Photon number statistics of multimode parametric down-conversion},}\ }\href {\doibase 10.1103/PhysRevLett.101.053601} {\bibfield  {journal} {\bibinfo  {journal} {Phys. Rev. Lett.}\ }\textbf {\bibinfo {volume} {101}},\ \bibinfo {pages} {053601} (\bibinfo {year} {2008})}\BibitemShut {NoStop}%
\end{thebibliography}%
\end{document}